\documentclass{jfm}

\usepackage{graphicx}
\usepackage{natbib}

\usepackage{amsmath}
\usepackage{bm}
\usepackage{hyperref}
\usepackage{mathtools} 
\usepackage{xcolor}
 \hypersetup{
  colorlinks,
  linkcolor={blue},
  citecolor={blue},
  urlcolor={blue}
}
\usepackage{upgreek} 
\usepackage{verbatim} 
\usepackage{xspace} 

\newcommand\ba{\skew2\bar{a}}
\newcommand\bb{\skew2\bar{b}}
\newcommand\bc{\skew2\bar{c}}
\newcommand\bDe{\skew2\bar{\De}}

\newcommand\bg{\skew2\bar{g}}

\newcommand\bH{\skew4\bar{H}}

\newcommand\bh{\skew2\bar{h}}
\newcommand\bk{\skew2\bar{k}}
\newcommand\bLa{\skew4\bar{\La}}
\newcommand\bp{\skew2\bar{p}}
\newcommand\bP{\skew4\bar{P}}
\newcommand\bpsi{\skew4\bar{\psi}}
\newcommand\bPsi{\skew2\bar{\Psi}}
\newcommand\br{\skew2\bar{r}}

\newcommand\bx{\skew2\bar{x}}
\newcommand\bX{\skew3\bar{X}}
\newcommand\by{\skew2\bar{y}}
\newcommand\bV{\skew2\bar{V}}
\newcommand\bVk{\bar{\mcV}_k}
\newcommand\bW{\overline{\mcW}}

\newcommand\ha{\skew2\hat{a}}

\newcommand\hd{\skew4\hat{d}}
\newcommand\hDe{\skew2\hat{\De}}
\newcommand\hg{\skew2\hat{g}}
\newcommand\hh{\skew2\hat{h}}

\newcommand\hp{\skew2\hat{p}}
\newcommand\hP{\skew4\hat{P}}
\newcommand\hpsi{\skew5\hat{\psi}}
\newcommand\hPsi{\skew2\hat{\Psi}}
\newcommand\hr{\skew2\hat{r}}
\newcommand\hx{\skew2\hat{x}}
\newcommand\hX{\skew4\hat{X}}
\newcommand\hy{\skew2\hat{y}}
\newcommand\hV{\skew2\hat{V}}
\newcommand\tg{\skew2\tilde{g}}
\newcommand\tH{\skew4\tilde{H}}

\newcommand\tL{\skew2\tilde{L}}
\newcommand\tQ{\skew4\tilde{Q}}
\newcommand\tU{\skew2\tilde{U}}
\newcommand\tnu{\skew3\tilde{\nu}}
\newcommand\trho{\skew4\tilde{\rho}}
\newcommand\ttau{\skew2\tilde{\tau}}
\newcommand\bVo{\bV_{\!0}}
\newcommand\Lao{\La_0}
\newcommand\Us{U_{\!s}}
\newcommand\al{\alpha}
\newcommand\be{\beta}
\newcommand\de{\delta}
\newcommand\ep{\epsilon}
\newcommand\ga{\gamma}
\newcommand\ka{\kappa}
\newcommand\la{\lambda}
\newcommand\om{\omega}
\newcommand\si{\sigma}
\newcommand\ze{\zeta}
\newcommand\vep{\varepsilon}
\newcommand\vka{\varkappa}

\newcommand\vth{\vartheta}
\newcommand\De{\Updelta}
\newcommand\vDe{\Delta}
\newcommand\Ga{\Upgamma}
\newcommand\vGa{\Gamma}
\newcommand\La{\Lambda}
\newcommand\Om{\Omega}
\newcommand\Si{\Sigma}

\newcommand\mcA{\mathcal{A}}
\newcommand\mcB{\mathcal{B}}
\newcommand\mcC{\mathcal{C}}
\newcommand\mcD{\mathcal{D}}
\newcommand\mcFk{\mathcal{F}_k}
\newcommand\mcI{\mathcal{I}}
\newcommand\mcS{\mathcal{S}}
\newcommand\mcT{\mathcal{T}}
\newcommand\mcU{\mathcal{U}}
\newcommand\mcV{\mathcal{V}}
\newcommand\mcW{\mathcal{W}}
\DeclareMathOperator\e{e} 

\DeclareMathOperator*{\res}{Res}

\DeclareMathOperator\sgn{sgn}
\newcommand\Ca{\mbox{\tit{Ca}}} 

\newcommand\deq{\!\vcentcolon=}
\newcommand\fhr{f_\mathrm{HR}}
\newcommand\phr{p_{\mathrm{HR}}}
\newcommand\fr{\frac}
\newcommand\lb{\left}
\newcommand\rb{\right}
\newcommand\m[1]{\mb{$#1$}} 
\newcommand\mb{\mbox}
\newcommand\mrm{\mathrm}
\newcommand\sqrb[1]{\sqrt{\smash[b]{#1}}}

\newcommand\ttmi{\ensuremath{\rightarrow -\infty}}
\newcommand\tst{\mrm{TST}}
\newcommand\ud{\mrm{d}}
\newcommand\ui{\mrm{i}} 

\newcommand\barray{\begin{array}}
\newcommand\earray{\end{array}}
\newcommand\bce{\begin{center}}
\newcommand\ece{\end{center}}
\newcommand\beq{\begin{equation}}
\newcommand\eeq{\end{equation}}
\newcommand\beqa{\begin{eqnarray}}
\newcommand\eeqa{\end{eqnarray}}
\newcommand\bseq{\begin{subequations}}
\newcommand\eseq{\end{subequations}}
\newcommand\bseqa{\begin{subeqnarray}}
\newcommand\eseqa{\end{subeqnarray}}
\newcommand\disp{\displaystyle}
\newcommand\q{\quad}

\newcommand\tbf{\textbf}
\newcommand\tit{\textit}
\newcommand\tsc{\textsc}
\newcommand\txt{\textstyle}
\newcommand\vs{\vspace*}
\newcommand\cf{cf.\xspace}
\newcommand\ie{i.e.\xspace}
\graphicspath{{figs/}}
\makeatletter
 \def\input@path{{figs/}}
\makeatother
\newcommand\bfig{\begin{figure}}
\newcommand\efig{\end{figure}}

\newcommand\tc{\textcolor}

\renewcommand{\theenumi}{\roman{enumi}} 
\newcommand\benum{\begin{enumerate}}
\newcommand\eenum{\end{enumerate}}

\newcommand\hl{\tc{black}} 

\title[Developed liquid film past a trailing edge: `teapot effect']
{Developed liquid film passing a \hl{smoothed and wedge-shaped} trailing edge: small-scale 
analysis and the `teapot effect' at large Reynolds numbers}

\author[B. Scheichl, R. I. Bowles and G. Pasias]%
{B. Scheichl\aff{1,2}%
  \corresp{\email{bernhard.scheichl@tuwien.ac.at}},\ns
 R. I. Bowles\aff{3}\ns\and\ns G. Pasias\aff{3}}

\affiliation{\aff{1} Institute of Fluid Mechanics and Heat Transfer, Faculty of Mechanical
             Engineering, Technische Universit\"at Wien, Tower~BA/E322, Getreidemarkt~9, 
             1060~Vienna, Austria
             \aff{2} AC2T research GmbH (Austrian Excellence Center for Tribology),
             Viktor-Kaplan-Stra{\ss}e~2/C, 2700~Wiener Neustadt, Austria
             \aff{3} Department of Mathematics, Faculty of Mathematical \& Physical Sciences, 
             \\ University College London (UCL), 25~Gordon Street, 
             London~WC1H~0AY, UK}

\begin{document}

\maketitle

\begin{abstract}
Recently, the authors considered a thin steady developed viscous \hl{free-surface flow} 
passing the sharp trailing edge of a horizontally aligned flat plate under surface tension and 
the weak action of gravity, acting vertically, in the asymptotic slender-layer limit 
(\emph{J.~Fluid Mech.}~\tbf{850}, pp.~924--953, 2018). We revisit the capillarity-driven 
short-scale viscous--inviscid interaction, on account of the inherent upstream influence, 
immediately downstream of the edge and scrutinise flow detachment on all smaller scales. We 
adhere to the assumption of a Froude number so large that choking at the plate edge is 
insignificant but envisage the variation of the relevant Weber number of $O(1)$. The main 
focus, tackled essentially analytically, is the continuation of the structure of the flow 
towards scales much smaller than the interactive ones and where it no longer can be treated as 
slender. As a remarkable phenomenon, this analysis predicts harmonic capillary ripples of 
Rayleigh type, prevalent on the free surface upstream of the trailing edge. They exhibit an 
increase of both the wavelength and amplitude as the characteristic Weber number decreases. 
Finally, the theory clarifies the actual detachment process, \hl{within a rational 
description} of flow separation. At this stage, the wetting properties of the fluid and the 
microscopically wedge-shaped edge, viewed as infinitely thin on the larger scales, come into 
play. As this geometry typically models the exit of a spout, the predicted wetting of the wedge 
is related to what in the literature is referred to as the teapot effect.
\end{abstract}



\section{Introduction}

We continue to \hl{analyse} a flow problem of fundamental importance as started in our 
forerunner study (Scheichl, Bowles \& Pasias 2018, hereafter referenced as SBP18).

Let a nominally steady and two-dimensional, developed, slender stream of a Newtonian liquid 
having uniform properties and at constant flow rate in an inertial frame of reference detach
from a horizontal, solid, impenetrable, perfectly smooth plate with a trailing edge that is 
initially considered as abrupt and sharp. Downstream, the \hl{resulting fluid jet} divides its 
gaseous environment, fully at rest and under constant pressures, into two parts. Here this 
picture is relaxed insofar as the upper one still defines the zero pressure level but we allow 
for a non-zero, constant support pressure prescribed at the downside of the detached layer. 
The body and interface forces crucially at play are the constant gravitational acceleration 
acting vertically towards the wetted side of the plate and surface tension. Based on the 
principle of least degeneration, our rigorous theoretical description of the detaching thin 
film under the assumption of very supercritical flow adopts a specific distinguished limit 
where the relevant Reynolds and Froude numbers are taken as asymptotically large but the 
corresponding Weber number as of $O(1)$. Hence, the details accompanying the detachment 
process are governed by a strong viscous--inviscid, shortened-scale interaction at the outset 
of our present study.

Subsequently, we refer to the sketch in figure~\ref{f:conf} throughout, illustrating the 
different flow regions considered when viewed on the global vertical scale \hl{defined by the 
height of the detaching layer. Specific interest is aroused by the so-called ``teapot 
effect'', here observed in the flow in the immediate vicinity of the trailing edge and thus 
strongly affected by its microscopic geometrical resolution. As a start, we critically review 
the prevailing, rather phenomenological view on this effect and its previous modelling.}

\bfig
 \centering
 \scalebox{0.5}{\input{conf.pdf_t}}
 \caption{\hl{Global view on detaching film (not to scale, variables introduced in 
  \S\,\ref{ss:ndg}): viscous sublayer (VSL), interactive flow comprising main deck (MD) and 
  lower deck (LD), flow on smaller scales captured by light-shaded region, near wake of 
  Hakkinen--Rott type (HRW).}}
 \label{f:conf} 
\efig

\subsection{The teapot effect: a digression}\label{ss:ted}

The frequently observed, at a first glance spontaneous (and often undesired) tendency of a 
liquid pouring from a spout to instead stick to its underside was originally reported by 
\citet[][also see the references therein]{Re56} \hl{and later by \cite{Wa84}}: see 
figure~\ref{f:tpe}(\tit{a}). More precisely, Reiner coined the notion ``teapot effect'' for 
pouring liquid along a rigid convex wall with a marked corner and adjoining to another (even 
liquid) fluid. He untangled the riddle of its occurrence experimentally: his observations 
ruled out the hitherto widely held belief that the wetting properties in terms of short-range 
inter-molecular adhesion forces, promoted by wetting agents, are its essential cause. However, 
his various experiments demonstrated that ``adhesion'' as the reaction force on the fluid 
flowing over a solid phase as well as surface tension at its common interface with the 
surrounding fluid play a decisive role. A recent survey of the various treatments of this 
scenario presented by \citet[][see the references therein]{JPetal19} spans the rigorous 
approach within the framework of classical fluid mechanics, outlined below, to the nowadays 
more common but less stringent approach. \hl{This proposes that the pivotal cause for the 
fluid sticking lies in the hydrophilic tendency of the liquid/wall pairing rather than the 
mechanisms of the pouring.} The latter authors provide new insight by coupling these ideas 
with classical arguments resorting to the first principles of continuum mechanics. Notably, 
\cite{Duetal10} indicate a significant reduction of the effect via the application of 
superhydrophobic substrates.

\bfig
 \centering
 \scalebox{0.5}{\input{teapot-effect.pdf_t}}
 \caption{(\tit{a}) Different realisations of the teapot effect for a low-momentum \hl{liquid
  film} typically strongly subject to gravity, described in and reprinted with permission from 
  \cite{Duetal10} (\copyright\ by the American Physical Society); (\tit{b}) its current 
  abstraction for a planar, horizontal high-momentum jet in fact passing a rounded wedge of 
  angle $\al$, detailing the flow around the trailing edge in figure~\ref{f:conf}, typical no 
  slip on the plate and free slip along the free streamlines, \tc{blue}{blue}: free and 
  internal streamlines and detachment point, \tc{red}{red}: plate and original (virtual) tip 
  in figure~\ref{f:conf}.}
 \label{f:tpe}
\efig

We advocate continuum mechanics for providing a satisfactory, rational unravelling of the 
effect. In agreement with the above mentioned early observations, we interpret it as a subtle 
interplay of inertia, capillarity and gravity in a two-dimensional setting. This is crucially 
tied in with the breakdown of viscous--inviscid interaction and thus the slender-layer 
approximation made on larger scales due to the assumed largeness of the globally defined 
Reynolds number of the oncoming attached flow. The significance of capillarity and inertia 
lies also in \hl{its proper adjustment immediately upstream of detachment}. Our asymptotic 
theory proposes a fully rational account of the onset of this phenomenon in the realistic 
situation of a developed incident flow. As a specific ingredient, the trailing edge is 
replaced by a tip, \ie a wedge formed by an acute cut-back angle or lip: this ``attracts'' the 
\hl{liquid film} such that it clings to it before the liquid sheet breaks away from it as a 
whole from its underside. This phenomenon of free rather than forced gross separation from a 
convex rigid surface, consequently referred to as the teapot effect from here onwards, does 
not yet have a satisfactorily rigorous and complete description. \hl{\cite{Duetal10} 
previously considered this ``inertial-capillary'' mechanism, investigated here in depth and 
breadth, as a crucial step towards a breakthrough in the explanation of the effect.}

\hl{An initial self-consistent clarification of the effect benefitted from} the quite 
restrictive assumption of irrotational free-surface flow of a weightless ideal fluid \hl{and 
the neglect of surface tension} past a horizontal plate, terminated by the aforementioned lip: 
remaining firmly attached both with the neglect of gravity \citep{Ke57} and under gravity 
\citep{VaKe86}; detaching grossly from the underside at zero gravity \citep{VaKe89}. \hl{In 
these investigations, the flow is \emph{stipulated} to cling to the wall and, due to the 
absence of viscosity, the position of detachment is also prescribed \citep{VaKe89}. However, 
the well-known Brioullin--Villat condition, met for vanishingly small effects of capillarity 
(and viscosity), fixes the physically admissible detachment point.}

Rather little is known when it comes to the rigorous inclusion of viscosity in this flow 
picture. At least, the passage of a layer over an asymptotically small convex wall corner 
(and in related situations) considered by \cite{Ga87} (and the refined numerical results by 
\citealp{Ya12}) is relevant. Specifically, there the unperturbed oncoming flow is fully 
developed (so as to model a real situation), as being already inclined towards gravity, and 
viscous--inviscid interaction of the double-layer structure in the high-Reynolds-number limit, 
adopted here, negotiates the slender obstacle \hl{which the corner forms}. However, the 
counteracting impact of surface tension in the resulting combined hypersonic- and 
wall-jet-type interaction law \citep[\cf][]{BoSm92} is ignored in the analysis although 
mentioned. Although the interactive flow considered by \cite{Ga87} is \emph{assumed} to remain 
grossly attached, it is \hl{certainly interesting} that the numerical solutions predict a 
closed separation bubble beyond the mild wedge for both sufficiently large turning angles and 
Froude numbers.

A seminal reference for the teapot effect in a realistic, \ie developed, flow is the numerical 
and partially analytical investigation of the full Navier--Stokes problem by \cite{KiSc94}. 
They unambiguously highlighted its viscous and capillary, \ie hydrodynamic, nature as 
underpinned by experimental evidence. This prompted them to conclude that ``the teapot effect 
is more than merely an issue of wetting''. Most remarkably, they pointed out how the 
restrictions of the microscopic wedge-type geometry of what is on larger scales viewed as an 
``infinitely sharp'' edge implies a contact-angle hysteresis, associated with non-unique flow 
states, but the point of flow detachment becomes the apex of the wedge when the jet Reynolds 
number, \ie the momentum it carries, becomes sufficiently large. The present asymptotic 
analysis corroborates this finding, where we deal with a horizontal oncoming flow past a wedge 
originally represented by a cut-back angle $\al$ (\m{0<\al<\upi}), using equal horizontal and 
vertical scales. However, here the wedge is no longer necessarily sharp as as we allow for its 
tip being realistically rounded: see the sketch in figure~\ref{f:tpe}(\tit{b}).

\subsection{Studied phenomena and open questions}

Our current concern is with the analytical/numerical challenges arising in the analysis of the 
free jet with particular emphasis placed on the description of its detachment at the abrupt 
plate edge on the smallest scales and the freely interacting flow immediately downstream of 
the trailing edge. As a key observation in SBP18, \hl{the free layer is strongly dependent on 
its history and therefore, of the no-slip condition satisfied upstream of its detachment}. 
Since the interaction mechanism is not alone capable of smoothing the flow quantities at the 
sharp edge, coping with this demand addresses the flow on still smaller and down to the 
smallest scales discernable and eventually the wetting properties of the plate as well as the 
\hl{detailed} geometry forming its edge. \hl{The threefold conclusions drawn from such an 
analysis attempt to shed light on some unsettled questions of fundamental interest}: 
\benum
 \item As a first cornerstone, it reveals the existence of (stationary) undamped capillary 
Rayleigh modes upstream of its break-away from the plate. 
 \item The \hl{multi-layer slenderness} of the flow, given the largeness of the Reynolds 
number, prevents its separation upstream of the trailing edge, which confirms the initially 
made assumption of detachment ``at the edge'' considered on larger scales. 
 \item As a second highlight, the implied wetting of the edge suggests a \hl{novel}, rational 
explanation of the teapot effect observed in a high-momentum \hl{liquid layer} when a convex 
corner provides~--~in a most simple but nevertheless sufficiently complex manner~--~the 
non-degenerate geometry modelling the plate edge.
\eenum

\hl{\subsection{Organisation of the paper, used notation and numerical software}}

\hl{The process of asymptotic scale separation, starting with the largest global scale down to 
the smallest ones where the teapot effect is at play, guides the structure of our study. 
Visualising this in figure~\ref{f:scal-sep} serves to illustrate and accompany the subsequent 
analysis of the individual flow regimes governed by those spatial scales.} Hence, 
figure~\ref{f:scal-sep}(\tit{f}) recovers the linkage to the teapot effect as in 
figure~\ref{f:tpe}.

\bfig
 \centering
 \scalebox{0.41}{\input{scal-sep.pdf_t}}
 \caption{Essential flow regions, shaded details zoomed-in consecutively from (\tit{a}) to 
  (\tit{f}) (not to scale, denotations provided in the course of the analysis): flow 
  detachment viewed on interactive down to smallest scales, where the detached streamline is no 
  longer elongated and the flow no longer slender; \hl{main deck (MD), lower deck (LD), inner  
  and outer Rayleigh stages (RSs), Hakkinen--Rott wake (HRW) as sublayer of LD (dashed 
  boundary); slip layer (SL) at bottom of LD below outer RS, Navier--Stokes (NS) regime; 
  \tc{blue}{blue}: free and internal streamlines and detachment point, \tc{red}{red}: plate 
  and original tip, coinciding with origin and detachment point in (\tit{a})--(\tit{e}), all 
  disparate in resolved situation~(\tit{f}).}}
 \label{f:scal-sep}
\efig

The paper is organised as follows.
\benum 
 \item[\S\,\ref{s:sep}:] We first pose the problem based on first principles in full. Our 
basic scaling arguments (\S\,\ref{ss:ndg} \hl{and appendix~\ref{a:a}}) justify the use of 
asymptotic analysis as the means of choice to study the flow, initiated by completing the 
formulation of the interaction problem, originally posed in SBP18. It then governs the 
continuation of the freely interacting jet downstream of the edge in a rigorous manner as long 
as the value of the appropriately rescaled Weber number does not fall below a certain 
threshold, so avoiding the onset of nonlinear stationary capillary waves even \hl{above the 
plate} (\S\,\ref{ss:fid} \hl{and appendix~\ref{a:b}}). \hl{Over the interactive streamwise 
scale, this brings into play the splitting of the film into the main deck (MD) and the lower 
deck (LD), this initiated by a viscous sublayer (VSL) adjacent to the plate.}
 \item[\S\,\ref{s:ids}:] \hl{A multi-structured small-scale flow, essentially controlled by 
capillarity only (\S\,\ref{ss:ioc}), supersedes locally the two-tiered interactive one. 
The Hakkinen--Rott-type near wake (HRW) forming at the base of the LD just downstream of the 
plate (\S\,\ref{ss:hrw}) is central for understanding the multi-structured small-scale flow 
locally superseding the two-tiered interactive one. Its thorough investigation reveals two 
nested square Euler regions (\S\,\ref{ss:oie}). These outer and inner Rayleigh stages (RSs) 
govern weak perturbations around the flow at detachment.} The exterior one extends vertically 
across most of the layer and is the source of phenomenon (i) above on the top free surface. 
\hl{Simultaneously, a viscous (passive) slip layer (SL) forms at the base of the predominantly 
inviscid flow.} 
 \item[\S\,\ref{s:fns}:] This essentially inviscid description of flow detachment paves the 
way for a full Navier--Stokes (NS) regime detected on even smaller streamwise and vertical 
scales, \hl{where the flow structure of \S\,\ref{s:ids} collapses. (\S\,\ref{ss:lop})}. Its 
analytical study leads to the implication (ii) above \hl{(\S\,\ref{ss:fcd})}. As a pivotal 
finding, achievement (iii), we also identify one or two interlaced Stokes regions resolving 
the smallest scales and the actual wedge-type resolution of the plate end 
\hl{(\S\,\ref{ss:dsl})}, until now seen as infinitely thin. Consequently, it is this flow 
regime where the break-away of the film, interacting with the larger-scale flow through the NS 
region, is finally controlled by both the \hl{effective} edge geometry and the static wetting 
angle. Thereby, an awareness of the close relationship of this situation to to the teapot 
effect is gained.
 \item[\S\,\ref{s:cfp}:] Surveying the current results and \hl{anticipating the inclusion of 
\eg unsteadiness and the aforementioned capillary undulations in our ongoing research} 
completes the study.
\eenum

\hl{So as not to distract attention away from the main arguments and their physical impact, 
the detailed steps of the asymptotic analysis, together with further technical side aspects, 
potentially of interest for the more mathematically orientated readership, are put forward as 
the accompanying ``Other supplementary material''. It consists of the individual Supplements 
\tit{A}--\tit{E}. Cross-references between these, its numbered subsections and the main 
document are conveniently employed. We add citations exclusively in the supplement to the list 
of references.}

\hl{In addition to the usual conventions for mathematical expressions, we adopt the following 
usage of accents and sub- and superscripts (\cf figure~\S\,\ref{f:conf}). Indices typically 
indicate orders in asymptotic expansions and partial derivatives unambiguously, and lowered 
``$-$'' and ``$+$'' refer to respectively the lower and upper boundaries of the liquid layer 
(\eg $h_-$ and $h_+$) or the states of the flow infinitely far upstream (``$-$'') and 
downstream (``$+$''). We endow dimensional quantities with tildes. Furthermore, we attempt a 
systematic as possible denotation of the dependent and independent $O(1)$-variables 
characteristic of the individual regimes: lowercase for the MD (\eg $x$), capitalised for the 
LD ($X$), capitalised with overbars for the outer RS ($\bX$), capitalised with hats for the 
inner RS ($\hX$), lowercase with overbars for the full NS region ($\bx$), lowercase with hats 
for the Stokes regions ($\hx$).}

All our numerical calculations used the widely-used, proprietary programming language and 
numerical-computing environment \cite{ML20}, supplemented with \cite{NATB20}. In particular, 
the computations benefit from its convenient handling of complex arithmetic and the, in 
principle, built-in arbitrarily high accuracy and precision.

\section{Statement of the extended problem}\label{s:sep}

It proves expedient to first reappraise the fundamental assumptions and the problem in full 
before revisiting the interactive limit.

\subsection{Non-dimensional groups and governing equations}\label{ss:ndg}

The problem has the following central ingredients. The slender layer of density $\trho$ and 
kinematic viscosity $\tnu$ and experiencing a tensile surface stress $\ttau$ and gravitational 
acceleration $\tg$ carries a volume flow rate per lateral unit width $\tQ$. It adjusts to a 
developed state over some sufficiently large distance $\tL$, serving as the basic length scale 
and measured along the plate from its trailing edge in the upstream direction. Simultaneously, 
$\tL$ is required to be so short that the vertical layer height has not grown sufficiently to 
allow for a significant impact of the hydrostatic pressure on streamwise convection. Then a 
layer height \m{\tH=\tL\tnu/\tQ} and flow speed \m{\tU=\tQ^2\!/(\tnu\tL)} representative of 
this near-supercritical film follow from conservation of the flow rate and the streamwise 
momentum, here expressed by the balance between convection and the shear stress gradient, 
respectively 
\beq
 \tQ=\tU\!\tH,\q \tU^2\!/\tL=\tnu\tU\!/\tH^2.
 \label{cl}
\eeq
In many applications, the vertical height and, accordingly, the speed of the layer have 
respectively increased and decreased so markedly over $\tL$ that it has almost attained its 
well-known perfectly supercritical, \emph{fully} developed or self-preserving state discovered 
by \cite{Wa64}: for related discussions see \cite{BoSm92}, \cite{Hi94} and, in the context of 
an axisymmetric and rotatory layer generated by vertical jet impingement, \cite{ScKl19}.

The flow is then controlled by the slenderness parameter or reciprocal Reynolds number $\ep$ 
and corresponding reciprocal Froude and Weber numbers $g$ and $\tau$:
\bseq
\beq
 \ep\deq\fr{\tH}{\tL}=\fr{\tnu}{\tU\!\tH}\ll 1,\q g\deq\fr{\tg\tH}{\tU^2}=O(\ep^{4/7}),\q 
 \tau\deq\fr{\ttau}{\trho\tU^2\!\tH}=O(1).
 \label{egt}
\eeq
Regarding the distinguished limit involving $g$, locally strong viscous--inviscid interaction 
describes the abrupt transformation of the wall-bounded flow \hl{on crossing the lip} towards 
the free liquid jet in a least-degenerate, self-consistent and sufficiently smooth manner. We 
remark that the conventionally defined capillary number 
\beq
 \Ca\deq\trho\tnu\tU\!/\ttau=\ep/\tau\ll 1
 \label{ca}
\eeq
 \label{ng}%
\eseq
or the alternative Ohnesorge number, here \m{\ep/\sqrb{\tau}\ll 1}, provide different albeit 
less preferable measures of the surface tension for a layer of slenderness \hl{expressed} by 
$\ep$: since the streamline curvature scales with \m{\tH/\tL^2=\ep^2/\tH}, the ratio of the 
viscous (deviatoric) stress, normal to a free surface and scaling with 
\m{\trho\tnu\tU/\tL=\trho\tU^2\ep^2}, to the capillary hoop pressure measured by 
\m{\ttau\tH/\tL^2=\tau\ep^2\trho\tU^2} is expressed by the augmented capillary number 
\m{\Ca/\ep=1/\tau=O(1)}, taking into account the aspect ratio of the flow. This indicates that 
in the limit provided by \eqref{egt} the surface jump of the total normal stress is fully 
retained in the dynamic boundary conditions (BCs) below. 

\hl{Order-of-magnitude arguments considering realistic flow situations support the above 
asymptotic scaling and demonstrate its applicability to the teapot phenomenon in typical 
settings: see appendix~\ref{a:a}.}

We introduce Cartesian coordinates $x$ and $y$ pointing respectively horizontally from the 
trailing edge and vertically towards the flow, the streamfunction $\psi$ and the pressure $p$, 
non-dimensional with $\tL$, $\tH$, $\tQ$ and $\trho\tU^2$ respectively. Then \m{u\deq\psi_y} 
is the horizontal and \m{v\deq -\ep\psi_x} the vertical flow component made dimensionless with 
$\tU$. These $O(1)$-quantities satisfy the NS equations in the form
\bseq
\begin{align}
 \psi_y\psi_{yx}-\psi_x\psi_{yy} &= -p_x+(\ep^2\p_{xx}+\p_{yy})\psi_y,
 \label{nsx}\\
 \ep^2(\psi_x\psi_{yx}-\psi_y\psi_{xx}) &= -p_y-(\ep^2\p_{xx}+\p_{yy})(\ep^2\psi_x)-g.
 \label{nsy}
\end{align}
Here and hereafter, the subscripts $-$ and $+$ indicate the evaluation along the lower- and 
the uppermost free streamline respectively. Accordingly, \m{y=h_-(x)} (\m{\equiv 0} for 
\m{x\leq 0}) and \m{y=h_+(x)} denote their positions, hence \m{h(x)\deq h_+ -h_-} the vertical 
film thickness and $p_\pm$ the given pressure levels along the free streamlines. Adopting the 
Heaviside step function $\theta$ then gives the kinematic boundary conditions including the 
conventional requirements of no slip at and no penetration through the plate as follows.
\beq
 y=h_-(x)\colon\;\; \psi=\psi_y\,\theta(-x)=0,\q y=h_+(x)\colon\;\; \psi=1.
 \label{kbc}
\eeq
The dynamic BCs express vanishing tangential stresses and total normal stresses equal to the 
capillary pressure jumps on the free surfaces of curvatures $\vka_\pm(x)$ and subject to the 
Young--Laplace equilibrium. Therefore, at
\beqa
 & y=h_-(x)\;\; \mb{if}\;\; x>0,\q y=h_+(x)\colon & 
 \nonumber\\[3pt]
 & (1-\ep^2 h_\pm'^2)(\psi_{yy}-\ep^2\psi_{xx})-4\ep^2 h_\pm'\psi_{yx}=0, & 
 \label{dbct}\\[3pt]
 & \lb.\barray{c}
 2\ep^2[\psi_{yx}(1-\ep^2 h_\pm'^2)+h_\pm'(\psi_{yy}-\ep^2\psi_{xx})]/(1+\ep^2 h_\pm'^2)+ 
 p-p_\pm=\tau\vka_\pm, \\[6pt]
 p_+=0,\q \vka_\pm=\mp\ep^2 h_\pm''/(1+\ep^2 h_\pm'^2)^{3/2}.
 \earray\!\!\!\rb\} &
 \label{dbcn}
\eeqa
 \label{gov}%
\eseq
This completes the problem \eqref{gov} as proper up- and downstream conditions will be 
condensed into requirements of continuity holding at the trailing edge \m{x=0}.

\subsection{Free interaction \hl{across} the trailing edge}\label{ss:fid}

The governing equations \eqref{gov} and \eqref{egt} immediately give rise to regular 
expansions valid for the flow above the plate on the original large streamwise scale, \ie for 
\m{1+x=O(1)}, \m{0>x=O(1)}:
\bseq
\begin{align}
 [\psi,h,p/g] &\sim [\psi^0(x,y),h^0(x),h^0(x)-y]+O(g)\q (\ep\ttz),
 \label{psihpr}\\[1pt]
 [\psi^0,h^0] &\sim [\psi_0(y),h_0]+O(x)\q (x\ttz-).
\end{align}
\eseq
In the leading order of this non-interactive limit, the classical parabolic shallow-water 
approximation of \eqref{gov} is recovered, predicting a pressure-free base flow described by 
$\psi^0$ and $h^0$. These quantities approach regularly some values $\psi_0$ and $h_0$ at the 
trailing edge. The higher-order contributions in \eqref{psihpr} control the modification by 
the hydrostatic pressure distributions and non-parallel-flow effects, the latter predominantly 
due to streamline curvature, capillary action and the viscous normal stresses 
$\pm\ep^2\psi_{yx}$, in the following iterative manner. At each level of improvement, the 
obtained approximation for $\psi$ feeds into \eqref{nsy} subject to \eqref{dbcn}. The 
resulting pressure correction then forces a problem that emerges from expanding \eqref{nsx} 
subject to \eqref{kbc} and \eqref{dbct} and governs a further correction for $\psi$, and so on.

Following SBP18, this hierarchy is singularly perturbed by weak irregular disturbances 
exhibiting exponential growth over a short streamwise scale measured by $\ep^{6/7}$. Thus, 
they are active in the VSL adjacent to the plate. Hence, subject to free viscous--inviscid 
interaction governed by streamline curvature, not accounted for in the classical shallow-water 
limit, they describe the intrinsic upstream influence in the film caused by both gravity and 
capillarity. Finally, the growth of these two effects renders the above hierarchy invalid 
around the trailing edge where \m{x=O(\ep^{6/7})} and they provoke a locally strong 
interaction over that scale in the limits \eqref{egt}. This typically involves a nonlinear 
distortion of the strongly viscosity-affected slow flow in the LD, here originating from the 
VSL, adjacent to the lowermost streamline where \m{y=O(\ep^{2/7})}. The latter exerts a linear 
response in the MD that comprises the bulk of the layer, beneath the upper free streamline.

The background flow enters the interactive scalings at leading order solely through two 
quantities condensing its upstream history: the momentum flux $J$ at the trailing edge and 
the shear stress $\la$ exerted on it,
\beq
 J\deq\int_0^{h_0}\!\psi_0'^2(y)\,\ud y,\q \la\deq\psi_0''(0) \;\;\;\mb{as}\;\;\;
 \psi_0\sim\fr{\la y^2}{2}+\fr{\la\om y^5}{60}+O(y^8)\;\;\; (y\ttz).
 \label{jl}
\eeq
The coeffcient $\om$ is only relevant in the small-scale analysis \hl{of \S\,\ref{sss:ors}}. 
We also note \eqref{kbc} and the free-slip condition resulting from \eqref{dbct}:
\beq
 \psi_0(h_0)=1,\q \psi_0''(h_0)=0.
 \label{psi0h}
\eeq
Usually, $\tH$ is definitely larger than the height of the film immediately downstream of its 
origin (as given by jet impingement) and where the flow starts to become developed: \hl{see
table~\ref{t:dat} in appendix~\ref{a:a}}. This prompts us to \hl{assume that the base flow is  
already described by Watson's (1964) self-similar solution and so to} neglect the small 
deviations from this due to the flow history, as in SBP18 and without any substantial loss of 
generality. In this idealisation, \m{h^0=\upi(x-x_v)/\sqrb{3}} provided some \m{x=x_v<0} 
indicates the virtual origin of the fully developed flow and $\psi^0$ is a universal function 
of $y/h^0$. At \m{x=0}, $\psi_0$ then satisfies
\beq
 \psi_0'^2(y)=x_v\psi_0'''(y)
 \label{weq}
\eeq
and has an exact representation given by \cite{ScKl19}\hl{:} writing 
\m{u_0^+\deq\psi_0'(h_0)} from here on, this implies the important \hl{canonical} results
\beq
 \lb.\barray{cc}
 h_0/|x_v|=\upi/\sqrb{3}\simeq 1.8138,\q
 \hl{|x_v|u_0^+=\bigl[\Ga(\txt\fr{1}{3})/\Ga(\txt\fr{5}{6})\bigr]^2\!\big/(2\upi)\simeq 
  0.89644,}
 \\[6pt]
 x_v^2\la=|x_v|J=\sqrb{2/3}\,\bigl(|x_v|u_0^+\bigr)^{3/2}\simeq 0.69301.
 \earray\rb\}
 \label{w}
\eeq

The interaction process itself is parametrised by suitably redefined reciprocal Froude and 
Weber numbers $G$ and $T$ and the rescaled support pressure $P_-$, all of $O(1)$. 
Specifically, $T$ is formed with the local momentum flux and thus measures the influence of 
capillarity relative to fluid inertia. We thus introduce
\beq
 (G,P_-)\deq (g h_0,p_-)/(M^2\la^6\ep^4)^{1/7},\q T\deq\tau/J,\q M\deq|T-1|J=|\tau-J|.
 \label{gptm}
\eeq
The above propositions enable us to reconsider the interaction problem, at first under the 
assumption that $T$ is not too close to unity. For the details of its numerical treatment by 
specifying $\psi_0$ as Watson's flow profile and marching downstream we refer to SBP18.

\hl{The given adjustment length $\tL$ serves to define} $\tH$ and $\tU$ via \eqref{cl}. Hence, 
for a given flow, we note the invariance of \eqref{cl} and thus of $\ep$, $\psi$, \eqref{gov} 
and $G$, $T$, $P_-$ under the affine transformation
\beq
 \bigl(\tL,\tH,\tU,x,y,h_\pm,p,g,\tau,J,\la\bigr)\mapsto\Bigl(
 a\tL,a\tH,\fr{\tU}{a},\fr{x}{a},\fr{y}{a},\fr{h_\pm}{a},a^2 p,a^3 g,a\tau,a J,a^2\la\Bigr)
 \label{a}
\eeq
with \m{a>0} being an arbitrary scaling factor. This confirms the independence to $\tH$ of the 
canonical formulation of the interaction problem below and thus on the specific choice of the 
streamwise length scale $\tL$ (for a sufficiently small \m{\ep=\tH/\tL}). In particular, its 
solution downstream of the edge does not depend on the scaling of the attached flow and, 
specifically, the position of the aforementioned virtual origin. For any subsequent numerical 
evaluation involving $\psi_0$ and $h_0$, however, we not only assume the flow as being fully 
developed but also adopt the natural standardisation \m{x_v=-1} from here on, \ie we specify 
$\tL$ to be the \hl{full} development length.

\subsubsection{Main deck}

Since the MD describes a predominantly inviscid flow in the long-wave limit, the central local 
expansion reads
\begin{align}
 \bigl[\psi,h,h_-,h_+\bigr] &\sim \bigl[\psi_0(z),h_0,0,h_0\bigr]+
 \ep^{2/7}m\bigl[A(X)\,\psi_0'(z),-A(X),H_-(X),H_+(X)\bigr] \nonumber\\[1pt]
 &+ O(\ep^{4/7}),\q H_+\deq H_- -A,\q m\deq(M\!/\la^4)^{1/7},\q z\deq y-h_-(x),
 \label{mdexp}
\end{align}
and \m{p=O(\ep^{4/7})}. The local streamwise variable \m{X=O(1)} is defined in 
\eqref{ldxy} below. The expansion \eqref{mdexp} induces the following hierarchy of equations 
resulting from the Euler operator in (\ref{gov}\tit{a},\tit{b}). The dominant viscous 
displacement exerted by the LD, \m{-A(X)}, generates typically the dominant perturbation of 
$\psi$ about $\psi_0$ in terms of the pressure-free eigensolution of the linearised streamwise 
momentum equation \eqref{nsx}, where we have conveniently introduced the Prandtl 
transposition. Entering \eqref{nsy}, this $O(\ep^{2/7})$-contribution to $\psi$ governs 
streamline curvature and, by virtue of integration with respect to $y$, \hl{supplements the 
hydrostatic portion of $p$ with the convective one, also of $O(\ep^{4/7})$}. The disturbances 
described so far account for the role of the MD for the interactive mechanism. The 
$O(\ep^{4/7})$-contributions to $p$ and to $\psi$, the latter induced subsequently by the 
streamwise pressure gradient, are specified in~SBP18.

\subsubsection{Lower deck}

In the LD, the expansion
\beq
 [\psi,p]/\ep^{4/7}\sim\bigl[(M^2\!/\la)^{1/7}\,\Psi(X,Z),(M^2\la^6)^{1/7}P(X)\bigr]+ 
 O(\ep^{6/7})
 \label{ldexp}
\eeq
employs the stretched coordinates 
\beq
 X\deq x l/\ep^{6/7},\q (Y,Z)\deq(y,z)/(\ep^{2/7}m),\q l\deq(\la^5\!/M^3)^{1/7}.
 \label{ldxy}
\eeq
To describe the flow up- and downstream of the plate edge, the \hl{variable $Z$ is preferred}
over $Y$ in the slender LD. In turn, (\ref{gov}\tit{a},\tit{b}) reduce locally to the boundary 
layer \hl{equation}
\bseq
\beq
 \Psi_Z\Psi_{Z\!X}-\Psi_X\Psi_{Z\!Z}=-P'+\Psi_{Z\!Z\!Z},
 \label{ldeq}
\eeq
and (\ref{gov}\tit{c},\tit{d}) to the mixed BCs expressing the downstream passage from no- to 
free-slip along
\beq
 Z=0\colon\;\; \Psi=\Psi_Z\,\theta(-X)=\Psi_{Z\!Z}\,\theta(X)=0.
 \label{ldbc}
\eeq 
To match \eqref{ldexp} and \eqref{mdexp} subject to \eqref{jl}, we require that for
\beq
 Z\tti\colon\;\; \Psi\sim[Z+A(X)]^2\!/2+[P(X)\hl{\,-\,G}+\tst].
 \label{ldfc}
\eeq
The \hl{rightmost} bracketed contribution herein is a consequence of \eqref{ldeq} \hl{and that 
the interactive flow branches off the unperturbed state given by 
\m{[\Psi,P]\equiv[Z^2\!/2,\hl{G}]} infinitely far upstream}; TST means transcendentally small 
terms.

Relating the displacement function $A$ to $P$ closes the interactive feedback loop and the 
weakly elliptic free-interaction problem. For \m{X<0}, that relationship is given by the 
jet-type interaction law \m{P-G=\sgn(T-1)(A''-H_-'')}, typically provoked by the streamline 
curvature in the MD (as introduced by \citealp{Sm77,SmDu77} and, for an \hl{unconfined wall} 
jet passing an abrupt edge, \citealp{Sm78}) and the (counteracting) capillary pressure jump 
across the uppermost streamline. For \m{X>0}, one eliminates $H_-$ from the interaction law via 
the representation of $P$ in terms of the pressure jump across the lowermost streamline to 
which \eqref{dbcn} reduces:
\beq
 \De P\,\theta(X)=TH_-''/|T-1|,\q \De P\deq P-P_-.
 \label{phl}
\eeq 
(in SBP18 only the case \m{P_- =0} was considered). We thus arrive at the $P$/$A$ law in the 
form
\beqa
 & \De P=C(T)(G+SA''-P_-),\q S\deq\sgn(T-1), &
 \label{pal}\\[3pt]
 & C(T)\deq\lb\{\!\barray{cl} 1 & (X\leq 0), \\[2pt] T/(2T-1) & (X>0). \earray\rb.             
 \label{cd}
\eeqa
We furthermore introduce \m{D(T)=1-C(T)}. The upstream case \hl{(\m{X\leq 0})} is included in 
this interaction law for the sake of completeness and clarity. Downstream of the edge, it 
accounts for a subtle interplay are referred to tacitly from here on. The pole of $C$ points 
to an interesting local increase of the capillary action for \m{T\sim 1/2}. The passage of $T$ 
over this threshold (where surface tension exactly compensates the streamwise momentum of the 
pressure-free base flow) is associated with an unbounded increase of $P$ and $H$ over $A$ and 
implies the onset of condensed interaction, which causes a breakdown of the existing flow 
description for the free jet. This requires the introduction of a streamwise scale relatively 
short as compared to the stretched interactive one and can be interpreted as choking of a 
capillary wave. A second critical value \m{T=1} (\m{S=0}) describes the cancelling of the 
counteracting effects of streamline curvature and capillarity on the transverse momentum 
transfer. Both are subsumed by $A''$ and thus actually originate in the viscous forcing of the 
LD. The absence of their net influence hampers the interaction pressure from becoming 
effective, where $H_-$ remains unspecified according to \eqref{phl}, unless $A$ grows 
significantly to allow for a proper regularisation over a suitably shortened scale. Both 
exceptional situations are skated over below (\S\,\ref{sss:sia}) and still subject of ongoing 
investigations.

\bfig
 \centering
 \scalebox{0.7}{\input{c.pdf_t}}
 \caption{$C(T)$ (solid) and $D(T)$ (dashed) by \eqref{cd} (\m{X>0}) with their asymptote 
  and poles (\hl{all} dotted), fixed point and zeros (\hl{all as} circles).}
 \label{f:c} 
\efig

The rescaled shear stress exerted at the plate, \m{\La(X)\deq\Psi_{Z\!Z}(X,0)}, plays a 
crucial role for the (unambiguous) formulation of the initial conditions (ICs) imposed at the 
plate edge \m{X=0} by SBP18 for the detached flow, controlling its upstream influence on the 
plate-bounded flow in a unique manner. The detailed rationale underlying these deserves to be 
clarified in terms of the following three steps.
\benum
 \item[(I)] The two original demands on the interaction mechanism were the simultaneous 
continuous approach of the overall pressure jump across the layer towards $-P_-$ and of $\La$ 
towards zero in the limit \m{X\ttz-}, but only the first of these typical edge conditions can 
be met.
 \item[(II)] If
  \beq
   \ep^{12/7}\ll T<1\q (S=-1,\;\; T\neq 1/2),
   \label{tc}
  \eeq
\hl{which is the case pursued here,} the conditions the flow has to meet at the edge can then 
be formulated without resorting to the analysis of smaller regions enclosing the edge.
 \item[(III)] Then a least-degenerate flow description that allows for a smooth gradual 
transition from attachment to detachment of the flow quantities on smaller streamwise scales 
requires continuity of $\Psi$ and $A'$ above the edge.
\eenum

\hl{The sought quantities $\Psi$ and $P$ satisfy the, with respect to $X$, first- and 
second-order equations \eqref{ldeq} and \eqref{pal}. In turn, three ICs are required to 
continue marching over the edge:}
\beq
 \Psi_0\deq\Psi(0+,Z)=\Psi(0-,Z)\;\;\; (Z>0),\;\;\; T[A'(0+)-A'(0-)]=0,\;\;\; P(0)=P_-
 \label{ic}
\eeq
 \label{iap}%
\eseq
\hl{(or, equivalently, \m{A''(0)=-SG}). These} complete the interaction problem \eqref{iap} 
for the free jet. Here the flow profile at  detachment $\Psi(0-,Z)$ and $A'(0-)$ are taken as 
obtained by the preceding sweep of numerical marching towards the edge. It is stressed that 
$\Psi$, $P$ behave regularly as \m{X\ttz-}. Moreover, these quantities are continuous across 
the edge except for the shear stress $\Psi_{Z\!Z}$ on \m{Z=0}, owing to \eqref{ldbc}. 

We also recall the behaviour, inferred from (\ref{iap}\tit{a},\tit{b}), for
\beq
 Z\ttz\colon\;\;
 \Psi\sim\lb\{\!\barray{lc} \La(X)Z^2\!/2+P'(X)Z^3\!/6+O(Z^5) & (X\leq 0), \\[5pt]
                            \Us(X)Z+[P'+\Us\Us'](X)Z^3\!/6+O(Z^5) & (X>0).
              \earray\rb.
 \label{psiz0}
\eeq
Hence, the finite slip emerging along the lower free streamline, $\Us$, supersedes the finite 
plate stress $\La$ upstream of the edge. We note that \eqref{psiz0} first implies
\beq
 \Psi_0\sim\Lao Z^2\!/2+P'(0-)Z^3\!/6+O(Z^5)\q (Z\ttz),\q \Lao(G,T)\deq\La(0-).
 \label{psie}
\eeq
The apparent non-uniformity of \eqref{psie} for \m{X=0+} is the topic of \S\,\ref{ss:hrw} 
below. The parameters $G$ and $P_-$, representing the freely chosen support pressure, enter 
the solution of the interaction problem only via \eqref{ic}, \ie $G$ in terms of the imposed 
momentum flux, and subsequent integration of $P'(X)$ found in the course of the marching 
procedure. The decoupled calculation of $H_-$ is finally provided by \eqref{phl}. Eliminating 
$P$ with the aid of \eqref{pal} gives the alternative relation
\beq
 H_-(X)=D(T)\bigl[A(X)-A(0-)-A'(0-)X+(G-P_-)SX^2\!/2\bigr],
 \label{h-}
\eeq
\ie \m{H_-(0)=H_-'(0)=0}. Evidently, the support pressure behaves as a body force 
counteracting gravity. 

\subsubsection{Some important aspects}\label{sss:sia}

To achieve the last requirement in \eqref{ic}, the interaction is initiated in the limit 
\m{X\ttmi} by a controlled branching from the oncoming base flow, here maintained as the 
trivial solution \m{\Psi\equiv Z} for \m{X\leq 0} if \m{G=P_-\geq 0}. Hence, the case 
\m{G>P_-} requires branching of expansive type as scrutinised by SBP18 (where \m{P_-=0} 
throughout) and the opposite one \m{0\leq G<P_-} compressive branching (unconsidered so far). 
However, since $A''(X)$ is the streamline curvature in the interactive limit, it becomes 
evident from \eqref{pal} that the interactive feedback loop triggers stationary capillary 
waves iff \m{SC>0}. Here this implies \m{0<T<1/2} or \m{T>1}; see the preceding studies by 
\cite{BoSm92} and SBP18 and the preliminary presentation of these interactive undulations by  
\cite{ScBoPa19}. \hl{Their revealing linkage to unsteady linear capillary waves is given in 
appendix~\ref{a:b}.}

Moreover, SBP18 demonstrated how the phenomenon of stationary waves up- and downstream of the 
edge for \m{T>1} is associated with pre-detachment and severely violates the considerations 
(I)--(III) and the notion of expansive branching. They finally disclosed non-uniqueness of the 
solutions due to an arbitrary phase shift far upstream, \hl{presumed  fixed by an as} yet 
missing further downstream condition. We are therefore still left with the two constraints 
\eqref{tc} in our consistent description of the flow continued downstream of the edge by dint 
of \eqref{iap}. The first states that not only $A(X)$ but also $A'(X)$ is continuous at 
\m{X=0}, so that we henceforth omit the signs in the arguments $0-$ and $0+$ of 
$A$, expressing one-sided limits. The second guarantees strictly forward interacting flow 
upstream of the edge, thus \m{\Lao>0} in \eqref{psie}. Since realistic values of $\tau$ and $J$ 
by \eqref{w} yields \m{T\lesssim 10}, assuming \m{T<1} seems acceptable: \hl{see 
table~\ref{t:dat} and the last comment in appendix~\ref{a:a}.}

However, $A$ becomes discontinuous at the edge in the limit \m{T\ttz} in \eqref{pal} 
and \eqref{ic}, implying the \hl{absence of} interaction (\m{P'\equiv 0}) for \m{X>0}. Here 
the possibility of free interaction exists but the conditions at \m{X=0} do not provoke it 
even upstream of the edge in the formal limit \m{G-P_-=T=0}. Then the classical Goldstein wake 
\citep{Go30} is recovered immediately downstream as the trivial solution 
\m{[\Psi,P]\equiv[Z^2\!/2,\hl{G}]}, representing the oncoming base flow, applies upstream of 
it. 

\section{Inviscid detachment at smaller scales}\label{s:ids}

As emphasised in more detail below, the interactive flow structure leaves us with a still 
singular transition from no- to free slip. It therefore initiates its own breakdown on scales 
much smaller than the interactive ones. The bottom line of the subsequent analysis is that of
demonstrating self-consistency of the interaction theory and a required smooth behaviour of 
all flow quantities at the edge demands a thorough analysis of the smaller scales 
\hl{(figures~\ref{f:scal-sep}b--d)}. This will also highlight the strikingly different 
characteristics of the gross break-away of the film, \ie the formation of a free streamline at 
the solid wall, in the present situation and (well-understood) steady internal separation. In 
the first, the flow quantities appear to undergo weak algebraic singularities, whereas in the 
second their behaviour is well-known to be regular at separation \citep{Go30}. 

\subsection{The influence of capillarity}\label{ss:ioc}

To advance further in completing the description of flow detachment, it proves useful to first 
summarise the analysis in SBP18 of the interplay of surface tension and the Goldstein wake in 
the non-interactive limit \m{x\ttz+}. Here the latter exerts a displacement $-a x^{1/3}$ with 
some constant \m{a>0} (\m{a\simeq 1.0079} if $\psi_0$ is given by Watson's profile \hl{on top 
of the wake), so that} \m{\psi\sim\psi_0(z)+a x^{1/3}\psi_0'(z)+O(x^{2/3})}. Accordingly, 
\eqref{kbc}, \eqref{psi0h} and the Prandtl shift in \eqref{mdexp} yield 
\m{[h_-,h_+]\sim[a_-,a_- -a]x^{1/3}+O(x^{2/3})} with some sought constant $a_-$, and 
\eqref{nsy} states that \m{p_y+g\sim\ep^2(a-a_-)(x^{1/3})''\psi_0'^2(y)}. By integration across 
the unperturbed layer, from \m{y=0} to \m{y=h_0}, one finally obtains from \eqref{dbcn} the 
limiting overall capillary pressure jump in the form \m{(a-a_-)x^{1/3}\sim -T(h_- +h_+)}, 
\ie \m{T(2a_- -a)=a_- -a}. This implies \m{[h_-,h_+]\sim a x^{1/3}[D,-C](T)}, \hl{\cf 
\eqref{cd}}. One draws the important conclusion that $h_-(x)$ is \hl{required to be 
regularised on the} interactive and again on smaller scales even for \m{T\geq 0}, whereas 
\m{h_+'(x)\;(>0)} remains continuous at \m{x=0} for \m{T=0} as the inverse Prandtl shift 
produces additional irregular terms in the core region for \m{x\ttz+} and a cuspidal 
distortion of $h_+(x)$ exists for \m{T>0} only. Even then, however, the \hl{complete} 
regularisation of $h_+(x)$ is left to higher orders over the \hl{interactive} $x$-scale, where 
it is accomplished by the introduction of a thin shear layer adjacent to the upper free surface 
in order to satisfy \eqref{dbct} (\cf SBP18, \S\,\ref{sss:do}).

It is noteworthy to highlight the difference to the related classical situation of the 
gravity- and capillarity-free axisymmetric flow exiting a pipe \citep{Ti68}. There symmetry 
cancels the leading-order displacement in the core region but the vorticity gradient of the 
Hagen--Poiseuille profile (as opposed to streamline curvature) provokes an higher-order 
displacement and vertical pressure, requiring a regularisation similar to that discussed 
below. 

Keeping in mind the above preliminary considerations \hl{operating for} arbitrarily small 
values of $T$, we consider the precise regularisation of $h_\pm$ for finite values of~$T$. To 
this end, we first reappraise the interaction under the first of the restrictions \eqref{tc}. 
The details of the detached flow in the close vicinity of the edge as reported by SBP18 
provide an insight into how the full interactive structure is recovered for 
\m{\ep^{9/14}\ll X=O(T^{3/8})}. In general, the so-called near-near wake, replacing the 
pressure-free Goldstein near-wake, emerges as a subregion split off the main portion of the LD 
to absorb the nonlinearity of the interaction immediately downstream of the trailing edge. 
Most importantly, it dictates the onset of free slip according to~\eqref{ldbc}.

\subsection{Extended Hakkinen--Rott wake}\label{ss:hrw}

As the second of the ICs \eqref{ic} requires \m{A-A(0)=O(X)} (\m{X\ttz}), the near-near wake 
must suppress any larger contribution to $A$, hence transferred passively through the core of 
the LD. As a consequence of this leading-order analysis, this wake itself then provides an 
example of condensed interaction through an interesting, capillarity-controlled specification 
of the pressure-driven Hakkinen--Rott wake (HRW, \citealp{HaRo65}): $P$ vanishes as \m{X\ttz} 
in an irregular manner such that the wake exerts zero displacement. Since the canonical 
pressure gradient in the HRW turns out to be adverse, the capillary pressure jump \eqref{phl} 
\hl{enforces the lower free streamline to be convex immediately downstream of detachment in 
\m{X=0} (where it is curvature-free). It thus bends} vertically upwards as $X$ grows. The 
strong pressure rise provokes an enhanced streamline curvature, and this in turn the 
aforementioned breakdown and required smoothing of the interaction theory for sufficiently 
small values of $X$, as already indicated in figure~\ref{f:conf}. In the LD, this 
\hl{behaviour may be} fully understood if one considers only the behaviour of the 
leading-order quantities $\Psi$ and $P$, \ie under the neglect of the vertical pressure 
variations.

The flow profile in the HRW matches that at detachment at its upper extent in its limiting 
form given by \eqref{psie}. As a result, the self-preserving flow in the HRW discerned for 
\m{X\ttz+} resolves the non-uniformity of \eqref{psie}. It is expressed as the inner limit
\beq
 \biggl[\fr{\Psi}{\Lao^{1/3}X^{2/3}},\fr{\De P}{\Lao^{4/3}X^{2/3}},
 \fr{H_-}{\Lao^{4/3}X^{8/3}}\biggr]\sim 
 \biggl[\fhr(\eta),\phr,\fr{9\,\phr}{40}\fr{1-T}{T}\biggr],\;\;\;
 \eta\deq\fr{\Lao^{1/3}Z}{X^{1/3}},
 \label{hr}
\eeq
with the pressure difference $\De P$ introduced in \eqref{phl}. Here the universal wake 
function $\fhr$ satisfying \m{\fhr'^2-2\fhr\fhr''=-2\phr+3\fhr'''}, \m{\fhr(0)=\fhr''(0)} and 
the matching condition \m{\fhr'\sim\eta+\tst} as \m{\eta\tti} is recalled. The absence of a 
constant \hl{displacement} term determines the eigenvalue $\phr$ and prevents $A$ from being 
of $O(X^{1/3})$ as \m{X\ttz+} and enforces continuity of $A'$ as required by \eqref{ic}. Our 
refined numerical study yields \m{\phr\simeq 0.61334} 
and a rescaled free slip \m{\fhr'(0)\simeq 0.89915} \hl{obtained with} \m{\max(\eta)=50} 
\citep[\cf][SBP18]{HaRo65}. This gives \m{\Us\sim\fhr'(0)X^{1/3}} (\m{X\ttz+}) in 
\eqref{psiz0} when rewritten in the limit \m{\eta\ttz}.

Next, we propose the regular/singular upstream/downstream behaviour including higher orders
\beq
 \De P\sim\lb\{\!\barray{ll} P'(0-)X+P''(0-)X^2\!/2+O(X^3) & (X\ttz-), \\[2pt]
 \phr\Lao^{4/3}X^{2/3}+c_1 X\ln{X}+c_2 X+O\bigl(X^{4/3}(\ln{X})^2\bigr) & (X\ttz+)
 \earray\rb.
 \label{p+}
\eeq
with the logarithmic variations and the constants $c_1,$ $c_2$ to be determined through a  
higher-order analysis of the HRW. Accordingly, from (\ref{iap}\tit{e}--\tit{g}) or \eqref{h-},
\beq
 A-A(0)\sim A'(0)X+(G-P_-)\fr{X^2}{2}+\lb\{\!\barray{l} O(X^3), \\[3pt]
 \disp \hl{-\fr{9\,\phr\Lao^{4/3}}{40\,C(T)}}\,X^{8/3}+O(X^3\ln{X}).
 \earray\rb.
 \label{a+}
\eeq
Our expectation of a more nonlinear theory superseding the current one when $T$ crosses $1/2$, 
at the pole of $C(T)$, complies with the sign change of the singular contribution to $A$ 
provided by the HRW. That weak downstream irregularity is also transferred to $H_+$, \cf 
\eqref{mdexp}, as
\beq
 H_+\sim -A(0)-A'(0)X-(G-P_-)\fr{X^2}{2}+\lb\{\!\barray{l} O(X^3), \\[5pt]
 \disp \fr{9\,\phr\Lao^{4/3}}{40}\,X^{8/3}+O(X^3\ln{X}).
 \earray\rb.
 \label{h+}
\eeq
By the expansive type of interaction for \m{S=-1}, $A(X)$ bends convexly but $P(X)$ concavely 
throughout (SBP18). That is, we can expect here \m{A(0)>0}, \m{A'(0)>0}, but \m{P'(0-)<0}. 

One infers from \eqref{ldfc} that the $i$-th (\m{i=1,2,\ldots}) contribution to the expansion 
for \m{\Psi-\Psi_0} as \m{X\ttz} attains the form \m{d_i(X)Z+e_i(X)+\tst} as \m{Z\tti} where 
the series of gauge functions $d_i$ and $e_i$ are determined by the expansions \eqref{p+} and 
\eqref{a+} and add up to respectively \m{A(X)-A(0)} and \m{[A(X)^2-A(0)^2]/2+P(X)}. Typically, 
\m{e_i(X)\Psi_0'(Z)} are the eigensolutions of the linearised convective operator in 
\eqref{ldeq}. By matching $\Psi$ in the LD and the MD, the solution of the inviscid version of 
\eqref{ldeq} indeed yields the accordingly refined form of the expansion for $\Psi$ given by 
SBP18 (as \eqref{p+}, correctly including the logarithmic terms). So, with $\De P$ expanded as 
in \eqref{p+}, we have for
\beq
 X\ttz\pm\colon\;\;
 \Psi-\Psi_0-A'(0)X\,\Psi_0'(Z)\sim\De P\,\Psi_0'\int_Z^\infty\!\fr{\ud t}{\psi_0'^2(t)}\sim
 \De P\lb\{\!\barray{ll} 1 &\! (Z\tti), \\[2pt] 1/\Lao &\! (Z\ttz). \earray\rb.
 \label{psi+}
\eeq
A detailed higher-order analysis of the HRW demonstrates self-consistency of the interactive 
asymptotic structure for \m{X\ttz}. Amongst other aspects, it fixes the dependence of the 
coefficients $c_1$, $c_2$ in \eqref{p+} on the parameters characterising the LD flow in the 
limit \m{X\ttz-}. Here we refer the interested reader to \ref{s:a}.

The breakdown and \hl{so a required} regularisation of the interactive flow structure for 
sufficiently small values of $X$ is due to an unbounded vertical flow component and vertical 
pressure gradient evoked by the $O(X^{2/3})$-term in \eqref{p+} and \eqref{psi+} and the 
associated $O(X^{3/8})$-term in \eqref{a+}. As a crucial observation, even then the pressure 
gradient in the HRW stays imposed by the flow on its top and must vary such that a potential 
singular displacement varying with $X^{1/3}$ is suppressed. Since the self-similar structure 
of the HRW already absorbs this type of condensed interaction and is recovered at its origin 
closer to the trailing edge, \eqref{hr} prevails even over an $x$-scale much smaller than the 
interactive one. As \hl{a result}, $h_-$ is still given by \eqref{hr} \hl{in \S\,\ref{ss:oie} 
below}.

\subsection{Outer and inner Euler regions}\label{ss:oie}

We here consider the two nested square outer and inner vortical-flow regions (when measured 
by the equally scaled global horizontal and vertical coordinates $x$ and $\ep y$) that 
supersede locally the MD (outer) and the LD (inner) but where \m{\psi\sim\psi_0} and 
\m{\Psi\sim\Psi_0} still govern the flow at leading order. The associated linearised Euler 
stages (outer and inner RS) account for the small-scale upstream influence, within that on the 
interactive scale, and serve to regularise the singular behaviour predicted in 
\S\,\ref{ss:hrw}; most importantly, $h_+(x)$ by virtue of $H_+$ (outer). It is furthermore 
noted that the aforementioned large-$Z$ representation of the expansion \eqref{psi+} 
accompanies a passive re-ordering of its hierarchy, so as to match the small-$X$ limit of 
\eqref{mdexp} provided by \eqref{a+}. Accordingly, the last expansion enforces a contribution 
of $O(X^{3/8})$ to \eqref{psi+} and this in turn a pressure-driven one of $O(X^{2/3})$ to the 
non-interactive disturbance of $O(\ep^{4/7})$ in \eqref{mdexp}. 

\subsubsection{Preliminaries}\label{sss:p}

Introductory considerations lay the foundation for the outer and the inner mechanism for the 
further regularisation of the HRW, as follows.
\benum
 \item[(a)] The interactive $u$- and $p$-variations, on account of streamline curvature via 
the vertical pressure variation in \eqref{nsy}, are of respectively $O(\ep^{2/7}X^{8/3})$ and 
$O(\ep^{4/7}X^{2/3})$ as \m{X\ttz+}. They and the non-interactive $u$-perturbation in 
\eqref{mdexp}, provoked by the streamwise pressure variation through \eqref{nsx}, all become 
of $O(\ep^{2/3})$ in the outer RS (\S\,\ref{sss:ors}) where, \cf \eqref{ldxy},
\beq
 \bX\deq x/\ep=X/(l\ep^{1/7})=O(1).
 \label{bx}
\eeq
 \item[(b)] Conversely, $v$ of $O(\ep^{5/7}X^{-1/3})$ grows significantly to become 
comparable in size to the $u$-perturbation of \m{\ep^{2/7}X^{2/3}} across most of the LD for 
\m{X=O(\ep^{3/7})}, \ie in the inner RS (\S\,\ref{sss:irs}) where 
\beq
 \hX\deq x/(m\,\ep^{9/7})=X/(lm\,\ep^{3/7})=O(1).
 \label{hx}
\eeq
However, as $p$ and $\psi$ of $O(\ep^{2/3})$ at its base and downstream of the edge are still 
prescribed by the HRW, the inner RS cannot regularise the associated singularity expressed by 
\eqref{p+} and \eqref{psi+}. Therefore, the analysis of inner RS is of only subordinate 
importance compared to that of the outer one.
 \item[(c)] A quick justification of the expansions of the flow quantities below for both 
square regions relies on the relevant inviscid-flow approximation of the elliptic vorticity 
transport equation, obtained from elimination of the pressure in (\ref{gov}\tit{a},\tit{b}):
\beq
 \psi_{yy}+\ep^2\psi_{xx}\sim -\Om(\psi)\deq\psi_0''\bigl(\psi_0^{-1}(\psi)\bigr).
 \label{vteoi}
\eeq
To express $\Om$ as the vorticity conserved along the streamlines, we use $\psi_0^{-1}$ to 
symbolise the inversion of the corresponding leading-order relationship \m{\psi\sim\psi_0(y)}. 
As a consequence, the contributions to those expansions are triggered by the vorticity 
imposed by the surrounding interactive flow and, in addition, the vorticity produced by the 
HRW and entering via non-trivial matching or BCs. These are provided by \eqref{psi+} with 
\eqref{p+} for \m{Z\tti} at the base of the outer RS and on top of the inner RS and by 
matching \eqref{psi+} for \hl{\m{Z\ttz}} and \eqref{hr} at the base of the latter. 
Consequently, eigensolutions of the linearised operator in \eqref{vteoi} are absent.
\eenum

It is noteworthy that \cite{St68} discovered the generic advent of a linearised Euler or 
Rayleigh stage when he solved the (non-rigorous) Oseen approximation of the NS problem 
governing the unconfined flow in a small region around a trailing edge, \hl{and prior to the
far-reaching rigorous appreciation of viscous--inviscid interaction on larger scales}
\citep{St69,Me70}.

\subsubsection{Outer Rayleigh stage: main deck}\label{sss:ors}

In the outer square region, $p$ is, as in the surrounding MD, of $O(\ep^{4/7})$, and the 
viscous terms in (\ref{gov}\tit{a},\tit{b}) become formally of $O(\ep)$ as all remaining ones 
can be scaled to $O(1)$. Following the comments (a) and (c) above, substitution of \eqref{a+} 
into \eqref{mdexp} suggests, in this domain, the expansion 
\m{\psi\sim\psi_0(y)+\ep^{2/7}\psi_1(y)+\ep^{3/7}\psi_2(y)+\ep^{4/7}\psi_3(y)+O(\ep^{4/6})}. 
The sought functions $\psi_{1,2,3}$ satisfy the hierarchy of Rayleigh equations
\beq
 (\p_{yy}+\p_{\bX\!\bX}-\psi_0'''\!/\psi_0')\psi_{1,2}=0,\q 
 (\p_{yy}+\p_{\bX\!\bX}-\psi_0'''\!/\psi_0')\psi_3=\psi_1^2(\psi_0'''/\psi_0')'/(2\psi_0')
 \label{res}
\eeq
resulting from expanding \eqref{vteoi}. According to the considerations following 
\eqref{vteoi} and the regularity of \eqref{mdexp} upstream of the trailing edge, $\psi_{1,2}$ 
consist just of the pressure-free disturbances given by the Taylor series of $A(X)$ up to 
second order, where \m{A''(0-)=G-P_-} from \eqref{pal} subject to \eqref{ic}. This and the 
inhomogeneity in the last equation in \eqref{res}, caused by the inertia-based nonlinearities, 
require an additional $y$-dependent component of $\psi_3$. 

Specifying these findings gives 
\begin{align}
 \bigl[\psi,h_+\bigr] &\sim \bigl[\psi_0(y),h_0\bigr]
 \nonumber\\[2pt]
 &+ m\bigl[\ep^{2/7}\!A(0)+\ep^{3/7}\!A'(0)\,l\bX+\ep^{4/7}(G\!-\!P_-)(l\bX)^2\!/2\bigr]
 \bigl[\psi_0'(y),-1\bigr]
 \nonumber\\[2pt] 
 &+ \ep^{4/7}m\bigl[\psi_\ast(y),-\psi_\ast(h_0)/u_0^+\bigr]+ 
 \ep^{4/6}\bigl[\bPsi(\bX,y),\bH(\bX)\bigr]+O(\ep^{5/7}) .
 \label{oexp}
\end{align}
Hence, $\psi_\ast$ denotes the limiting value of the corresponding $O(\ep^{4/7})$-contribution 
to the expansion \eqref{mdexp} of $\psi$ in \m{X=0}. That quantity satisfies 
$\psi_\ast''-(\psi_0'''/\psi_0')\psi_\ast=(\psi_0'''/\psi_0')'A(0)^2\!/2-A''(0-)\psi_0 '$, 
where the last inhomogeneity reflects the action of the streamwise pressure gradient. We 
furthermore expand
\beq
 p\sim\ep^{4/7}m l^2\biggl[(G\!-\!P_-)\int_0^y\!\psi_0'^2(t)\,\ud t-
 M\Bigl(\fr{Gy}{h_0}\!-\!P_-\Bigr)\biggr] +\ep^{4/6}\bP(\bX,y)+O(\ep^{5/7}).
 \label{op}
\eeq
The $X$-independent leading-order term in \eqref{op} is again just the dominant contribution 
to $p$ in the MD up- and downstream of the trailing edge (\cf SBP18) evaluated at \m{X=0} and 
rewritten with the aid of \eqref{gptm}. Here the irregular terms in \eqref{a+} play no role. 
It follows from inserting \eqref{oexp} into \eqref{nsy} and integrating its thereby reduced 
form \m{p_y\sim\ep^{4/7}m l^2\psi_0'^2 -g} subject to \m{p\sim p_-} as \m{y\ttz}, to match $p$ 
in the LD. Moreover, \eqref{oexp} fulfils \eqref{kbc} supplemented with \eqref{psi0h} and, 
together with \eqref{op}, complies with the capillary pressure jump at \m{y=h_+} in 
\eqref{dbcn} up to $O(\ep^{4/7})$ for \m{\bX=O(1)}. The $O(\ep^{4/6})$-contributions to 
\eqref{oexp}, \eqref{op} serve to regularise the flow quantities in the MD. As the subsequent 
analysis of $\bPsi$, $\bH$, $\bP$ makes clear, those expansions do not contain lower-order 
eigenfunctions having sufficiently strong decay for \m{|\bX|\tti}, consistent 
with~\eqref{mdexp}.

Invoking the inverse Prandtl shift in \eqref{oexp} gives 
\m{\psi_0(y)\sim\psi_0(z)+h_-\psi_0'(z)} for \m{h_-=O(\ep^{2/3})}, see \eqref{h+}, and brings 
to mind matching $\psi$ up to $O(\ep^{2/3})$ in \eqref{mdexp} and also in the LD, according to 
\eqref{psi+} and \eqref{p+}. Furthermore, $\bPsi$, $\bP$ are seen to satisfy the linearised 
Euler equations
\beq
 \psi_0''\bPsi_{\!\bX}-\psi_0'\bPsi_{\!y\bX}=\bP_{\!\bX},\q 
 \psi_0'\bPsi_{\!\bX\!\bX}=\bP_{\!y}.
 \label{lee}
\eeq
To separate the influence of the shear stress at detachment, $\Lao$, effective in the LD and 
of a potential $\bX$-independent contribution to $\bPsi$ arising from integration of 
\eqref{lee} (\ie no $O(\ep^{2/3})$-contribution to $\Om$, \cf \eqref{vteoi}, in the 
surrounding MD), we advantageously consider the scaled vertical flow perturbation
\beq
 \bV\deq -\bPsi_{\!\bX}/\bLa,\q \bLa\deq 2\,\phr\,\la^{1/3}\!\Lao^{4/3}\!\big/3.
 \label{bv}
\eeq
Equations \eqref{lee} yield the Rayleigh equation governing $\bV$ in accordance with 
\eqref{res}:
\bseq
\beq
 (\p_{yy}+\p_{\bX\!\bX}-\psi_0'''\!/\psi_0')\bV=0.
 \label{re}
\eeq
Matching $\psi$ and $p$ in the outer RS and the LD with the support of \eqref{ldexp} and $m$, 
$l$ given by \eqref{mdexp}, \eqref{ldxy} requires for
\beq
 y=0\colon\;\; \bV=-\theta(\bX)\bX^{-1/3}.
 \label{bv0}
\eeq
Furthermore, expanding \eqref{kbc} and \eqref{dbcn} gives
\beq
 \bH=-\bPsi(\bX,h_0)/u_0^+
 \label{bh}
\eeq
and \m{\bP(\bX,h_0)=-\tau\bH''(\bX)} respectively. By the same token, inspection of 
\eqref{lee} with the help of \eqref{psi0h}, \eqref{gptm} and \eqref{bh} gives for
\beq
 y=h_0\colon\;\; u_0^+{}^2\,\bV_{\!y}=-TJ\bV_{\!\bX\!\bX},
 \label{bvh}
\eeq
 \label{bvpr}%
\eseq
\ie the explicit dependence of $\bV$ on $T$. Also, matching \eqref{oexp}, \eqref{op} with 
\eqref{mdexp} and $p$ in the MD subject to \eqref{a+} and \eqref{p+} yields \m{\bPsi\ttz} and 
\m{\bP\ttz}, thus \m{\bV\ttz} and \m{\bH\ttz} by \eqref{bh}, as \m{\bX\ttmi}. In contrast, 
\m{\ep^{2/3}\bPsi} for \m{\bX\gg 1} must match the dominant singular behaviour of 
\m{\psi-\psi_0\sim\ep^{2/7}m(A-H_-)(X)\psi_0'(y)} for \m{X\ll 1} as inferred from 
\eqref{mdexp}. The expansions \eqref{hr}, \eqref{a+} and 
\m{\psi-\psi_0\sim -(9\,\phr/40)\ep^{2/3}m\Lao^{4/3}(l\bX)^{8/3}} imply
\m{\bV\!/\psi_0'(y)\sim 9\la\bX^{5/3}\!/(10M)+O(\bX^{-1/3})} (\m{\bX\tti}). Likewise, 
\eqref{bv} and \eqref{bh} give \m{\bH/\bLa\sim 27\la\bX^{8/3}\!/(80M)+O(\bX^{2/3})}. This 
proves consistent with the interplay of the two free surfaces in \S\,\ref{ss:hrw}. 

It is illuminating to demonstrate that the up- and downstream asymptotes are already intrinsic 
to the problem (\ref{bvpr}) governing $\bPsi$ and $\bH$. To this end, we consider the weakest 
admissible, \ie first algebraic, decays of $\bV$ for \m{\bX\to\pm\infty} with unknown dominant
corresponding rates $\ba_\pm(\bX)$, say. We obtain from (\ref{bvpr}\tit{a},\tit{b}), using 
\m{(\p_{yy}-\psi_0'''/\psi_0')\bV\equiv(\psi_0'\bV_{\!\!y}-\psi_0''\bV)_y/\psi_0'} and
standard methods and \eqref{jl}, the long-wave approximation of $\bV$
\beq
 \fr{\bV}{\psi_0'(y)}\sim\ba_\pm+\ba_\pm''
 \biggl[\bb_\pm-\!\int_0^y\!\fr{\ud t}{\psi_0'^2(t)}\int_0^t\!\psi_0'^2(s)\,\ud s\biggr]-
 \fr{\la\theta(\bX)}{\bX^{1/3}}\!\int_y^{h_0}\!\fr{\ud t}{\psi_0'^2(t)}+ 
 O(\ba_\pm''''\!,\bX^{-7/3})
 \label{bvi}
\eeq
where $\ba_\pm$ and the constants $\bb_\pm$ are determined by solvability conditions of the 
inhomogeneous problems governing the $O(\ba_\pm'')$- and the $O(\ba_\pm'''')$-term 
respectively. The small-$y$ behaviour of $\psi_0$ in \eqref{jl} grants a corresponding 
regularity of the right-hand side of \eqref{bvi}. Substitution of \eqref{bvi} into \eqref{bvh} 
using \eqref{jl} and \eqref{psi0h} gives, after division by $u_0^+$, the solvability relation 
\m{\ba_\pm''J-\la\theta(\bX)\bX^{-1/3}\sim\ba_\pm''\tau}. In the upstream case, this statement 
can only be met in the limit \m{T\to 1-}, \cf \eqref{gptm}. Consequently, \m{\ba_-\equiv 0}, 
\m{\bb_-=0}, and the upstream decay is indeed exponential, although the limit of an undamped 
(neutral or harmonic) oscillation may also be taken into consideration and an unbounded 
increase of $\bV$ is expected for \m{T\to 1-}. In contrast, 
\beq
 \ba_+=9\la\bX^{5/3}/[10 J(1-T)]
 \label{ba+}
\eeq 
confirms the aforementioned leading-order asymptote involving $M$ defined in \eqref{gptm}. 
This shows that matching \eqref{oexp} and \eqref{mdexp} requires \m{T<1}.

As a further result, \eqref{lee} yields
\beq
 \bP=\psi_0''\bPsi-\psi_0'\bPsi_{\!y},
 \label{bp}
\eeq
and \m{\bP\sim 3\la\bLa\bX^{2/3}\!/2} (\m{\bX\tti}) provides the match of $p$ in the MD, 
according to \eqref{p+}, \eqref{psi+} and \eqref{bv0}. This and 
\m{\bP(\bX,0)=3\la\bLa\,\theta(\bX)\bX^{2/3}\!/2} make evident how $\bPsi$ and $\bP$ resort to 
these behaviours originating in the HRW and why the inner RS is required to complete the 
regularisation closer to the trailing edge. Since the coefficient $\psi_0'''/\psi_0'$ in 
\eqref{re} becomes, from \eqref{jl}, \m{\om y} for \m{y\ll 1}, \eqref{bv0} allows $\bV$ to 
attain an undesired potential-flow pole in the origin, as described by the singular 
eigensolutions of the Laplacian \m{r^{-N}\sin(N\vth)} where
\beq
 r\deq\sqrb{\bX^2+y^2}\ttz,\q 0\leq\vth\deq\arctan(y/\bX)\leq\upi
 \label{r}
\eeq
and \m{N>0} is some integer \citep[\cf][]{Sc14}. Its occurrence has to be avoided in the 
further treatment of \eqref{bvpr}. Rather, \eqref{bv0} and the vorticity term provoke a weaker 
singularity as one readily finds that
\beq
 \bV\sim\bVo+\bc_1 y+\bc_2 xy+O(r^{8/3})\q (r\ttz),\q 
 \bVo\deq 2 r^{-1/3}\sin(\upi/3-\vth/3)/\sqrb{3},
 \label{bvo}
\eeq
and \eqref{bp} recovers the pressure induced by the HRW as \m{\bP=O(r^{2/3})}. The first three 
contributions to $\bV$ in \eqref{bvo} are of potential-flow type, and the coefficients 
$\bc_{1,2}$ of the homogeneous ones are determined by the overall solution for $\bV$. The 
(lengthy expression of the) $O(r^{8/3})$-term in \eqref{bvo} solves the Poisson problem to 
which \eqref{re} reduces to with \m{\psi_0'''\bV/\psi_0'\sim\om y\bVo} forming the 
inhomogeneity. The singularity described by $\bVo$ is pivotal in \S\,\ref{sss:irs} where it 
comes to its regularisation by the inner~RS.

For what follows, we introduce the Fourier transform of a function $f(\bX,y)$ for complex 
wavenumbers~$k$,
\beq
 \phi\{f\}(k,y)=\fr{1}{2\upi}\int_{-\infty}^\infty f(\bX,y)\e^{-\ui k\bX}\ud\bX.
 \label{ft}
\eeq
We first assume that $\bV$ decays exponentially far upstream. Since it grows with 
$O(\bX^{5/3})$ as $\bX$ becomes large, \eqref{ft} defines $\phi\{\bV\}$ first in the open 
strip \m{-\mu_1(T)<\Imag\,k<0} where \m{-\mu_1} denotes the imaginary coordinate of the pole 
in 
the lower half-plane \m{\Imag\,k\leq 0} lying closest to the real axis. The analytic 
continuation of $\bV$ into the entire $k$-plane excluding the locations of singularities is 
provided by the convenient decomposition
\beq
 \phi\{\bV\}(k,y)=\mcB(k)\mcV(k,y),\;\;\;
 \mcB(k)\deq\phi\bigl\{\theta(\bX)\bX^{-1/3}\bigr\}(k)=
 1\big/\bigl[\sqrb{3}\,\Ga({\txt\fr{1}{3}})(\ui k)^{2/3}\bigr].
 \label{vb}
\eeq
The last expression is understood in connection with a branch cut along the positive imaginary 
$k$-axis. Absorbing \eqref{bv0} and accommodating the non-integer growth with $\bX$ in 
\eqref{ba+}, it captures the influence of the HRW and gives a non-trivial $\bV$. Poles of 
$\mcV$ on the real $k$-axis allow for relaxing the original assumption of exponential decay by 
the inclusion of harmonic modes surviving far upstream. From \eqref{bvpr} we deduce the 
Rayleigh equation
\bseq
\beq
 (\p_{yy}-k^2-\psi_0'''\!/\psi_0')\mcV=0 
 \label{veq}
\eeq
subject to the then inhomogeneous lower and the homogeneous upper BC,
\begin{align}
 y=0\colon\;\; & \mcV=-1,
 \label{v0}\\
 y=h_0\colon\;\; & \psi_0'^2\mcV_y=T J k^2\mcV, 
 \label{vh}
\end{align}
 \label{vpr}%
\eseq
\cf \eqref{jl}. The solution of the two-point boundary value problem \eqref{vpr}, parametrised 
by $k$, facilitates the semi-analytical inversion of \eqref{ft} so as to determine $\bV$, 
parametrised by $\psi_0(y)$ and $T$, in an elegant manner, avoiding the above-mentioned 
Laplacian eigensolutions; all the more, as our focus lies on $\bH(\bX)$ given by \eqref{bh}. 
For the numerical implementation of \eqref{vpr}, we recall that $\psi_0$ is \hl{typically} 
specified by Watson's (1964) flow profile. In turn, the properties \eqref{psi0h}, \eqref{weq} 
and the closed form of $\psi_0$ in \cite{ScKl19} and the values for \m{J=\la} and $u_0^+$ given 
by \eqref{w} are employed. Detailing the properties of \eqref{vpr}, especially the behaviours 
of $\mcV$ for \m{k\ttz} and \m{|\Real\,k|\tti} and the analysis of its poles, which select the 
discrete spectrum of $\bV$ out of the continuous one (and where \eqref{vpr} does not have a 
solution but its homogeneous form does), is relegated to \ref{s:b}. These findings enable the 
representation of $\bV$ in most efficient manner as envisaged next.

The poles of $\mcV$ lie symmetrically with respect to both the real and the imaginary 
$k$-axes. There are a double pole at \m{k=0}, exactly two real simple poles where 
\m{k=\pm k_u(T)} with \m{k_u>0} (\S\,\ref{ss:b1}) and an infinite number of simple poles lying 
on \m{k=\pm\ui\mu_i(T)} (\m{i=1,2,\ldots}) with \m{\mu_i>0} (\S\,\ref{ss:b3}). Since 
\m{\mcV(-k,y)\equiv\mcV(k,y)}, \m{\res_{k=-k_u}(\mcV)=-\res_{k=k_u}(\mcV)} and real, and 
\m{\res_{k=-\ui\mu_i}(\mcV)=\res_{k=\ui\mu_i}(\mcV)} and imaginary. We then have
\beq
 \biggl[\bV,\fr{\bPsi}{\bLa}\biggr](\bX,y)=\int_\mcC \mcB(k)\mcV(k,y)\e^{\ui k\bX} 
 \biggl[1,\fr{\ui}{k}\biggr]\ud k
 \label{ift}
\eeq
where all possible paths of integration $\mcC$ stretch from \m{\Real\,k=-\infty} to 
\m{\Real\,k=+\infty} and originate from one another through a continuous deformation as they 
divide the $k$-plane in two portions: the origin and all poles \m{k=\ui\mu_i(T)} lie in the 
upper and all poles \m{k=-\ui\mu_i(T)} in the lower part. We furthermore anticipate that both 
real poles are located either in the upper or the lower part to guarantee $\bV$ being real. 
Indeed, as will be argued below to render $\bV$ unique, $\mcC$ must bypass both real poles 
such that they lie in the lower part. This situation is sketched in figure~\ref{f:kpl} with 
the path $\mcC$ specified for the numerical calculation of $\bH(\bX)$ by means of \eqref{ift} 
and \eqref{bh} for \m{\bX\geq 0}. There the branch cut prevents a more efficient treatment of 
\eqref{ift} using Cauchy's residue formula: to avoid accuracy issues associated with complex 
integration, we specified $\mcC$ to follow the real axis apart from small squares of lengths 
$2\vep$ with the midpoints \m{k=\pm k_u} and of length $\vep$ with the midpoint in the origin. 
Consistency of the results is confirmed for values of $\vep$ ranging from 0.1 to 0.3. On the 
other hand, applying Cauchy's residue theorem to \eqref{ift} yields with \eqref{vb}, the fact 
that \m{\res_{k=-k_u}(\mcV)=-\!\res_{k=k_u}(\mcV)} and Euler's reflection formula after some 
algebra 
\beq
 \fr{\bPsi}{\Ga(\fr{2}{3})\bLa}=2\res_{k=k_u}(\mcV)\,\fr{\cos(k_u\bX-\upi/3)}{k_u^{5/3}}+
 \ui\sum_{i=1}^{\infty}\res_{k=-\ui\mu_i}(\mcV)\,\fr{\exp(\mu_i\bX)}{\mu_i^{5/3}}\q 
 (\bX\leq 0)
 \label{psiu}
\eeq
\citep[\cf][]{Ti68}. This series of residues converges (uniformly) for any \m{\bX<0}. The full 
evaluation of \eqref{ift} and smoothness of $\bPsi$ for \m{y>0} in \m{\bX=0} confirms that 
\eqref{psiu} holds even there although the decay of the exponentials has disappeared. 

\bfig
 \centering
 \scalebox{0.5}{\input{kpl.pdf_t}}
 \caption{Sketch of $k$-plane: double-symmetric singular points (circles), actual path $\mcC$
  and direction of integration.}
 \label{f:kpl}
\efig

Finally, $\bH/\bLa$ for \m{\bX\leq 0} follows from \eqref{bh} and directly from \eqref{psiu} 
in a convenient manner. This approach allows us to check the accuracy of the full integration 
according to \eqref{ift}. It is definitely preferred for resolving most accurately the novel
discrete undamped capillary Rayleigh modes, forming a wave crest upstream of the edge. 
\hl{These are revealed, as arising from} the real poles, with wavenumbers \m{k=k_u}, found to 
strictly increase as $T$ decreases. Here we point to the classical dispersion relation of 
small-amplitude capillary waves in a finite-depth layer of uniform parallel flow with uniform 
speed scaled to unity over a flat bed (see \citealp[][p.~30]{DrRe04}; 
\citealp[][\S\,2.4.2]{VaBr10}). \hl{We can infer it directly from that of symmetric Squire 
modes \citep{Sq53} as discussed in appendix~\ref{a:b}, hence with $k/2$ therein replaced by 
$k$ here. Such stationary modes then exist for the two wavenumbers \m{k=\pm k_u} satisfying 
\m{1=T k_u\tanh{k_u}}. In the current setting,} we extract from \eqref{psiu} the neutral 
amplitude normalised with $\bLa$
\beq
 \ba_u\deq 2\,\Ga({\txt\fr{2}{3}})\res_{k=k_u}[\mcV(k,h_0)]\big/
 \bigl(u_0^+ k_u^{5/3}\bigr).
 \label{bau}
\eeq
\hl{A linchpin of the analysis in \ref{s:b} is the asymptotic representation of $k_u$ and 
$\res_{k=k_u}(\mcV)$ as $k_u$ vanishes and $\ba_u$ diverges for \m{T\to 1-} (\S\,\ref{ss:b1}) 
and the the qualitatively reciprocal behaviour for \m{T\ttz} (\S\,\ref{ss:b2}). In combination 
with \eqref{w} (for \m{x_v=-1}), this boils down to the following, numerically valuable, 
finite limits obtained with high accuracy:
\bseqa
 k_u/\sqrb{1-T}\simeq 1.8046,\;\;\; \ba_u(1-T)^{7/3}\simeq 0.2805 &&\;\;\; (T\to 1-),
 \slabel{to}\\[2pt]
 k_u T\simeq 1.1596,\;\;\; \ba_u k_u^{2/3}\exp(k_u h_0)\simeq 6.0422 &&\;\;\; (T\ttz).
 \slabel{tz}
 \label{toz}
\eseqa}

To compute \eqref{ift} (for \m{y=h_0}), we restrict the numerical integration to the interval 
\m{|\Real\,k|\leq 20}, which in view of the exponential large-$k$ tails of $\mcV$ 
(\S\,\ref{ss:b2}) gives satisfactorily accurate results. Specifically, we find
\m{\mcV(k,h_0)=O(\exp[-|k| h_0]/k)}. The evaluation of the integrand employs a cubic-spline 
interpolation of the solution $\mcV$ of the Rayleigh problem \eqref{vpr} for discrete values 
of $k$. We advantageously mitigated the singularity at \m{k=0}, circumvented at a small 
distance (see figure~\ref{f:kpl}), by splitting off the first two terms in the small-$k$ 
expansion of $\mcV(k,h_0)$ (\S\,\ref{ss:b1}) and finally adding their inverse Fourier 
transform, which results in the reciprocal large-$\bX$ representation of $\bV$ and $\bH$ via 
\eqref{ift}. We skip the details of this alternative derivation of \eqref{bvi} in its more 
complete form, supplemented with \eqref{ba+}, also yielding the corresponding asymptote of 
$\bH$ by integration. To evaluate \eqref{psiu} (for \m{y=h_0}) and discrete $\bX$-values, the 
poles of $\mcV$ are detected as the roots \m{k=k_p}, say, of $\mcV^{-1}(k,y)$. Since 
\m{\mcV\sim\res_{k=k_p}(\mcV)/(k-k_p)} as \m{k\to k_p}, the according residuals (given by a 
homogeneous solution to \eqref{vpr}, see above) are computed as \m{1/[\p_k\mcV^{-1}(k_p,y)]} 
(\m{y=h_0}). For \m{i>7} and $\bX$ lying not too close to zero, the values of the exponentials 
\hl{in \eqref{psiu}} have already fallen below the round-off error; a few more modes 
calculated using the asymptotic behaviour of the residuals (\S\,\ref{ss:b3}) were, however, 
added.

The resulting plots in figures~\ref{f:kuvst} and \ref{f:bhvsx} also employ cubic-spline 
interpolation of the pointwise data sets. Figure~\ref{f:bhvsx}(\tit{a}) displays the results 
obtained by summation of residuals. As one expects, these are slightly more accurate for very 
negative values of $\bX$ and for small values of $T$ than those found by the direct evaluation 
of \eqref{ift}. Figure~\ref{f:bhvsx}(\tit{b}) indicates that excellent agreement with the 
asymptotes found analytically can be achieved. It is seen that $\bH$ undergoes a trough 
immediately downstream of the edge before it recovers to rapidly assume \hl{the} algebraic 
far-downstream growth governed by \eqref{bvi}, \eqref{ba+} (see also \S\,\ref{ss:b1}). \hl{The 
second result in \eqref{tz}} corroborates the extremely rapid upstream decay of the Rayleigh 
modes found numerically as \m{T\ttz}. Even the maximum value of $k_u$ shown lies on the part 
of $\mcC$ considered for the numerical integration, but the suppression of exponentially 
growing terms in the calculation of $\mcV$ and the residuals becomes a numerically delicate 
task when $|k|$ becomes sufficiently large. In the long-wave limit \m{k_u\ttz} as \m{T\to 1-}, 
$\bPsi$ diverges both immediately upstream of the trailing edge, as $\ba_u$ grows like 
$k_u^{-14/3}$, and for constant but sufficiently large positive values of $\bX$. Also these 
findings compare favourably with the curves in figure~\ref{f:bhvsx}. The intriguing further 
implications of the long-wave limit are addressed in~\S\,\ref{s:cfp}.

\bfig
 \centering
 \scalebox{0.7}{\input{ku+a-vs-t.pdf_t}}
 \caption{Wavenumber $k_u$ (solid) and amplitude $\ba_u$ (dashed), see \eqref{bau}, of the 
  neutral capillary mode vs.\ inverse Weber number $T$, asymptotes \hl{for \m{T\to 1} (dotted) 
  and \m{T\ttz} (dash-dotted) from \eqref{toz}}.}
 \label{f:kuvst} 
\efig

\bfig
 \centering
 \scalebox{1.}{\input{h-vs-x.pdf_t}}
 \caption{$\bH$ vs.\ $\bX$ and $T$ (\tit{a})~upstream, (\tit{b})~downstream of trailing edge: 
  labels indicate $T$-values; multiples \m{\neq 1} of $\bH$ (in parentheses) shown for 
  enhanced visibility; plot resolution of strongly augmented oscillations for \m{T=0.1} 
  discerned in (\tit{a}); two-terms downstream asymptotes (dashed) from \eqref{bvi} with 
  \eqref{ba+}.}
 \label{f:bhvsx} 
\efig

\hl{We complete the numerical analysis by demonstrating the excellent agreement between the 
computed wavelengths \m{2\upi/k_u}, see figure~\ref{f:bhvsx}(\tit{a}), and their leading-order 
asymptotes: for \m{T=0.95} and \m{T=0.8}, \eqref{to} predicts those as about $15.570$ and 
$7.785$ respectively; for \m{T=0.1}, \eqref{tz} gives a wavelength of about $0.542$. In 
addition, \m{\ba_u\simeq 8.654\times 10^{-10}}. The details of this case displayed in 
figure~\ref{f:bhvsx01} shows that our numerical method resolves even the rapid oscillations of 
exponentially small amplitude for rather small $T$-values with surprisingly high accuracy.}

\bfig
 \centering
 \scalebox{0.58}{\begin{picture}(0,0)%
\includegraphics{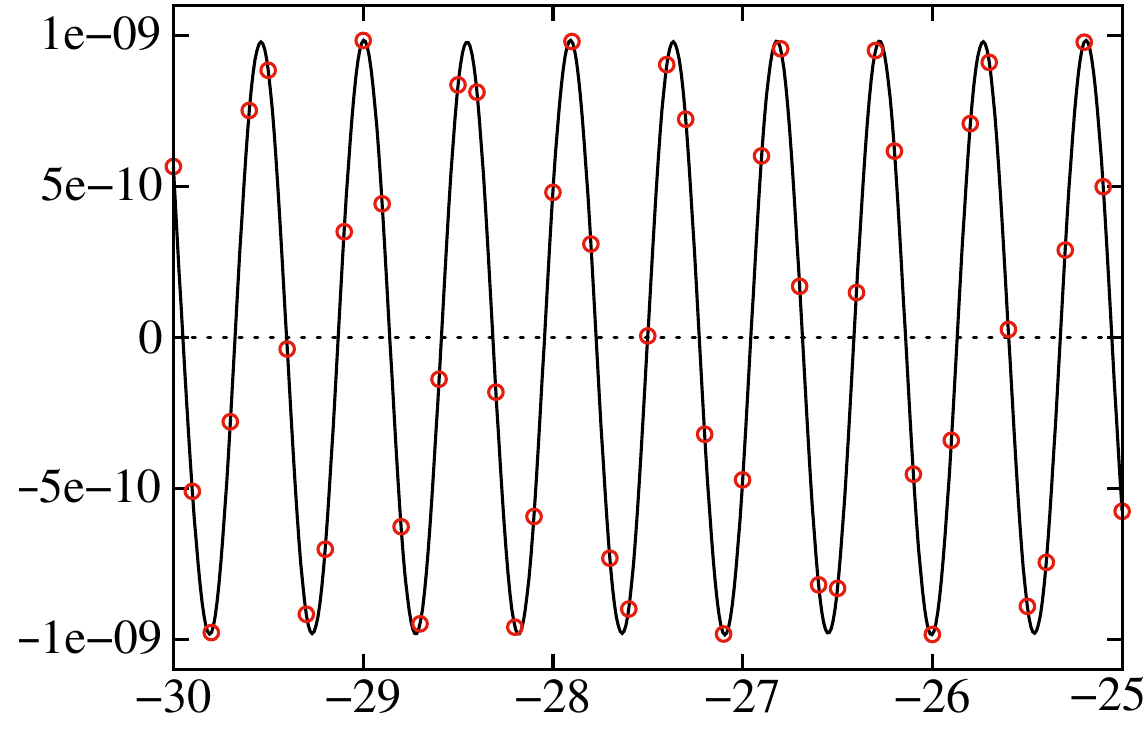}%
\end{picture}%
%
%
\setlength{\unitlength}{3947sp}%
\begingroup\makeatletter\ifx\SetFigFont\undefined%
\gdef\SetFigFont#1#2#3#4#5{%
  \reset@font\fontsize{#1}{#2pt}%
  \fontfamily{#3}\fontseries{#4}\fontshape{#5}%
  \selectfont}%
\fi\endgroup%
\begin{picture}(5495,3529)(1283,-4006)
\put(4307,-3924){\makebox(0,0)[lb]{\smash{{\SetFigFont{14}{16.8}{\rmdefault}{\mddefault}{\updefault}{\color[rgb]{0,0,0}$\bX$}%
}}}}
\put(1524,-2184){\makebox(0,0)[lb]{\smash{{\SetFigFont{14}{16.8}{\rmdefault}{\mddefault}{\updefault}{\color[rgb]{0,0,0}$\disp\fr{\bH}{\bLa}$}%
}}}}
\end{picture}%
}
 \scalebox{0.58}{\input{h-vs-x_T=0.1_wavelen.pdf_t}}
 \caption{$\bH$ vs.\ $\bX$ for \m{T=0.1} upstream of the edge at different resolutions: data 
  points (red circles) interpolated by cubic splines (solid).}
 \label{f:bhvsx01} 
\efig

\subsubsection{Why capillary undulations exist only upstream of the trailing edge} 
\label{sss:wcu}

In fact, the decision whether the oscillatory capillary modes occur either up- or downstream 
of the trailing edge, which depends on whether the real poles are within the lower or upper 
part of the $k$-plane divided by $\mcC$, cannot be left to the present steady-flow analysis. 
In both cases, these small-scale Rayleigh waves are also manifest above the MD of the 
interactive flow, modulating their amplitude over the interactive streamwise length scale. We 
now \hl{return to} a convincing (although not rigorous) argument restricting their presence to
upstream of the edge, as already anticipated in figure~\ref{f:conf}.

As inferred from the long-wave limit of \eqref{re}, the Rayleigh-type perturbation of the 
streamfunction $(\ep^{2/3})$ in \eqref{oexp} morphs into the pressure-free one 
\m{\ep^{2/3}\bPsi_{\!y}(\bX,0)\Psi_0'(Z)} in the LD. It exhibits a rapid (harmonic) streamwise 
variation, either far up- or far downstream. Inspection of \eqref{nsx} shows that typically a 
further viscous sublayer or SL (figure~\ref{f:scal-sep}\tit{b}) where \m{Z=O(\ep^{1/21})} is 
required on account of the no-slip BC. For very negative values of $\bX$, this shear layer is 
of the type provoked by the rapid small-scale disturbances considered in SBP18. For 
\m{\bX\ll\ep^{1/7}} (\m{X\ll 1}), it becomes absorbed into the HRW, there serving as the 
viscous correction of the LD; for larger values of $\bX$, an additional perturbation in the 
expansion of \m{-\ep^{2/3}\bPsi_{\!y}(\bX,0)\Us(X)} of $h_-$ serves to satisfy the free-slip 
condition \m{\psi_{zz}\sim 0} on \m{z=0} to which \eqref{dbct} reduces:
see \eqref{psiz0}.

These observations allow for the existence of the undular modes up- or downstream of the edge, 
\hl{\ie without preferring one of these alternatives}. That is, a steady-flow analysis 
\emph{cannot} rule out one of these two possibilities. We therefore justify our choice by 
making a recourse to the detection of capillary modes exclusively upstream of a wall-mounted 
obstacle, serving as a compact forcing, by \cite{BoSm92} and Rayleigh's celebrated radiation 
principle, which exploits the anomalous dispersion relation for small-amplitude capillarity 
waves. Acknowledging their essentially inviscid nature in both situations (despite their 
amplitude of $O(\ep^{2/3})$ here), we consider this analogy as reasonable.

As a serious objection, however, we have to admit that this principle applies strictly only to 
a uniform (potential) background flow, where it was adopted by \cite{CuNo79}. The last authors 
also point to the rigorous justification of \hl{the above} observation by solving the 
signalling problem, following \cite{DeWu57}. \hl{When} applied to the current situation 
\hl{(in a separate study)}, this demands the solution of the unsteady extension of \eqref{re} 
subject to an artificial, spontaneous introduction of the trailing edge in the unperturbed flow 
described by $\psi_0$. That is, one expects a pertinent neutral mode for zero frequency in the 
long-time response in the spectrum, to occur upstream rather than downstream of the edge. An 
easier modification of this ideal, rigorous approach is the introduction of artificial 
viscosity and tracing that particular wavenumber in the $k$-plane when the then complex 
frequency tends to zero \citep[\cf][\S\,3.4]{HuMo90}. This plausibility argument serves to 
single out the mode upstream as the physically meaningful alternative.

\subsubsection{Diffusive overlayer}\label{sss:do}

Expansion \eqref{oexp} accounts for the second dynamic BC \eqref{dbct}, requiring vanishing 
shear stress on the top free surface, up to $O(\ep^{3/7})$, \ie as long as \eqref{dbct} 
reduces to \m{\psi_{yy}\sim 0} on \m{y=h_+}. Moreover, it was indicated in SBP18 how 
\eqref{dbct} alters the highest-order contribution of $O(\ep^{4/7})$ to the inviscid flow 
described by \eqref{mdexp} in a thin layer adjacent to the upper free streamline accounting 
for viscous diffusion of weak perturbations around the base flow. From inspection of 
\eqref{nsx}, it penetrates to values of \m{h_+ -y} measured by the square-root of its 
horizontal extent and thus of $O(\ep^{3/7})$. Since the flow therein itself becomes inviscid 
over the shortened Rayleigh scale, a further diffusion layer of reduced vertical depth arises 
where $\bX$ and \m{\xi\deq(y-h_+)/\ep^{1/2}} are of $O(1)$ and \eqref{dbct} is formally 
retained in full. A comprehensive completion of the present self-consistent theory requires a 
brief examination of this overlayer meeting \eqref{dbct}: see \ref{s:c}.

\subsubsection{Inner Rayleigh stage: lower deck}\label{sss:irs}

Following the outline (b) at the end of \S\,\ref{ss:oie}, the inner square region regularises 
$P$ by taking into account the transverse variation of $p$, which becomes of 
$O(\ep^{6/7})$ according to \eqref{bvo} and \eqref{r} with \eqref{hx}. Then \eqref{oexp}, 
\eqref{op} yield the relevant expansion
\beq
 [\psi,p-p_-]\sim\ep^{4/7}[(M^2/\la)^{1/7}\Psi_0(Y),0]+\ep^{6/7}[\hPsi,\hP](\hX,Y)+
 O(\ep^{20/21}),
 \label{iexp}
\eeq
where we advantageously revert to the inverse Prandtl transposition in \eqref{ldxy}. Again, 
the quantities $\hPsi$, $\hP$ describe a linearised Euler flow, now with $\Psi_0$ providing 
the base profile. Therefore, \m{\hV(\hX,Y)\deq -\hPsi_{\!\bX}} satisfies a Rayleigh problem of 
the type \eqref{vpr} except for \eqref{bh}, \eqref{bvh} being replaced by the required decay 
for large values of \m{R\deq r/(m\,\ep^{2/7})=(\hX^2+Y^2)^{1/2}}, where the displacement of the 
HRW controls $\hV$ by virtue of a $R^{-1/3}$-variation matching \eqref{bvo}. Since the absence 
of a free surface at play renders the Rayleigh operator here self-adjoint, all poles lie on the 
imaginary axis of the corresponding wavenumber plane, which suppresses oscillations of 
wavenumbers much smaller than those detected in \S\,\ref{sss:ors}. Moreover, following the 
analysis leading to \eqref{bvo} recovers the far-field singularity also for \m{R\ttz}. 

This shows that the inner RS is unable to fulfil its original task of regularising the 
pressure provoked by the HRW in the outer RS across the LD, and the associated Rayleigh 
problem does not therefore merit a more detailed analysis as it proves physically 
insignificant.

Since the scaled slip $\bPsi_{\!y}(\bX,0)$ exerted by the outer RS becomes of $O(\bX^{1/3})$ 
as \m{\bX\ttz-}, the vertical extent of the associated SL introduced in \S\,\ref{sss:wcu} 
shrinks typically to \m{Y=O(\ep^{1/21}\bX^{1/3})}. It is continued as a sublayer covering the 
inner region where \m{Y=O(\ep^{1/7})} (figure~\ref{f:scal-sep}\tit{c}). There the driving slip 
is replaced by $\hPsi_{\!Y}(\hX,0)$, which again attains a $\hX^{1/3}$-behaviour as 
\m{\hX^{1/3}\ttz-}. We are therefore driven to consider a collapse of the inner RS, the SL and 
the HRW into a single region (figure~\ref{f:scal-sep}\tit{d}), addressed next.

\section{Full Navier--Stokes and Stokes regions}\label{s:fns}

As the conditions \eqref{ic} take into account the detachment of the lowermost streamline but 
not the edge as a geometric restriction or even its micro-geometry on the length scales 
considered so far, the prior analysis does not determine whether detachment occurs actually at 
the edge or further upstream. Therefore, this question is taken up first through an 
examination of even smaller scales, governing first a full NS regime. This ensues from a 
breakdown of \eqref{iexp} initiated by the unresolved singularity of $\hP$, just discussed, and 
the associated unbounded growth of the vertical flow component, $-\hPsi_{\hX}$. The associated 
growth of $v$ evaluated in the HRW shows the emergence of the NS region. We will see that it 
in turn contains at least one Stokes region around detachment so that the flow can accommodate 
the wetting properties controlling the emerging meniscus and defined by the thermodynamic 
three-phase equilibrium holding in the detachment point.
 
\subsection{Leading-order problem in an upper half-plane}\label{ss:lop}
 
The slender-flow approximation underlying \eqref{hr} ceases to be valid where both $u$ and $v$ 
become of $O(\ep^{1/2})$ as \m{(\bx,\by)\deq(x/\ep^{3/2},y/\ep^{1/2})} and, see \eqref{r}, 
\m{\br\deq r/\ep^{1/2}=(\bx^2+\by^2)^{1/2}} are of $O(1)$. In this half-plane 
\m{0\leq\vth\leq\upi}, we expand
\beq
 \bigl[\psi/\ep,\,(p-p_- +gy)/\ep,\,h_-/\ep^2\bigr]\sim
 \bigl[\bpsi(\br,\vth),\,\bp(\br,\vth),\,\bh(\bx)\bigr]+O(\ep^{3/2})
 \label{bexp}
\eeq
with the sought quantities $\bpsi$, $\bp$, $\bh$ of $O(1)$ as \m{\ep\ttz}. Due to the 
sufficiently smooth variation of the detached streamline beneath the HRW, this remains slender 
in the present NS region where
\beq
 \by\sim\ep^{3/2}\bh(\bx) \q\mb{or}\q \vth\sim\ep^{3/2}\bh(\bx)/\bx.
 \label{byh}
\eeq
Consequently, $\bpsi$, $\bp$ satisfy the full NS equations (\ref{gov}\tit{a},\tit{b}) 
describing a perfectly supercritical flow in the upper half plane. This is subject to mixed, 
linear, homogeneous BCs implied by (\ref{gov}\tit{c}--\tit{e}) and a \hl{far-field} condition 
which 
accounts for the externally imposed shear flow. From the reference capillary number in 
\eqref{ca}, the reduced velocity scale $\sqrb{\ep}\,\tU$ and the relative flatness of the 
detaching streamline given in \eqref{byh}, the currently relevant capillary number 
\m{\ep^{1/2}\,\Ca/\ep^{3/2}=1/\tau} of $O(1)$ implies the leading-order balance
\m{2\ep^2\psi_{yx}+p-p_-\sim\tau\vka_-} retained in \eqref{dbcn}. However, here the 
normal-stress jump across the fluid--gas interface evaluated at \m{\by=0} determines its small 
curvature \m{\vka_-\sim\ep\bh''(\bx)}, which then has only a passive, higher-order effect on 
the flow. Accordingly, the weak vertical displacement of the former provokes the
$O(\ep^{3/2})$-correction in \eqref{bexp}, for \m{\bx\tti} matching the displacement by the 
HRW provided by the inverse Prandtl shift. The neglected lower-order contributions to 
\eqref{bexp} consist of eigensolutions of the linearised NS operator that exhibit asymptotic 
growth as \m{\br\tti} so as to match the $O(\ep^{5/7})$-term in \eqref{iexp} and higher-order 
terms apparent in the expansion of $\hPsi$, $\hP$ for \m{R\ttz}.

With \m{\bDe\deq\br^{-1}\p_\br(\br\p_\br)+\br^{-2}\p_{\vth\vth}} being the Laplacian, the 
leading-order NS problem reads
\bseq
\begin{align}
 \bpsi_\vth(\bpsi_\vth/\br)_\br-\bpsi^2_\br-
 \bpsi_\br\bpsi_{\vth\vth}/\br &= -\br\bp_\br+\bDe\bpsi_\vth,
 \label{bnsx}\\[1pt]
 \bpsi_\br\bpsi_{\br\vth}-(\br\bpsi_\br)_\br\bpsi_\vth/\br &= -\bp_\vth-\br(\bDe\bpsi)_\br,
 \label{bnsy}
\end{align}
supplemented with (\ref{gov}\tit{c}--\tit{e}) when evaluated for \m{\by=\vth=0},
\begin{align}
 \vth=0\colon\;\; & \bpsi=0,\;\; \bpsi_{\vth\vth}=0,\;\; 2(\bpsi_\vth/\br)_\br+\bp=\tau\bh'',
 \label{bbc0}\\[1pt]
 \vth=\upi\colon\;\; & \bpsi=\bpsi_\vth=0.
 \label{bbcpi}
\end{align}
Matching $\psi$ and $p$ in the NS and the surrounding inner Rayleigh region, \ie \eqref{bexp} 
and \eqref{iexp}, completes the problem \eqref{bgov} governing $\bpsi$, $\bp$ and $\bh$. We 
have for
\beq
 \br\tti\colon\;\; \bpsi\sim(\Lao/2)(\br\sin{\vth})^2+o(\br)\q 
 (\vth\gg\br^{-2/3},\;\; \upi-\vth\gg\br^{-2/3}),\q \bp\ttz.
 \label{bmc}
\eeq
 \label{bgov}%
\eseq
The smallness of the remainder term imposed on $\bpsi$ provides the required second kinematic 
far-field BC. Since we are dealing with the full NS equations, \eqref{bgov} already captures 
the inner Rayleigh region and its subregions both upstream (SL) and downstream (HRW, 
\m{\by\sim\br\vth=O(\br^{1/3})} there) of detachment; \cf figure~\ref{f:scal-sep}(\tit{d}). 
That is, \eqref{bmc} already implies \m{(\bpsi,\bp)=O(\br^{2/3})} and \m{\bh=O(\br^{8/3})} at 
the onset of the HRW. The BCs for \m{\vth=0} in \eqref{bbc0} describe zero tangential stress 
along the detached streamline and the net normal-stress jump across it. Eventually, 
eliminating $\bp$ from (\ref{bgov}\tit{a},\tit{b}) yields the vorticity transport equation
\beq
 (\bpsi_\vth\,\p_\br-\bpsi_\br\,\p_\vth)\bDe\bpsi=\br\bDe^2\bpsi,
 \label{vte}
\eeq
to be solved subject to the first two BCs in \eqref{bbc0} and (\ref{bgov}\tit{d},\tit{e}). 
Hence, $\bpsi$ is solely induced and parametrised by the externally exerted shear rate~$\Lao$.
\hl{We recall that this is determined by the solution of the viscous--inviscid interaction 
problem on a larger scale and accounts for the upstream momentum flux, gravity and 
capillarity.}

The variation of $\bh$ with $\br$ is then found from integrating the capillary normal-stress 
jump in \eqref{bbc0} and, given the identical match of $\bh$ and $H_-$ according to 
\eqref{hr}, two ICs to be imposed as \m{\br\ttz}. Before tackling their determination, \hl{we 
first identify} the flow topology near detachment, solely based on the information extracted 
from the NS problem posed above in the limit \m{\br\ttz}. The importance of this insight by 
far outweighs the perspective of obtaining the full numerical solution. Therefore, we have 
refrained from tackling this considerable challenge. (The considerations below suggest 
spectral collocation in the $\vth$-direction as the method of choice.)

\subsection{Flow close to detachment}\label{ss:fcd}

As $\bpsi$ must satisfy four BCs in (\ref{bgov}\tit{c},\tit{d}), the viscous terms are 
retained in the limiting forms of (\ref{bgov}\tit{a},\tit{b}) as \m{\br\ttz} and 
\m{\vth\in[0,\upi]}. Requiring strict forward flow in the immediate vicinity of 
detachment, 
\beq
 \bpsi>0\q (\br\ttz),
 \label{ff}
\eeq 
is initially seen as a natural additional constraint. It is supported by the extensive 
numerical investigation by \cite{KiSc94} of the full NS problem for a flow passing a 
wedge-shaped lip, see figures~\ref{f:tpe}(\tit{b}) and \ref{f:scal-sep}(\tit{f}): this 
predicts an eddy at its underside in some situations associated with rather low to moderate 
Reynolds numbers but strictly forward flow detaching at its tip in the present 
high-Reynolds-number limit. The analysis below, however, demonstrates that \eqref{ff} is 
only met in the least singular situation chosen from the initial alternatives.

\subsubsection{The full inertial--viscous limit}

The convective--viscous balance in \eqref{vte} is restored in full if $\bpsi$ varies 
essentially with $\ln{\br}$:
\beq
 \bpsi\sim\bg(\vth)-\vGa\ln{\br}/(2\upi)\;\;\; (\br\ttz),\q 
 \vGa\bg'''/(2\upi)-2\bg'\bg''=(4\bg+\bg'')''.
 \label{jh}
\eeq
We are thus concerned with a spiralling extension of a special type of a radial 
Jeffery--Hamel (JH) flow described by $\bg(\vth)$ \citep[see][]{Fr62}, exhibiting the 
vorticity \m{-\bDe\bpsi=-\bg''\!/\br^2} and an outwards flow speed $\bg'(\vth)/\br$ as 
collapsing in a line source of strength $\bg'(\vth)$, due to a superimposed potential vortex 
of some strength $\vGa$. Here the homogeneous BCs \m{\bg(0)=\bg''(0)=\bg(\upi)=\bg'(\upi)=0} 
originating in (\ref{bgov}\tit{c},\tit{d}) require \m{\vGa=0}, and $\bg$ represents an 
eigensolution of the full NS problem. Nevertheless, the case \m{\vGa\neq 0} and 
\m{\bg''\not\equiv 0}, apparently unconsidered before now, might be of interest in a different 
context. We also remark that for an inviscid flow, removing the Stokes operator in \eqref{jh}, 
$\bg(\vth)$ varies sinusoidally in general but linearly in the case of a potential flow.

An analytical--numerical study shows that there exist two eigensolutions $\bg$. Each describes 
a distinctly different canonical flow topology as both exhibit a dividing streamline \m{\bg=0} 
for \m{\vth=\vth_0\simeq 1.12777} and thus violate the premise \eqref{ff} and point to the 
existence of a closed reversed-flow eddy. This is located either adjacent to the plate 
(\m{\bg<0} for \m{\vth_0<\vth<\upi}) or fully detached as bounded by the free streamline 
(\m{\bg<0} for \m{0<\vth<\vth_0}): see figure~\ref{f:jh}(\tit{a}). In the first case, the flow 
undergoes pre-separation to reattach in the origin \m{\br=0}; in the second, the free 
streamline attaches rather than detaches there from the plate. These flow pictures are the 
immediate consequence of including azimuthal higher-order corrections to the purely radial JH 
flow and extending the streamline pattern over the full NS scales: see 
figure~\ref{f:jh}(\tit{b}). However, our scrutiny of the related literature does not inform 
about what, at first sight, is a rather pathological situation. In particular, the conception 
of a detached eddy with a stagnation point forming at the free and material streamline, to 
which the fluid particles stay attached, raises serious concerns.

We therefore rule out the JH solution as the local limit of the full NS solution. 
Notwithstanding its apparent shortcoming, however, we refer the interested reader to the 
higher-order corrections and some of the further impact of this limit in \ref{s:d}. These 
findings are not required for the core arguments at present but are potentially of interest 
for pursuing the study of this flow structure in a related context.

\bfig
 \centering
 \scalebox{0.71}{\input{jh.pdf_t}}
 \caption{(\tit{a}) Eigensolutions of \eqref{vte} referring to a JH flow given by \eqref{jh}; 
  (\tit{b}) sketched flow patterns for the two cases in (\tit{a}): reversed-flow bubble 
  upstream of detachment or dictating attachment of free streamline.}
 \label{f:jh}
\efig

\hl{\subsubsection{An extended Stokes limit as the alternative}}\label{sss:aes}

Discarding the possibility of a full inner NS problem, \eqref{jh}, leaves us with the 
degenerate situation of the dominant Stokes balances
\beq
 0\sim\bDe^2\bpsi,\;\;\; \bp_\br\sim\bDe\bpsi_\vth/\br,\;\;\; \bp_\vth\sim -\br(\bDe\bpsi)_\br
 \label{bps}
\eeq
and \m{\bpsi\ttz} as the origin \m{\br=0} is approached along any path from within the flow. 
We then expand $\bpsi$ into the eigensolutions $\bpsi_i$ of the biharmonic operator 
in \eqref{bps} when supplemented with the homogeneous BCs in (\ref{bgov}\tit{c},\tit{d}) found 
by separation in the polar variables, following \cite{Mo64} and the references therein. 
\hl{However, here the subordinate convective terms in \eqref{vte} control their admissibility 
and thus the form of the dominant eigensolution. This straightforward but long-winded 
selection process is detailed in \ref{s:e}. As the most significant result, it predicts
regular behaviours for \m{\br\ttz} towards a separating flow (\m{\bpsi_{\by\by}=0}):}
\bseqa
 & \bpsi\sim -4a_5\by^3+o(\br^3),\q \bp-\bp_0\sim -24a_5\bx+o(\br)\q (a_5<0,\;\; \br\ttz), &
 \slabel{bpexp}\\[2pt]
 & \bh\sim\bh(0)+\bh'(0)\bx+\bp_0\bx^2\!/(2\tau)-4 a_5\bx^3\!/\tau+o(\bx^3)\q (\bx\ttz+). &
 \slabel{bhexp}
 \label{bphexp}
\eseqa
The constant $a_5$ and the offset pressure $\bp_0$ are part of the solution to the full NS 
problem, \hl{in turn, forced by the value of of $\Lao$. Let us first indicate how to fix the 
unknown coefficients $\bh(0)$ and $\bh'(0)$, governing the local elevation of the just detached 
streamline,} and complete our analysis at this stage, \ie without taking into consideration 
any smaller length scale.

As an obvious geometrical requirement, \m{\bh(0)=0} then. In full agreement with the current 
status of the theory, the position of flow detachment not only defines the origin \m{x=y=0} 
but an arbitrary point of the upper side of the plate rather than necessarily coinciding with 
the trailing edge, as a genuine geometrical constraint. Detachment further upstream then 
requires the actual static wetting or contact angle, observed in the NS region, as an input 
quantity being so close to $\upi$ that it is approximated by \m{\upi-\ep^{3/2}\bh'(0)}. This 
determines a positive value of $\bh'(0)$. However, and as an immediate consequence of the 
slenderness of the lower free streamline, this thereby resulting distinguished limit refers to
the quite exceptional break-away of an almost perfectly hydrophobic liquid. Additionally, such 
a scenario demands for the geometrical constraint \m{\bh>0} for \m{\bx>0}, which admittedly 
cannot be guaranteed as long as the numerical solution of the above NS problem is not 
available. It is also not likely to occur in reality, where unavoidable (though here 
neglected) surface imperfections already affect the flow described on the vertical NS scale. 
It is a natural step, therefore, to identify the location of flow detachment indeed at the 
trailing edge. However, then $\bh'(0)$ remains undetermined as long as its microscopic shape 
remains unresolved.

The outcome of these considerations is threefold. Firstly, we expect both $\bh(0)$ and 
$\bh'(0)$ to be fixed by conditions of matching the full NS and a Stokes flow in a hidden 
region of an extent much smaller than that of the encompassing NS region. Secondly, as we 
raised in the introduction to \S\,\ref{s:fns}, the description of that creeping flow must take 
into account the meniscus formed by the actual slope of the free streamline at its detachment 
point of three-phase contact as a hitherto unconsidered physical input. And third, that new 
length scale must resolve the microscopic contour of the plate with sufficient accuracy. \\

\subsection{Distinguished Stokes limits and wetting properties}\label{ss:dsl}

Although possibly not satisfied in a particular realisation of the flow, let us treat the 
surface of the plate as locally chemically heterogeneous and ignore distributed roughness on 
all scales for the sake of clarity. Then the so-called quasi-static apparent contact angle, 
$\be$, is observed between the wetted plate and the tangent to the free streamline at its 
point of detachment and formation, where three phases (locally) at rest meet under the 
Young--Dupr\'e equilibrium: for its precise conceptual foundation we refer to \cite{TeDaSc88}, 
\cite{KiSc94}, \cite{WhBoSt08} and \cite{Boetal09}. Since this macroscopic contact angle 
summarises all related submicroscopic phenomena \citep[see][and references therein]{KiSc94} 
and shall apply even to the smallest scales identified in the flow, we have consistently used 
the notion ``microscopic'' in the context of the resolved geometry of the trailing edge.

To progress further, we introduce the new length scale \m{\ell\ll\ep^{1/2}}, non-dimensional 
with the nominal film thickness $\tH$. In the new flow region, 
\m{[\hx,\hy]\deq[x/(\ep\ell),y/\ell]} and \m{\hr\deq r/\ell=(\hx^2+\hy^2)^{1/2}}, see 
\eqref{r}, are of $O(1)$ as \m{\ell\ttz}. Hence, supplementing \eqref{bexp} with 
\eqref{bphexp}, the associated increase of \m{\bp-\bp_0} and \eqref{bhexp} suggests the 
two-parameter expansion
\beq
 \biggl[\fr{\psi}{\ell^3\,\ep^{-1/2}},\,\fr{p-p_- +gy-\ep\bp_0}{\ell\,\ep^{1/2}},\, 
 \fr{h_-}{\ell}\biggr]\sim
 \biggl[\hpsi(\hx,\hy),\,\hp(\hx,\hy),\,\hh_0(\hx)+\fr{\ell}{\ep^{1/2}}\hh_1(\hx)\biggr].
 \label{hexp}
\eeq
The $O(1)$-quantities $\hpsi$, $\hp$, $\hh_0$ and the only first-order correction of interest 
$\hh_1$ are to be found. The scaled elevation \m{\hh(\hx)\deq h_-/\ell} of the detaching 
streamline \hl{remains} to be determined by the capillary normal-stress jump in \eqref{dbcn}. 
All terms on its left side are retained to leading order, and that becomes of 
$O(\ell\ep^{1/2})$. However, \m{\vka_-\sim\ell^{-1}\hh''/(1+\hh'^2)}, stating that the 
capillary number at play, $\ell^2\ep^{1/2}\!/\tau$, is small. This is also inferred from 
reducing $\Ca$ in \eqref{ca} by the small relative velocity scale $\ell^2/\ep^{1/2}$. In turn, 
\m{\hh''\equiv 0}, and matching \eqref{hexp} and \eqref{bhexp} shows that the lower free 
streamline remains horizontally inclined under an angle no larger than of \hl{$O(\ep^{3/2})$.
Accordingly,}
\beq
 \hh_0=\bh(0)/\vDe,\q \hh_1=\bh'(0)\,(\hx-\hx_d)/\vDe.
 \label{hh}
\eeq
Here the parameter $\vDe$ measures the strength of the required distinguished limit,
\beq
 \ell=\vDe\,\ep^2,\q 0<\vDe=O(1),
 \label{de}
\eeq
and \m{(\hx,\hy)=(\hx_d,\hh_0)} denote the position of the actual detachment point, $\mcD$, 
taken \hl{initially to be known}.

Let $\Si$ denote the resolved surface of the plate. Inspection of \eqref{gov} and the 
behaviour \eqref{hh} confirm that the leading-order quantities $\hpsi$, $\hp$ satisfy
\bseq
\beq
 \hDe^2\hpsi=0,\;\;\; \hp_\bx=\hDe\hpsi_\by,\;\;\; \hp_\hy=-\hDe\hpsi_\hx,\;\;\;
 \hDe\deq\p_{\hx\hx}+\p_{\hy\hy},
 \label{hseq}
\eeq
subject to mixed boundary conditions in the limit of zero capillary number as,
\begin{align}
 \hr\tti\colon\;\; & \hpsi\sim -4a_5\hy^3+o(\hr^{5/2}),
 \label{hmc}\\[1pt]
 \mb{on}\;\; \Si\;\;(\hx<\hx_d)\colon\;\; & \hpsi=\hpsi_\hy=0,
 \label{hbcu}\\[1pt]
 \hy=\hh_0\;\;(\hx\geq\hx_d)\colon\;\; & \hpsi=\hpsi_{\hy\hy}=0.
 \label{hbcd}
\end{align}
 \label{hpr}%
\eseq
Once $\hpsi$ is found, one can calculate $\hp$ by integration, giving 
\m{\hp\sim -24 a_5\hx+O(1)}. This matches identically the small-$\br$ form of $p$ in 
\eqref{bpexp} as the remainder term negotiates a constant of integration found from the 
$O(\ep^{3/2})$-contribution to \eqref{bexp}. In accordance with the above results and 
likewise, the neglected remainder term in \eqref{hmc} expresses the second necessary far-field 
condition and the absence of an eigensolution of the NS problem of $O(\ep^{3/4})$ that would 
enter the right-hand side of \eqref{bexp}. This seems to be a natural choice, as \eqref{iexp} 
would require it to die out for large values of $\hr$. Rather, \eqref{hbcd} enforces an 
$O(1)$-correction $a_5\hg(\vth)$, say, in the large-$\hr$ form of $\hpsi$. The function $\hg$ 
is then governed by 
\beq
 (4\hg+\hg'')''=0,\;\;\; 
 \hg(0)=-4a_5\hh_0^3,\;\;\; \hg''(0)=-24a_5\hh_0,\;\;\; \hg(\upi)=\hg'(\upi)=0,
 \label{hg}
\eeq 
\cf \eqref{jh}. Eventually, 
\beq
 \fr{\hpsi}{a_5}\sim-4\hy^3\cdots+\hg(\vth)+o(1),\;\;\;
 \hg=-6\hh_0(\sin{\vth})^2-4\hh_0^3\biggl[1+\fr{\sin(2\vth)}{2\upi}-\fr{\vth}{\upi}\biggr],
 \label{hmca}
\eeq
where the dots indicate potential eigensolutions of $o(\hr^{2})$. The behavior \eqref{hmca} 
provides the match of \eqref{hexp} with \eqref{bexp} supplemented with an 
$O(\ep^{9/2})$-contribution, hence also excited by the displacement \eqref{byh} of the 
interface. While that of $O(\ep^{3/2})$ is controlled by the linearisation of $\bpsi$ as 
\m{\by\ttz}, this is due to the corresponding third-order terms. As these dominate as 
\m{\br\ttz} where \m{\bpsi\sim 4 a_5\by^3}, evaluating $\hg$ for \m{\vth\ttz} describes the 
feedback of the displacement on the flow near detachment.

The case of a perfectly flat surface associated with \m{\hh_0\equiv 0} and the trivial 
solution \m{\hpsi=-4 a_5\hy^3} of \eqref{hpr} recovers the dominant Stokes limit of the full 
NS solution for \m{\br\ttz} and the aforementioned pathological case of fully hydrophobic 
dewetting with both $x_d$ and $\ell$ then remaining unspecified. This situation is therefore 
ruled out, and we are indeed left with flow detachment in a vicinity of the originally sharp 
plate edge covered by the Stokes region, where the microscopic resolution of the edge dictates 
the definition of $\ell$. We henceforth refer to the sketch of the flow around the resolved 
edge in figure~\ref{f:nose}, detailing figures~\ref{f:tpe}(\tit{b}) and 
\ref{f:scal-sep}(\tit{f}) on the new scale for various values of $\be$ 
\hl{\citep[\cf][]{Duetal10}. As previously discussed,} the edge is, without substantial loss 
of 
generality, assumed to be given by a smoothed but at first ideal wedge of cut-back angle $\al$ 
and with an apex lying at the coordinate origin. Then the curvature radius typical of the 
rounded nose conveniently defines $\ell$; the degenerate situation of a wedge still found sharp 
when viewed on the scale $\ep^2$ is assumed in the limit \m{\vDe\ttz}. The case of specific 
relevance \m{\al=0} can be interpreted as a plate-type thin tip formed by a semi-circle and of 
local thickness~$2\ell$ (figure~\ref{f:nose}\tit{b}).

\bfig
 \centering
 \scalebox{0.5}{\input{nose.pdf_t}}
 \caption{Stokes flow around resolved smoothed trailing edge: (\tit{a}) wedge-type 
  (\m{\al>0}), inner region emerging for \m{\be<\al} (green); (\tit{b}) plate-type and 
  semi-circular (\m{\al=0}), no inner region.}
 \label{f:nose}
\efig

Assuming \m{\vDe=1} in \eqref{de} and \m{\tH=1\,\mb{mm}} (table~\ref{t:dat} \hl{in 
appendix~\ref{a:a}}) typically gives a quite small physical scale 
\m{\ell\tH\simeq 0.01}$-$\m{0.04\,\upmu\mb{m}}. However, it is large enough to consider the 
asymptotic theory applicable to curvature radii achieved in manufacturing practice.

Completing our flow model at this stage is indeed possible for a non-degenerate, smoothed 
wedge tip and a sufficiently large apparent wetting angle $\be$ as the wedge geometry imposes 
a closure condition on \eqref{hpr}. This fixes the location of $\mcD$
\beq
 \mb{on}\;\; \Si\colon\;\; \ud\hy/\ud\hx\sim\tan{\be}+o(\ep^{3/2}).
 \label{be}
\eeq
This describes the general, non-degenerate case where $\bh(0)$ is found in virtue of 
\eqref{hh}. Evidently, then also \m{\hh'(0)=0} as the linear follow-up problem to \eqref{hpr} 
governing disturbances of $O(\ell/\ep^{1/2})$ in \eqref{hexp} has the \hl{zero} solution. 
Higher-order perturbations, already affected by the curvature of the detache\hl{d} streamline, 
control the (physically insignificant) remainder term in \eqref{be}. Proceeding in this manner 
determines successively the two initial conditions that each term arising in the expansion of 
$h_-$ in \eqref{bexp} has to meet as \m{\br\ttz}. This consideration confirms self-consistency 
of the proposed theoretical framework. As a crucial result, the flow wets the underside of the 
wedge as $\hh_0$ represents a (strictly) monotonic function of $\be$, which decreases from $0$ 
as $\be$ decreases from $\upi$. This justifies our reference to the teapot effect. The 
pathological limit \m{\be\to\upi-} or \m{(\hx_d,\hh_0)\to(0,0)}, however, leads to a 
non-trivial value of $\hh'(0)$. Here we only note that the above analysis by inspection gives 
\m{\ell=O(\ep)} in the degenerate case \m{\bh(0)=0}, \m{\bh'(0)>0}. On the other extreme, 
$\mcD$ has reached the point on the nose where its curvature vanishes once $\be$ has become as 
small as $\al$. All together, we arrive at the geometrical constraint
\beq
 \upi-\ep^{3/2}\bh'(0)\geq\be\geq\al.
 \label{bec}
\eeq
The variation of $\hh_0$ with $\be$ is more and more squeezed towards the edge as this gets
sharpened. Finally, $\mcD$ is seen as \emph{pinned} to the edge as \eqref{bec} is interpreted 
as the well-known Gibbs inequality: see \cite{OlHuMa77,Dy88,KiSc94}. In accordance with the 
last authors, we find that the distance of $\mcD$ from the apex decreases with both 
increasing values of $\be$ and the Reynolds number.

The formidable task of solving the Stokes problem \eqref{hpr}, parametrised by $\al$ and 
$\be$, has not yet been accomplished. Most importantly, in the situation sketched in 
figure~\ref{f:nose}(\tit{b}), mastering this challenge will establish a comprehensive flow 
description in the entire range \m{\upi>\be>0} of physical significance. If, however, 
\m{\be\geq\al}, determining the actual position of $\mcD$ requires the introduction of a 
further, inner Stokes region, as indicated in figure~\ref{f:nose}(\tit{a}). Contrasting with 
its counterpart \eqref{hpr}, there the governing problem is of non-degenerate free-surface 
type, thus controlled by a capillary number of $O(1)$, to accommodate to the necessary local 
bending of the detaching streamline. We expect $\mcD$ to be found the further away from the 
apex the smaller is $\be$, with its position fixed by a constraint arising of the interplay 
of these nested Stokes regions. This is a topic of our future activities.

As the final step, we focus on the flow properties in the immediate vicinity of detachment, 
specified on condition \eqref{bec}. Here we again follow \cite{Mo64} in his analysis of local 
eigensolutions of \eqref{hpr} varying algebraically with distance from a singular point at a 
rigid wall. These suggest that the streamlines are locally pushed away from the nose. 
\citet[][\S\,3.2]{Mo64} also showed that a related class of eigensolutions controls the 
behaviour of $\hpsi$ at small distances \m{\hd=[(\hx-\hx_d)^2+(\hy-\hh_0)^2]^{1/2}} from the 
detachment point: using (\ref{hpr}\tit{a},\tit{c},\tit{d}) and reusing the azimuthal angle,
\m{\vth\deq\arctan[(\hy-\hh_0)/(\hx-\hx_d)]}, yields for \m{0\leq\vth\leq\be} and
\beq
 \hd\ttz\colon\;\;
 \fr{\hpsi}{\ha\hd^{\,\si}}\sim\lb\{\!\barray{ll}
 \sin(\si\vth)\sin[(\si-2)\be]-\sin(\si\be)\sin[(\si-2)\vth]+c.c. &\! (\si\neq 2), \\[3pt]
 \sin(2\vth)-\vth/\be &\! (\si=2). \earray\rb.
 \label{me}
\eeq 
The constant $\ha$ is determined by the full solution to \eqref{hpr}, and $\si$ appears to be
a (complex) eigenvalue related to $\be$ by
\beq
 (\si-1)\sin(2\be)=\sin[2\be(\si-1)]\q (\si\neq 2),\q \tan(2\be)=2\be\q (\si=2).
 \label{besi}
\eeq
A continuous relationship requires \m{\be=3\upi/4} for \m{\si=2}. One readily confirms that 
the eigensolutions considered in \S\,\ref{sss:aes} are recovered in the limit \m{\be\to\upi}. 
Equation \eqref{besi} is symmetric with respect to \m{\Real\,\si-1}. However, physically 
admissible solutions require the flow speed, of $O(\hd^{\,\si-1})$, to vanish and the shear 
and the normal stress (the pressure), both of $O(\hd^{\,\si-2})$, on $\Si$ being integrable as 
\m{\hd\ttz} (and not to compromise the validity of the Young--Dupr\'e equilibrium). Thus only 
values of $\si$ having \m{\Real\,\si>1} are permitted, anticipated by the requirement 
\m{\hpsi=0} at detachment in \eqref{hbcd}. The plot of the real branches of \eqref{besi} in 
figure~\ref{f:be-vs-si} illustrates the infinite multiplicity of $\si$, not considered by 
\cite{Mo64}, the asymptotes \m{\be\to\upi/2,\,\upi} as \m{\Real\,\si\tti} and the local 
extrema of $\be$. There \eqref{besi} is continued to complex values of $\si$, via \eqref{me} 
associated with Moffat's (1964) \hl{prominent and} exceptional infinite sequence of eddies. 
Hence, our flow model does not predict a single eddy as do the calculations by \cite{KiSc94} 
for moderately large Reynolds numbers but this series of eddies if the value of $\be$ falls 
below its absolute minimum. \cite{Mo64} predicted this well-established threshold as 
\m{\simeq 78^\circ}; here we recompute it as $\simeq 79.557^\circ$ for \m{\si\simeq 3.7818}.

\bfig
 \centering
 \scalebox{0.7}{\begin{picture}(0,0)%
\includegraphics{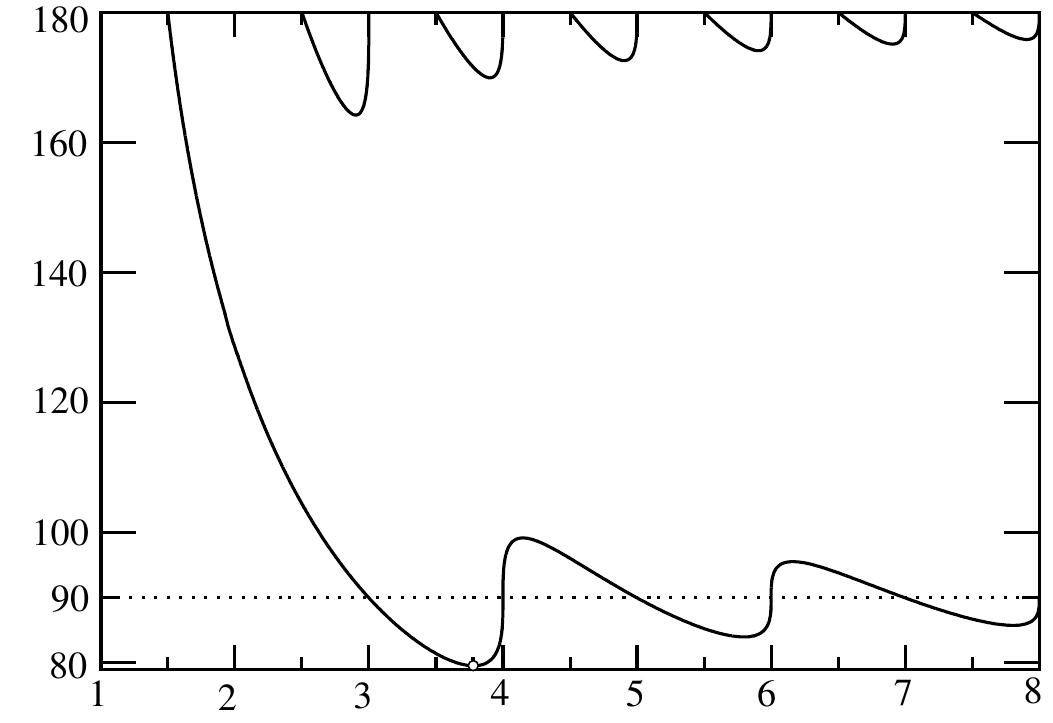}%
\end{picture}%
%
%
\setlength{\unitlength}{3947sp}%
\begingroup\makeatletter\ifx\SetFigFont\undefined%
\gdef\SetFigFont#1#2#3#4#5{%
  \reset@font\fontsize{#1}{#2pt}%
  \fontfamily{#3}\fontseries{#4}\fontshape{#5}%
  \selectfont}%
\fi\endgroup%
\begin{picture}(5042,3459)(1465,-3993)
\put(4151,-3929){\makebox(0,0)[lb]{\smash{{\SetFigFont{12}{14.4}{\rmdefault}{\mddefault}{\updefault}{\color[rgb]{0,0,0}$\si$}%
}}}}
\put(1480,-2188){\makebox(0,0)[lb]{\smash{{\SetFigFont{12}{14.4}{\rmdefault}{\mddefault}{\updefault}{\color[rgb]{0,0,0}$\be\;(^\circ)$}%
}}}}
\end{picture}%
}
 \caption{Contact angle $\be$ vs.\ real $\si$ in the limit of zero capillary number, from 
  \eqref{besi}, having absolute minimum (circle).}
 \label{f:be-vs-si} 
\efig

A more elaborate discussion of these details and their consequences requires the yet pending 
full numerical solution of \eqref{hpr}.

\section{Conclusions and further perspectives}\label{s:cfp}

As an unexpected extension of the interactive flow structure around flow detachment at the 
free plate edge, we report neutral capillary Rayleigh modes on the upper free surface solely 
and immediately upstream of the edge. Demonstrating this confidently calls for solving a 
signalling problem where typically a compact forcing dividing the flow into an upstream and an
downstream part. Here this is provided by a delta functional describing the transition of the 
vertical flow component over the geometric discontinuity formed by the trailing edge but tied 
in with an additional non-compact excitation by the displacement of the Hakkinen--Rott wake, 
necessary to provoke the non-trivial Rayleigh state.

The small-scale/small-amplitude ripples differ markedly in their origin from those already 
predicted by \cite{BoSm92} upstream of a wall-mounted obstacle over the interactive length 
scale. Accordingly, they are separated a streamwise extent of $O(\ep)$ from those of much 
larger wavelength found in the solutions of the interaction problem and set off by that wake 
in the downstream direction on both free surfaces in phase for \m{0<T<1/2} 
\hl{(Scheichl \etal\ 2019)}. On the other hand, since these rather long waves on the upper 
free surface are observed even upstream of the trailing edge for \m{T>1} \citep{ScBoPa18}, 
they collapse there with the short Rayleigh modes when \m{T\sim 1}, as the long-wave limit of 
the latter indicates (figure~\ref{f:kuvst}). This heralds how the introduction of a reduced 
streamwise length scale paves the way for an Euler stage to regularise the breakdown of 
viscous--inviscid interaction in a more general setting when the measure \m{T-1} of the 
typical counteracting dispersive effects, namely capillary versus convective streamline 
curvature, of classical Korteweg-de-Vries-type changes sign. \hl{Although already identified} 
in related studies \citep{Ga87,BoSm92,Kletal10}, this has not been investigated in due detail 
\hl{so far}. Having in mind the anomalous dispersion for classical linear capillary waves, we 
find it appropriate to speak of ``choking'' when both the wavelength and the amplitude of the 
capillary ripples, controlled by the dominant eigensolution $\psi_0'(y)$ of the Rayleigh 
operator and triggered by the displacement of the Hakkinen--Rott wake, diverge for 
\m{T\to 1-}. This consideration highlights the identical nature of the threshold \m{T=1} in 
this long-wave limit as for the interactive flow. \hl{For the current state of our research on 
the interactive stationary waves we refer to Scheichl \etal\ (2019) and appendix~\ref{a:b}.}

Neither the onset of the interactive, long waves above the plate for \m{T\to 1+} nor the 
formulation of additional conditions imposed at the plate edge to account correctly for the 
upstream influence that render them unique have yet been clarified satisfactorily 
\citep[\cf][]{ScBoPa18}. This and other exciting related phenomena attributed to the 
solution of the interaction problem downstream of the edge, such as its sound regularisation 
when \m{T-1/2} changes sign and attracting attention through (\ref{iap}\tit{d}--\tit{f}) and 
\eqref{h-}, are topics of our current research. A stability analysis of the detached 
interactive flow, where unsteadiness of the streamwise momentum balance becomes typically 
explicit in the lower deck, should clarify the analogy of the \hl{capillary waves with the 
classical linear Squire modes \citep{Sq53}}.

As a major conclusion of our analysis, the layer undergoes its break-away from the trailing 
edge at its underside when this is geometrically resolved in a least-degenerate but most 
simple manner as a (cut-back) wedge having a rounded nose. As a rule of thumb, the higher the 
wettability, the more the fluid ``sticks'' on the underside and the more the point of 
three-phase contact or detachment is remote from the plane wall on top. In the authors' mind, 
the present analysis rationalises for the first time how a high-Reynolds-number flow 
negotiates the formation of free streamline with due rigour. As the vital idea, any physically 
viable flow always ``feels'' a small reference length (the nose radius $\ell$) that resolves 
the abstraction of geometrical perfection (the sharp trailing edge). This then defines the 
smallest scales at play and hence controls the thereby arising Stokes limits and local 
dewetting or film rupture. As an interesting aspect, the convective influence and thus the 
flow profile stretching towards the upper free surface is only felt through a single 
coefficient of the dominant eigensolution of the Stokes operator. Pinpointing the flow on 
those smallest, geometrically induced length scales provides a self-consistent and 
qualitatively reasonable explanation of the teapot effect observed in the detachment of a 
high-momentum \hl{liquid layer}. The underlying continuum hypothesis is admissible as long as 
the smallest scales are so large that the liquid/gaseous interface can be taken as infinitely 
thin. \hl{We hope} that this appealing and promising \hl{approach} stimulates future research 
in this direction.

The Stokes problems governing the steady, \hl{capillarity-dominated} free-surface flow on the 
smallest scales \hl{constitute} the central building block for completing the rigorous 
examination of the teapot effect. This appears as an essentially hydrodynamic phenomenon, but 
the adjustment of the flow to the three-phase equilibrium defining the wetting properties in 
terms of the apparent contact angle represents the most salient ingredient. More will be able 
to be said and further progress sparked once the inner Stokes problem is established and the 
outstanding solutions of these core problems are available.


Last although not least, we feel an urgent need for careful and systematic laboratory 
experiments, with the ultimate goal to corroborate the theoretical findings on all scales. 
Here the values in the tables~\ref{t:dat} and \ref{t:par} \hl{in appendix~\ref{a:a}} might be 
helpful.

\backsection[Acknowledgements]{The authors express their thanks to Dr.\ S. N. Timoshin 
(Department of Mathematics, UCL) and the referees for fruitful discussions and helpful 
comments.}

\backsection[Funding]{Financial support from the Austrian Research Promotion Agency (FFG, grant 
no.: 849109, COMET K2 program: \emph{XTribology}) and from a UCL Mathematics Teaching 
Assistantship is greatly acknowledged.}

\backsection[Declaration of interest]{The authors report no conflict of interests.}

\backsection[Author ORCIDs]{\\
B. Scheichl 
\href{https://orcid.org/0000-0002-5685-9653}{https://orcid.org/0000-0002-5685-9653}.}

\backsection[Author contributions]{All authors contributed equally to this paper.}

\appendix

\section{Orders of magnitudes and their physical relevance}\label{a:a}

Even though left unspecified here, a horizontal nozzle or the impingement of a vertical jet
represent the \hl{most likely methods} of generating a flow configuration of engineering 
concern \hl{and of the type considered here. The work then is certainly relevant for a} 
variety of physical scenarios. However, one might question the validity of the 
order-of-magnitude \hl{requirements} made in \eqref{egt} in a conceivable situation of 
industrial importance or even of observations in daily life -- such as the \hl{falling jet} 
generated by wielding a teapot. Such settings are characterised by feasible geometrical and 
flow conditions and an aqueous, viscous fluid under the action of gravity and surface tension. 
Indeed, the chosen largeness of the Froude number at a moderate Weber number deserves some 
comment. The following arguments yield the values, \hl{presented in table~\ref{t:par}}, of 
the non-dimensional groups in \eqref{ng} and \eqref{ca}, relevant to a \hl{film} of pure water 
under standard conditions \hl{and based on the reference values of the input quantities as 
well as $\tH$ and $\tU$, following from \eqref{cl}, presented in table~\ref{t:dat}.}

With \m{\ttau\lesssim 100\,\mb{mN/m}} throughout (water as a polar liquid, and of potential 
interest, has a comparatively high surface tension), an adequately small Bond number 
\m{g/\tau=\tg\trho\tH^2\!/\ttau} (\hl{allowing the} neglect of gravity over surface tension) 
is, however, definitely not smaller than \m{10^5\,\mb{m}^{-2}\!\times\!\tH^2}. This requires 
\m{\tH\approx 1\,\mb{mm}}; for much smaller film heights, effects originating in technically 
unavoidable geometric imperfections of the plate surface might no longer be negligible (but 
worthy of study). Likewise, \m{g\ll 1} (\hl{allowing the} neglect of gravity over inertia) is 
achieved if \m{\tU\gg 0.1\,\mb{m/s}}. Given their rather narrow range of physically acceptable 
values and prediction of an extremely thin and fast film, these estimates have admittedly to 
be adopted with some caution. As an essential finding, the Reynolds number $\ep^{-1}$ proves 
to be indeed large but not to the extent that laminar--turbulent transition becomes an issue. 
However, the accompanying rather large value of $\tau$ \hl{alleviates} these doubts as it 
points to a numerically rather high sensitivity of the key parameters to slight variations of 
the input data. Also, \hl{the requirement} \m{\ep=O(g^{7/4})} for \eg \m{g=0.1} implies a 
reference or effective plate length $\tL$ of about 5 to 6\,cm, which seems sensible, and 
\m{\ep\approx 0.018}. We may check the reliability of the last estimate on the basis of the 
second relationship defining $\ep$ in \eqref{egt}: the above estimate for $\tU$ predicts values 
for $\ep$ barely smaller than 0.01. Given the potential variety of the input data, we achieve 
a satisfactorily good agreement. Our prerequisites, \hl{summarised} in~\eqref{egt}, can then 
be considered as self-consistent.

Most critically, the validity and sensitiveness of the scalings originate in a sufficiently 
small typical film height $\tH$ rather than in the values of the remaining physical 
parameters. Nonetheless, the subsequent asymptotic analysis of \eqref{gov} in the 
distinguished limits provided by~\eqref{egt} remains valuable even if the underlying 
order-of-magnitude estimates should be interpreted with a greater flexibility. In particular, 
the actual value of $\tau$ is taken as definitely smaller than its upper bound stated in 
table~\ref{t:par}.
\begin{table}
 \begin{center}
 \begin{tabular}{cccccccc}
  \;$\trho$\,(kg/m$^3$)\; & \;$\tnu$\,(mm$^2$\!/s)\; & \;$\ttau$\,(mN/m)\; & 
  \;$\tg$\,(m/s$^2$)\; & \;$\tQ$\,(l/min)\; & \;$\tL$\,(mm)\; & \;$\tH$\,(mm)\; & 
  \;$\tU$\,(m/s)\; \\[3pt]
  998.20 & 1.00 & 72.75 & 9.81 & $\gtrapprox 6$ & $\approx$ 50$-$60 & $\approx 1$ & $>0.1$ \\
 \end{tabular}
 \caption{Typical input data (water at standard conditions) \hl{and output $\tH$, $\tU$}.}
 \label{t:dat} 
 \end{center}
\vs{2mm}\hrule\vs{5mm}\hrule\vs{2mm}
 \begin{center}
  \begin{tabular}{ccccc}
   $\ep$ & $g$ & $\tau$ & $\Ca$ & $\Ca/\ep$ \\[3pt]
   \;\;0.01$-$0.02\;\; & \;\;0.1\;\; & \;\;$\lessapprox 7$\;\; & \;\;$\gtrapprox 0.00137$\;\; 
   & \;\;0.0686$-$0.137\;\; \\
  \end{tabular}
 \caption{Typical key parameters resulting from table~\ref{t:dat}.}
 \label{t:par}
 \end{center}
\end{table}

\section{Small-amplitude waves}\label{a:b}

For the following instructive analogy to (unconditionally stable) linear \hl{Squire modes, 
perturbing weakly a planar, uniform jet having constant speed in the $x$-direction and two free 
surfaces \m{y=h_-=0} and \m{y=h_+=h_0}, we refer the reader to \cite{Sq53}, 
\citet[][p.~30]{DrRe04} and \citet[][\S\,A.4]{Vi20}}.

Let $k$ denote their wavenumber, non-dimensional with $\tH$, and $c$ the ratio of their phase 
speed relative to the unperturbed jet speed. Using the definition of $J$ \hl{in \eqref{jl} 
yields} the classical anomalous dispersion relation in the form
\beq
 (c-1)^2=T k\times\lb\{\!\barray{ll}
 \coth(k/2) & (\mb{skew-symmetric modes}), \\[3pt] \tanh(k/2) & (\mb{symmetric modes}). 
 \earray\rb.
 \label{dr}
\eeq
Here the symmetry refers unambiguously to the $u$-perturbation with respect to the centreline 
\m{y=h_0/2}. Hence, the antisymmetric modes give the picture of a sinusoidally meandering 
\hl{or flapping} jet as \m{h_+\sim h_0+h_-} and $h_-$ are in phase. On the contrary, they are 
in antiphase as \m{h_+\sim h_0-h_-} for the symmetric modes, producing a ``varicose'' or 
symmetrically looking jet. These latter modes appear visually as the classical axisymmetric 
Rayleigh--Plateau modes, \hl{thus forming their counterpart on a circular jet (see 
\citealp[][p.~22 f\-f.]{DrRe04}; \citealp[][\S\,A.5]{Vi20})}. There exists a single stationary, 
choked mode (\m{c=0}) for each value of $T$ in the symmetric case \hl{but} only for \m{T<1/2} 
in the antisymmetric one, where indeed \m{T\to 1/2-} in the long-wave limit \m{k\ttz}, 
resembling the interactive limit. Moreover, our first numerical solutions of \eqref{iap} 
predict a sinusoidal modulation only of the detached jet if \m{0<T<1/2} and of varicose kind in 
the yet poorly-understood case \m{T>1}, where the onset of the waviness of the upper free 
surface approaches the edge from upstream as $T$ tends to $1$ from above (see SBP18).

These results allow for the following interpretation. The undulations for \m{0<T<1/2} 
represent a nonlinear, viscosity-affected variation of their classical counterpart, also 
strongly impacted by the background vorticity or the reduced fluid velocity at the lower 
interface. Like the classical ones, these vanish only for vanishing capillarity. For $T$ 
sufficiently exceeding $1$, the predominance of capillarity over both vorticity and the 
symmetry-breaking displacement effect implements a nonlinear modification of steady varicose 
modes. This analogy becomes evident from inspection of \eqref{mdexp}, \eqref{h-} and 
figure~\ref{f:c}: for sufficiently large $|A|$, we have \m{H_+\sim(D-1)H_-}; thus 
\m{\sgn(H_+)=\sgn(\hl{H_-})} for \m{T<1/2} and \m{\sgn(H_+)=-\sgn(\hl{H_-})} for \m{T>1/2}, 
where the symmetry of the varicose waves downstream of the plate allows also for their 
emergence above the plate; their failure occurring for \m{T\to 1+} is \hl{again} associated 
with an unbounded LD displacement~\hl{\m{-A}}.


\newpage%
\oneappendixfalse\appendixfalse%
\setcounter{figure}{0}%
\setcounter{page}{1}%
\setcounter{section}{0}%
\renewcommand{\theequation}{{\itshape\Alph{section}}\,\arabic{equation}}%
\renewcommand{\thefigure}{S\,\arabic{figure}}%
\renewcommand{\thepage}{S\,\arabic{page}}%
\renewcommand{\thesection}{Supplement\,{\itshape\Alph{section}}}%
\renewcommand{\thesubsection}{{\itshape\Alph{section}}.\arabic{subsection}}%
\renewcommand{\thesubsubsection}{\thesubsection.\arabic{subsubsection}}%

\vs{0mm}
\section*{\Large Other supplementary material}
\vs{2mm}

In this supplement (Supplements \tit{A}--\tit{E}), we present the technical details required 
to reproduce the details of our analysis. 

\section{Hakkinen--Rott wake: higher-order scheme}\label{s:a}

To elucidate the structure of \eqref{p+} and \eqref{psi+} \hl{in more detail}, we have to 
match the latter \hl{in the limit \m{Z\ttz}} to the appropriate expansion
\beq
 F\deq\fr{\Psi}{\Lao^{1/3}X^{2/3}}\sim\fhr(\eta)+
 \fr{c_1 X^{1/3}}{\Lao^{4/3}}[\ln{X}F_1(\eta)+F_2(\eta)]+O\bigl(X^{2/3}(\ln{X})^2\bigr)
 \label{f+}
\eeq
\hl{holding} for \m{\eta=O(1)} as \m{X\ttz+} and with $F_{1,2}$ to be found. The logarithmic 
contributions to \eqref{p+}--\eqref{h+} \hl{and} \eqref{f+}, by anticipating 
\m{c_{1,2}\neq 0}, comply with the representation 
\beq
 \Psi\sim\Lao Z^2\!/2+\hl{P'}(0-)Z^3\!/6+(P-P_-)[\Lao^{-1}+P'(0-)\Lao^{-2}Z\ln{Z}+O(Z)]
 \label{psiol}
\eeq
of \eqref{psi+}. \hl{This is} found with the help of \eqref{psie}, in the overlap 
\m{1\gg Z\gg X^{1/3}} or \m{X^{-1/3}\gg\eta\gg 1}, see \eqref{hr}, with the expansion 
\eqref{f+}. There they cause a reordering of its terms arising for large values of $\eta$, as 
seen from expressing $Z$ in \eqref{psiol} in terms of $\eta$:
\beq
 F\sim\fr{\eta^2}{2}+\fr{P-P_-}{\Lao^{4/3}X^{2/3}}+\fr{P'(0-)X^{1/3}\eta}{3\Lao^{4/3}}
 \biggl\{\fr{\eta^2}{2}+\fr{P-P_-}{\Lao^{4/3}X^{2/3}}
 \bigl[\ln{X}+3\ln{\eta}+O(1)\bigr]\biggr\}.
 \label{fol}
\eeq
As it is the HRW where the inertia terms in \eqref{ldeq} are fully restored, the 
$Z$-independent contribution to $\Psi$ in \eqref{psiol} is again in agreement with 
\eqref{ldfc}. Thus, matching the $X$-independent terms in \eqref{f+} and \eqref{fol} confirms 
that \m{\fhr-\eta^2\!/2\sim\phr+\tst} (\m{\eta\tti}); matching the higher-order terms requires 
the successive emergence of the logarithmic terms.

To this end, we substitute \eqref{f+} into \eqref{ldeq}--\eqref{ldfc} and exploit \eqref{fol} 
with \m{\eta\gg 1} kept fixed. This reveals two inhomogeneous linear problems governing $F_i$ 
(\m{i=1,2}):
\bseqa
 & \fhr'F_i'-\fhr''F_i-2\fhr F_i''/3=F_i'''-G_i(\eta),\q F_i(0)=F_i''(0)=0, &
 \slabel{Fi}\\[2pt]
 & G_1\deq 1,\q G_2\deq 1+c_2/c_1+\fhr'F_1'-\fhr''F_1. &
 \slabel{Gi}
 \label{FGi}
\eseqa
The well-known behaviour of $\fhr$ entails \hl{the essentially algebraic growth}
\beq
 F_1\sim\ga_1(\eta^3+6\,\phr\eta\ln{\eta})+\de_1\eta-6\ga_1+G_1+O(\eta^{-1})\q (\eta\tti)
 \label{gd1}
\eeq
with some constants $\ga_1$, $\de_1$. This is confirmed by expanding \eqref{fol} up to 
$O(X^{1/3}\ln{X})$ provided that \m{\ga_1=0} and \m{c_1=\phr P'(0-)/(3\de_1)}. Then 
\m{F_1''(\infty)=0}, and \eqref{FGi} has a unique non-trivial solution in the case \m{i=1}. 
The numerical method and discretisation we used to compute $\fhr$ gives 
\m{\de_1\simeq -2.6110}, 
implying \m{c_1>0}. (We also note that \m{F_1'(0)\simeq -1.8422}.) 
In turn, \m{G_2=1+c_2/c_1+\tst} (\m{\eta\tti}) so that \eqref{gd1} holds also for the 
corresponding quantities having the subscript $2$. Accordingly, expanding \eqref{fol} up to 
$O(X^{1/3})$ then gives \m{\ga_2=P'(0-)/(6c_1)=\de_1/(2\,\phr)}. This fixes the missing BC as 
\m{F_2''\sim 3\de_1\eta/\phr} (\m{\eta\tti}). Adjusting $F_2'(0)$, however, allows this 
condition to be satisfied for any $c_2$ as this determined by proceeding in this manner and 
considering the abbreviated remainder terms in \eqref{f+} and~\eqref{fol}.

\section{Outer Rayleigh problem: wavenumber space}\label{s:b}

We investigate the Rayleigh problem \eqref{vpr} in the distinguished limit \m{T\to 1-} and 
\m{k\ttz}, for \m{|\Real\,k|\tti} and the eigenspace (poles) of $\mcV$ in detail.

\subsection{Singular limits\, $k\ttz$, $T\to 1-$}\label{ss:b1}

The analysis of the long-wave limit \m{k\ttz} is closely related to the discussion of 
\eqref{bvi} with \eqref{ba+}. We substitute the expansion 
\m{\mcV\sim\sum_{i=0}^\infty \ka_i(k)\mcV_i(y)} (\m{\ka_{i+1}/\ka_i\ttz}) with, at first, 
unknown gauge functions $\ka_i$ and $O(1)$-functions $\mcV_i$ into \eqref{vpr}. At leading 
order, the problem \eqref{vpr} then only permits the homogeneous solution parametrised by 
$\ka_0$, thus \m{\mcV_0=\psi_0'}. To allow for deviations from this, the analysis of $\mcV_1$ 
requires \m{\ka_1=O(k^2\ka_0)}. Specifying \m{\ka_1\deq k^2\ka_0} yields the inhomogeneous 
problem
\beq
 \mcV_1''-(\psi_0'''\!/\psi_0')\mcV_1=\psi_0',\q \mcV_1(0)=-\ka_1^{-1},\q 
 u_0^+\mcV_1'(h_0)=T J,
 \label{v1pr}
\eeq
where we have anticipated that \m{\ka_1=O(1)}, including the alternative \m{\ka_1\tti} as 
\m{k\ttz} as a limiting case. By using the first BC here and after some rearrangements,
\beq
 \fr{\mcV_1(y)}{\psi_0'(y)}=\al_1-\la\,\ka_1^{-1}\int_y^{h_0}\!\fr{\ud t}{\psi_0'^2(t)}+
 \int_0^y\!\fr{\ud t}{\psi_0'^2(t)}\int_0^t\!\psi_0'^2(s)\,\ud s.
 \label{v1}
\eeq
The initially arbitrary constant $\al_1$ indicates again a homogeneous solution and the first 
BC in \eqref{v1pr} is met in the limit \m{y\ttz}. The second BC in \eqref{v1pr} represents a 
solvability condition for \eqref{v1pr} as it gives
\beq
 \la\,\ka_1^{-1}=(T-1)J
 \label{sc1}
\eeq
to fix \m{\ka_0=\ka_1/k^2} with \m{\ka_1=\la/[(T-1)J]}.
 
This analysis reveals a double pole of $\mcV$ at \m{k=0} the strength of which becomes 
unbounded as \m{T\to 1}. It refers to an apparent solution $|\bX|\psi_0'(y)$ (\m{\bX\neq 0}) 
to the homogeneous problem formed by (\ref{bvpr}\tit{a},\tit{c}). Since $k^2$ enters 
\eqref{vpr} linearly, the above expansion is now specified as
\beq
 \mcV\sim\ka_0(k)\sum\nolimits_{i=0}^\infty k^{2i}\mcV_i(y),\q \mcV_0=\psi_0'.
 \label{vexp}
\eeq
and breaks down passively where \m{y=O(k^2)}. We solve the resulting hierarchy of the 
inhomogeneous problems
\beq
 \mcV_i''-(\psi_0'''\!/\psi_0')\mcV_i=\mcV_{i-1},\q \mcV_i(0)=0,\q
 u_0^+{}^2\,\mcV_i'(h_0)=T J\mcV_{i-1}(h_0)\q (i>1)
 \label{vipr}
\eeq
using the approach that leads to \eqref{v1}:
\beq
 \fr{\mcV_i(y)}{\psi_0'(y)}=\al_i+
 \int_0^y\!\fr{\ud t}{\psi_0'^2(t)}\int_0^t\!\psi_0'(s)\mcV_{i-1}(s)\,\ud s
 \label{vi}
\eeq
where the last BC in \eqref{vipr} fixes the constant $\al_{i-1}$ as a function of $T$ in terms 
of a solvability condition for the problem governing $\mcV_{i-1}$. We finally write this 
constraint after some manipulations as the recursive relationship
\beq
 \fr{\mcV_i(h_0)}{u_0^+}=\fr{\mcI_i}{1-T},\q 
 \mcI_i\deq\fr{1}{J}\int_0^{h_0}\!\psi_0'^2(y)\,\ud y\int_y^{h_0}\!\fr{\ud t}{\psi_0'^2(t)} 
 \int_0^t\!\psi_0'(s)\mcV_{i-1}(s)\,\ud s.
 \label{vih}
\eeq
Also, $\mcV_i$ (\m{i>0}) is inversely proportional to \m{1-T}.

Inspection of \eqref{bvpr} reveals immediately the expansion 
$\bV\sim\ba_+\psi_0'(y)-\sum_{i=1}^\infty \ba_+^{(2i)}\mcV_i(y)$ (\m{\bX\tti}), which 
completes \eqref{bvi} subject to \eqref{ba+} as the reciprocal form to \eqref{vexp}. From 
this, one infers that \m{\bb_+=-\al_1}.

The above analysis ceases to be valid when \m{1-T} is so small that \m{\ka_1^{-1}=O(k^2)} and 
thus no longer enters \eqref{v1pr} but, instead, the problem governing $\mcV_2$. However, the 
asymptotic series \eqref{vexp} captures this shift of the lower BC formally when we introduce 
the (positive) parameter
\beq
 \mcT\deq(1-T)/k^2=O(1)
 \label{mct}
\eeq
to quantify the consequent least-degenerate distinguished limit. Then $\ka_1^{-1}$ is replaced 
by $0$ and $T$ by $1$ in \eqref{v1pr}, \eqref{v1} such that \eqref{sc1} is satisfied 
identically. Most importantly, the BCs in \eqref{vipr} are modified to
\beq
 \mcV_2(0)=-\ka_2^{-1},\;\;\; \mcV_i(0)=0\;\; (i>2),\;\;\; 
 u_0^+{}^2\,\mcV_i'(h_0)=J[\mcV_{i-1}-\mcT\mcV_{i-2}](h_0)\;\; (i>1)
 \label{vibc}
\eeq
where we take \m{\ka_2=k^4\ka_0} as of $O(1)$. Hence,
\beq
 \fr{\mcV_2(y)}{\psi_0'(y)}=\al_2-\la\,\ka_2^{-1}\int_y^{h_0}\!\fr{\ud t}{\psi_0'^2(t)}+
 \int_0^y\!\fr{\ud t}{\psi_0'^2(t)}\int_0^t\!\psi_0'(s)\mcV_1(s)\,\ud s.
 \label{v2}
\eeq
The special form of the dynamic BC in \eqref{vibc} determines the value of the constant 
$\al_{i-2}$ for \m{i>2}. It is sufficient for our purposes to concentrate on this BC for 
\m{i=2}. As $\mcV_1$ is given by \eqref{vi} for \m{i=1}, this solvability condition for 
\eqref{v2} yields with $\mcV_0$ specified in \eqref{vexp} and the definitions of $\mcI_1$ in 
\eqref{vih} and $\mcT$ in \eqref{mct}, and after some rearrangements, an expression for 
$\ka_2$ and thus $\ka_0$, independent of the value of $\al_1$:
\beq
 J\ka_0/\la=k^{-2}[k^2\mcI_1-(1-T)]^{-1}.
 \label{sc2}
\eeq

We find that the distinguished limit \eqref{mct} is rich enough to disclose the behaviour of 
$\mcV$ near the critical point \m{k=0} and \m{T=1} (\hl{but the last \hl{situation is 
excluded} in this study}). First, one readily finds that $\mcV$ admits a regular expansion in 
\m{T-1} as \m{T\to 1} for \m{k\neq 0} or \m{\mcT\ttz}. It is seen that $\mcV$ attains a 
fourth-order pole in \m{k=0}, \m{T=1} which morphs into a double pole for \m{T<1}, here 
recovered in the limit \m{\mcT\tti} where the two forms of \eqref{vexp} considered match. This 
behavior is associated with the bifurcation of a simple pole for $k$ becoming positive, whose 
location we trace in the $(T,k)$-plane as
\beq
 k=k_u(T),
 \label{kut}
\eeq 
indicating the existence of undamped capillary oscillations, where \m{\mcT\sim\mcI_1} or
\beq
 T\sim 1-\mcI_1 k_u^2+O(k_u^4)\q (k_u\ttz).
 \label{tku0}
\eeq 
As suggested by these asymptotic findings, our numerical study predicts exactly one value of
\m{k^2=k_u^2} for each value of $T$ in the relevant interval \m{0<T<1}. We also infer from 
\m{\mcV\sim\res_{k=k_u}(\mcV)/(k-k_u)} (\m{k\to k_u}), \eqref{vexp}, \eqref{sc2} and 
\eqref{tku0} that
\beq
 \res\nolimits_{k=k_u}(\mcV)/\psi_0'(y)\sim\la/(2 J\mcI_1 k_u^3)\sim
 \la\,\mcI_1^{1/2}\!\big/\bigl[2 J(1-T)^{3/2}\bigr]\q (k_u\ttz).
 \label{res0}
\eeq
The amplitude $\ba_u$, see \eqref{bau}, varies predominantly with $k_u^{-14/3}$ or 
\m{(1-T)^{-7/3}} in this limit. To exploit the above results numerically, we specify $\psi_0'$ 
by Watson's flow, using \eqref{weq} \hl{with \m{x_v=-1}}. Then $\mcI_1$ can be transformed 
into a single integral, and we add
\beq
 J=\la,\q \mcI_1=\fr{2}{u_0^+}-\fr{1}{\la h_0}\hl{-\fr{1}{\la}}+\la\int_0^{h_0}\,
 \biggl[\fr{1}{\psi_0'^2(y)}-\fr{1}{(\la y)^2}\biggr]\ud y\simeq 0.307059.
 \label{iw}
\eeq
\hl{The relationships \eqref{tku0}--\eqref{iw} set the basis for \eqref{to}.}

The above results can be nicely captured on the basis of \eqref{mct} and \eqref{sc2} cast in 
the normal form \m{\ka_c=1/[k_c^2(k_c^2-1)]} employing the canonical variables
\m{k_c\deq[\mcI_1/(1-T)]^{1/2}k}, \m{\ka_c\deq J(1-T)^2\ka_0/(\la\mcI_1)}: see 
figure~\ref{f:kvst}.

\bfig
 \centering
 \scalebox{0.592}{\input{k-vs-t+ka-vs-k.pdf_t}}
 \caption{Plots of poles and variation of $\mcV$ near \m{k=0}, \m{T=1}.}
 \label{f:kvst} 
\efig 

\hl{\subsection{Singular short-wave limit\, $|\Real\,k|\tti$}}\label{ss:b2}

For \m{|\Real\,k|\tti}, $\mcV$ obviously varies exponentially weakly with $k$. We first 
identify a viscous sublayer \m{\ze\deq ky=O(1)} where we take $\mcV$ as a function of $\ze$ 
and 
$k$. There \eqref{veq} and \eqref{jl} give 
\m{\p_{\ze\ze}\mcV-\mcV\sim\om\ze\mcV/k^3+O(k^{-6})}. In turn,
\beq
 \mcV\sim e\e^{-\ze}\biggl[1-\fr{\om}{4k^3}(\ze+\ze^2)+O(k^{-6})\biggr]-
 (1+e)\e^\ze\biggl[1-\fr{\om}{4k^3}(\ze-\ze^2)+O(k^{-6})\biggr]
 \label{ves}
\eeq
where $e$ is some function of $k$ satisfying \m{e(-k)\equiv -e(k)-1} as \eqref{vpr} enforces 
symmetry of $\mcV$ with respect to $k$. For \m{y=O(1)}, the exponential variation in 
\eqref{ves} is morphed into a rapid one, typically captured by a 
Wentzel--Kramers--Brillouin--Jeffreys (WKBJ) ansatz. Inserting this into 
(\ref{vpr}\tit{a},\tit{c}) yields after some manipulations and exploiting the above symmetry 
property intrinsic to \eqref{vpr}
\beq
 \fr{\mcV}{E(k)}\sim
 \e^{-k(y-h_0)}\biggl[1-\fr{K_0-K(y)}{2k}+O(k^{-2})\biggr]-
 \e^{k(y-h_0)}\biggl[1+\fr{K_0-K(y)}{2k}+O(k^{-2})\biggr],
 \label{vec}
\eeq
skew-symmetric in $k$ and depending solely on the homogeneous BC \eqref{vh}. The asymptotic 
relationship \m{E(k)\equiv -E(-k)} follows from matching \eqref{ves} and \eqref{vec} as 
$e(k)$, and
\beq
 K(y)\deq\int_y^{h_0} \fr{\psi_0'''(t)}{\psi_0'(t)}\,\ud t,\q 
 K_0\deq \fr{2 u_0^+{}^2}{T J}.
 \label{k}
\eeq
The last constant of integration ensures that \eqref{vh} is satisfied with the accuracy 
specified in \eqref{vec} and \eqref{ves}, where the terms varying algebraically with $k$
originate in the vorticity of the base flow. 

Expanding $K$ for \m{y\ttz} with the help of \eqref{jl} confirms the match of \eqref{ves} and 
\eqref{vec}, up to contributions of respectively $O(k^{-3})$ and $O(k^{-1})$ in the brackets. 
This first yields two relationships involving $e$ and $E$:
$e\sim E\e^{kh_0}[1-K_0/(2k)+O(k^{-2})]$, \m{1+e\sim E\e^{-kh_0}[1+K_0/(2k)+O(k^{-2})]}. 
From these we infer
\beq
 e^{-1}\sim -1+\e^{-2k h_0}\!\mcA(-k)/\mcA(k),\q 
 E^{-1}\sim\e^{-kh_0}\!\mcA(-k)-\e^{kh_0}\!\mcA(k)
 \label{e}
\eeq
where we abbreviate the algebraic variations with $k$ in terms of
\beq
 \mcA(k)\sim 1-K_0/(2k)+O(k^{-2})\q (k\to\pm\infty).
 \label{mca}
\eeq 
The first of the relations \eqref{e} verifies that \m{\mcV=O(1)} in \eqref{ves} and the second 
the exponential smallness of $\mcV$ for \m{y\gg 1/|k|}. From \eqref{vec}, there
\beq
 \mcV\sim -\!\e^{\mp ky}\biggl[1-\fr{K(y)}{K_0\mp 2k}+O(k^{-2})\biggr]+\,\e^{\mp k(2 h_0-y)} 
 \biggl[1-\fr{2 K_0-K(y)}{K_0\mp 2k}+O(k^{-2})\biggr]
 \label{vki}
\eeq
where the exponentially weak remainder term is of $O\bigl(\e^{\mp k(y+2h_0)}\bigr)$. 
Therefore,
\beq
 \mcV(k,h_0)\sim -2\e^{\mp k h_0}\bigl[K_0/(K_0\mp 2k)+O(k^{-2})\bigr]+
 O\bigl(\e^{\mp 3k h_0}\bigr)
 \label{vkih}
\eeq
with the aid of \eqref{k}.

As seen from \eqref{e} and \eqref{mca}, the $O(k^{-2})$-terms in \eqref{vki} \hl{and 
\eqref{vkih}} are correct as long as \m{E\sim -\e^{-k h_0}\!/\mcA(k)} or 
\m{|\mcA(k)|\gg|\e^{-2k h_0}\!|} for \m{\Real\,k>0} and \m{E\sim\e^{k h_0}\!/\mcA(-k)} or 
\m{|\mcA(-k)|\gg|\e^{2k h_0}\!|} for \m{\Real\,k<0}. However, they increase up to $O(k^{-1})$ 
when these constraints are violated, that is, when both contributions to $E^{-1}$ in \eqref{e} 
become of the same order of magnitude or
\beq
 k\sim\pm K_0/2+O(T).
 \label{kk0}
\eeq
The expression for $K_0$ in \eqref{k} heralds this possibility when $T$ is so small that
\m{Tk\sim\pm u_0^+{}^2/J+O(T^2)}. Moreover, $E^{-1}$ might change sign then, which reveals the 
emergence of a real pole of $\mcV$ that represents the small-$T$ asymptote of \eqref{kut}. 
Hence, the trace of the pole is known in a first approximation as weak deviations must account 
for its weak straining due to the higher-order corrections in \eqref{vki}:
\beq
 T k_u[1+O(k_u^{-1})]=u_0^+{}^2/J 
 =\sqrb{3/4\upi}\,\Ga({\txt\fr{1}{3}})/\Ga({\txt\fr{5}{6}})\simeq 1.15960\q (k_u\tti)
 \label{tkui}
\eeq
when $u_0^+$ is evaluated for Watson's flow. Then \eqref{vki} yields with 
\m{K_0\sim 2k_u+O(1)} from \eqref{kk0}
\beq
 \res\nolimits_{k=k_u}(\mcV)\sim -\e^{-k_u y}\bigl[K(y)/2+O(k_u^{-1})\bigr]+
 \e^{k_u(y-2 h_0)}[2 k_u+O(1)].
 \label{resi}
\eeq
In turn, and as also obtained directly from \eqref{vkih},
\beq
 \res\nolimits_{k=k_u}[\mcV(k,h_0)]\sim\e^{-k_u h_0}\bigl[2 k_u+O(k_u^{-1})\bigr]+ 
 O\bigl(\e^{-3k_u h_0}\bigr).
 \label{resih}
\eeq

These delicate consequences of matching exponentially varying terms verify a-posteriori the 
inclusion of the algebraically varying ones. \hl{The asymptotic behaviours \eqref{tkui} and 
\eqref{resih} are finally condensed into \eqref{tz}.} \\

\subsection{Eigenspace and poles}\label{ss:b3}

As an important aspect \hl{shown next}, for any \m{T\geq 0}, the homogeneous version of 
\eqref{vpr} is solvable only for a countable, infinite set of real eigenvalues of $k^2$ 
bounded from above.

We let $\bVk(y)$ symbolise the space of eigenfunctions,
\beq
 \bVk''=(k^2+\psi_0'''\!/\psi_0')\bVk,\q y=0\colon\;\; \bVk=0,\q y=h_0\colon\;\;
 \psi_0'^2\bVk'=T J k^2\bVk.
 \label{vk}
\eeq
In \S\,\ref{ss:b1}, we considered the eigenvalue \m{k=0} and the associated double pole. All 
other eigenvalues are expected to define the simple poles of $\mcV$ in the $k$-plane: an 
infinite number of conjugate imaginary and an, at most finite, number of real ones, these 
associated with isolated neutral capillary modes. To demonstrate these fundamental properties, 
we first \hl{consider two twice differentiable functions $\mcU(y)$, $\mcW(y)$ and the 
typical inner product \m{\int_0^{h_0}\mcU\,\bW\,\ud y}; in the following, overbars 
unambiguously indicate complex-conjugates. One readily confirms that the operator 
\m{(\ud^2\!/\ud y^2-\psi_0'''\!/\psi_0')\,\mcU} subject to \m{\mcU(0)=\mcU'(h_0)=0} is 
self-adjoint, but the appearance of the eigenvalue $k$ in the BC at \m{y=h_0} impedes proving 
its typically expected properties in standard fashion if \m{T>0}: $k^2$ is real; members of 
$\mcV_k$ for different eigenvalues are orthogonal with respect to the above inner product. 
Rather, if $\mcU$ and $\mcW$ now denote eigenfunctions for different eigenvalues $k_1$ and 
$k_2$, say, we obtain from \eqref{vk} via integration by parts}
\beq
 \hl{(k_1^2-\bk_2^2)
 \biggl[\fr{TJ(\mcU\,\bW)(h_0)}{\psi_0'^2(h_0)}-\int_0^{h_0}\!\mcU\,\bW\,\ud y\biggr]=0.}
 \label{ivk}
\eeq
\hl{This prompts us to seek a transformation of \eqref{vk}} such that $k$ no longer enters the 
BC for \m{y=h_0}. To this end, we introduce the transformed eigenfunctions 
\m{\mcFk\deq\psi_0'\mcV_k'-\psi_0''\mcV_k}. We then obtain from \eqref{vk}
\beq
 [\psi_0'^2(\bVk/\psi_0')']'\equiv\mcFk'=k^2\psi_0'\bVk
 \label{feq}
\eeq
and, since \m{\psi_0''(h_0)=0}, \m{\psi_0'^2\mcFk=T J\mcFk'} for \m{y=h_0}. Differentiation of 
\eqref{feq} after division by $\psi_0'^2$ casts \eqref{vk} into the form 
\beq
 (-\mcFk'/\psi_0'^2)'=-k^2\mcFk\!/\psi_0'^2,\q y=0\colon\;\; \mcFk=0,\q y=h_0\colon\;\; 
 \psi_0'^2\mcFk=T J\mcFk'.
 \label{sl}
\eeq

Adopting the signs in the usual notation, \eqref{sl} represents a traditional self-adjoint 
Sturm--Liouville eigenvalue problem with the (for \m{y\ttz} singular) weight function 
$\psi_0'^{-2}$ for the eigenvalues of \m{-k^2}. According to classical results, these indeed 
form a discrete set \m{k^2=k_i^2(T)} (\m{i=0,1,\ldots}) bounded from below and satisfying the 
Weyl asymptotics \m{-k_i^2\sim(\upi i/h_0)^2+O(i)} (\m{i\tti}), controlled by the right-side 
of the BC for \m{y=h_0} \citep[\cf][]{Te12}. Here \m{k_0^2=k_u^2>0}, referring to the single 
neutral mode considered in \S\,\ref{ss:b2}, so that $k_i^2$ is set to \m{-\mu_i^2(T)<0} for 
\m{i>0}. It is also noteworthy that this sequence $\mu_i$ does not collapse in the limiting 
case \m{T=1} as \m{\mu_1(1)\simeq 0.015569}.

\section{Outer Rayleigh problem: diffusive overlayer}\label{s:c}

Let us take $\psi$, $p$, $h_+$ as functions of $\bX$, $\xi$, $\ep$. We rectify the Maclaurin 
expansion of $\psi$ and $p$ for \m{\xi=O(1)} justified by \eqref{kbc}, \eqref{oexp}, 
\eqref{op} by adding an $O(\ep^{5/3})$-term (and resultant higher-order corrections) that
involves the $O(1)$-functions $\Psi^\ast$, $P^\ast$ so as to account for \eqref{dbct}:
\beq
 [\psi,p]\sim[1,P_0(\bX)]+\sum_{i=1}^\infty
 [\Psi_i,P_i](\bX;\ep)\,\fr{\ep^{i/2}\xi^i}{i!}+\ep^{5/3}[\Psi^\ast,P^\ast](\bX,\xi)+
 O(\ep^{12/7}).
 \label{mlr}
\eeq
The structure of \eqref{mlr} is explained in the following. 

The bounded coefficient functions $\Psi_i$, $P_i$ (\m{i\geq 0}) ensue from expanding 
\eqref{oexp}, \eqref{op} together with \eqref{psi0h} and the resultant property 
\m{\psi_0''''(h_0)=0} (see SBP18). One obtains
\bseq
\begin{align}
 \Psi_1 &\sim u_0^+ +\ep^{4/7}m[\psi_\ast-m\psi_0'''(h_0)A^2(0)/2]+
 \ep^{2/3}\,\bPsi_{\!y}(\bX,h_0) +O(\ep^{5/7}),
 \label{psi1}\\[1pt]  
 \Psi_2 &\sim \ep^{4/7}m[\psi_\ast''-\psi_\ast\psi_0'''/u_0^+](h_0)+
 \ep^{2/3}[\bpsi_{\!yy}(\bX,h_0)+\bH\psi_0'''(h_0)]+O(\ep^{5/7}),
 \label{psi2}\\[1pt]
 [P_0',P_1'] &\sim \ep^{2/3}[\bP_{\!\bX},\bP_{\!y\bX}](\bX,h_0)+O(\ep^{5/7}),\;\;\; 
 P_{1,2}=O(\ep^{4/7}),\;\;\; h_{+,\bX}=O(\ep^{3/7}).
 \label{pi}
\end{align}
 \label{qi}%
\eseq
Substituting \eqref{mlr} into the shear-layer approximation of \eqref{nsx},
\beq
 \psi_\xi\psi_{\xi\bX}-\psi_\bX\psi_{\xi\xi}\sim -\ep\,p_\bX-\ep^{1/2}h_{+,\bX}\,p_\xi+ 
 \ep^{1/2}\psi_{\xi\xi\xi}
 \label{nsxs}
\eeq
(the Prandtl shift preserves the convective operator), and collecting powers of $\xi$ yields a 
hierarchy of relations involving $\Psi_i$ and $P_i$. Insertion of \eqref{qi} into the first 
two,
\beq
 \Psi_1\Psi_{1,\bX}\sim -P_{0,\bX}-h_{+,\bX}P_1-\ep\,\Psi_3,\q 
 \Psi_1\Psi_{2,\bX}\sim -P_1'-h_{+,\bX}P_2-\ep\,\Psi_4,
 \label{psipi}
\eeq
just confirms the two-terms expansion of the streamwise momentum equation in \eqref{lee} for 
\m{\xi=O(1)}. On the other hand, the left-hand side of \eqref{dbct} reduces to
\m{\Psi_2-u_0^+ h_{+,\bX\!\bX}+O(\ep^{23/21})} within the accuracy provided by 
\eqref{oexp}, \eqref{op} and \eqref{qi}. Evaluating \eqref{dbct} by using \eqref{psi2} and 
\eqref{oexp} shows that this BC is satisfied up to $O(\ep^{4/7})$ once 
$m[\psi_\ast''-\psi_\ast\psi_0'''/u_0^+](h_0)+m l^2 u_0^+(G-P_-)=0$. This constraint for 
$\psi_\ast$ must already be provided by the surrounding shear layer addressed in SBP18. 
However, the follow-up contributions of $O(\ep^{2/3})$ to \eqref{dbct} yield in connection 
with \eqref{bh} the residual \m{(\bPsi_{\!yy}-\bPsi_{\!\bX\!\bX})(\bX,h_0)}. Compensating for 
this requires the perturbation stream function $\Psi^\ast$ to enter \eqref{mlr} at the same 
order of approximation as the $O(\ep^{2/3})$-contribution to $P_2$. In turn, \eqref{nsxs} 
subject to \eqref{kbc} and \eqref{dbct} yields with the aid of \eqref{qi} the diffusion problem
\beq
 u_0^+\Psi^\ast_{\xi\bX}=\Psi^\ast_{\xi\xi\xi},\q \xi=0\colon\;\;
 \Psi^\ast=0,\;\; \Psi^\ast_{\xi\xi}=\bPsi_{\!\bX\!\bX}-\bPsi_{\!yy},\q \xi\ttmi\colon\;\; 
 \Psi^\ast_{\xi\xi}\ttz.
 \label{prast}
\eeq
This also implies, for \m{\xi\ttmi}, a vanishing velocity perturbation $\Psi^\ast_\xi$ but 
finite viscous displacement exerted in the bulk flow, measured by $\Psi^\ast(\bX,\infty)$. The 
far-upstream and far-downstream asymptotes of $\Psi^\ast$ are found to be forced by the 
inhomogeneous BC. Therefore, $\Psi^\ast$ dies out exponentially for \m{\bX\ttmi} and grows 
algebraically for \m{\bX\tti}. We then use \eqref{bv}, \eqref{bvi} and \eqref{ba+} to describe 
the merge with the original overlayer (see SBP18) in this limit by
\m{\Psi^\ast\sim 27\la\psi_0'''(h_0)/(80M u_0^+)\bX^{11/3}F_\ast(\eta_\ast)+O(\bX^{5/3})}. 
Herein, the typical Rayleigh variable \m{\eta_\ast\deq\xi\sqrb{u_0^+/\bX}} is of $O(1)$ as 
$F_\ast$ satisfies 
\beq
 19 F_\ast'/6-\eta_\ast F_\ast''/2=F_\ast''',\q F_\ast(0)=0,\q F_\ast''(0)=1,\q 
 F_\ast''(-\infty)=0.
\eeq 
The solution to this problem can be expressed in terms of Kummer's confluent hypergeometric 
function, $M$:
\beq
 F_\ast(\eta_\ast)=
 (3/11)[M({\txt -\fr{11}{3}},{\txt\fr{1}{2}},-\eta_\ast^2/4)-1]-
 \eta_\ast M({\txt -\fr{19}{6}},{\txt\fr{3}{2}},-\eta_\ast^2/4)\big/\Ga({\txt\fr{25}{6}}).
 \label{fast}
\eeq
Far downstream, the viscous displacement is quantified by \m{F_\ast(-\infty)=-3/11}. Adopting 
the results of \S\,\ref{sss:ors}, one may expresses the full solution to \eqref{prast} as the 
Fourier integral
\beq
 \Psi^\ast=\int_\mcC \fr{[k^2\mcV+\mcV_{\!yy}](k,h_0)}{u_0^+ k^2}
 \Bigl(1-\e^{\sqrb{\ui u_0^+ k}\,\xi}\Bigr)\e^{\ui k\bX}\ud k.
 \label{psiast}
\eeq

A final remark on the higher-order corrections in \eqref{mlr} reinforces the self-consistency 
of the above flow description. The kinematic BC \eqref{kbc} induces an 
$O(\ep^{5/3})$-disturbance in the expansions of $h^+$ in \eqref{oexp} and thus the capillary 
pressure jump in \eqref{dbct} such one in \eqref{op}. This produces the non-zero $P^\ast$ in 
\eqref{mlr}. In addition, \eqref{nsy} gives \m{P^\ast_\xi\equiv 0}.

\section{Extended Jefferey--Hamel limit}\label{s:d}

We address briefly two formal aspects of the JH limit. 

At first, inspection of \eqref{vte} suggests the local expansion
\beq
 \bpsi\sim\bg(\vth)+\br^\si G(\vth)+o(\br^\si)+c.c.,\q \Real\,\si>0.
 \label{jhexp}
\eeq
Here $\si$ denotes the eigenvalue and $G$ the corresponding eigenfunction satisfying the 
resulting eigenvalue problem
\bseqa
 & \bigl[\si^2(\si-2)\bg'-\si\bg'''\bigr]G-2\bg''G'+(\si-2)\bg'G''= &
 \nonumber\\[1pt]
 & \bigl[(\si-2)^2+\ud^2\!/\ud\vth^2\bigr](\si^2 G+G''), &
 \slabel{geq}\\[1pt]
 & G(0)=G'(0)=G(\upi)=G'(\upi)=0. &
 \slabel{gbc}
 \label{gpr}
\eseqa
In \eqref{jhexp}, $\si$ then specifies the member of the discrete series of eigenvalues with
minimum positive real part. The validity of the JH solution and thus the local representation 
\eqref{jhexp} of the full NS solution depends on the existence of this value of~$\si$.

Secondly, we envisage $\bp$ and $\bh$, related via the dynamic BC in \eqref{bbc0}, near 
\m{\br=0}. The pressure gradient ensues from the momentum equations 
(\ref{bgov}\tit{a},\tit{b}) in the form  
\m{\bp_\br\sim(\bg'''+\bg'^2)/\br^3=[\bg'''(\upi)-4\bg']/\br^3}, where the last equality 
follows from \eqref{jh} upon integration, and \m{\bp_\vth\sim 2\bg''/\br^2}. Finally,
\beq
 \bp\sim\fr{4\bg'-\bg'''(\upi)}{2\br^2}+o(\br^{-2})\;\;\; (\br\ttz),\q
 \bh\sim\fr{\bg'''(\upi)}{2\tau}\ln{\bx}\;\;\; (\bx\ttz+)
 \label{bh0}
\eeq
with \m{g'''(\upi)\simeq 87.9545} (\m{\bg'''(\upi)\simeq 19.6983}) for the attached (detached) 
eddy and for any finite value of $\tau$. These singularities are much stronger than those 
found for the alternative, preferred Stokes limit elucidated \hl{in \S\,\ref{sss:aes} and 
below in \ref{s:e}}. Accordingly, their resolution would take place in a further NS region 
defined by the smallest scales, describing the microscopic resolution of the trailing edge.

\section{Extended Stokes limit}\label{s:e}

Separation of variables in \eqref{bps} yields
\beq
 \bpsi\sim\sum\nolimits_{i=0}^\infty \bpsi_i+O(\br^{\si_n+\si_q})+c.c.,\q
 \bpsi_i\deq\br^{\si_i}f_i(\vth),\q \Real\,\si_{i+1}>\Real\,\si_i>0.
 \label{fexp}
\eeq
Herein, $\si_i$ denotes the $i$-th eigenvalue, $f_i$ the corresponding eigenfunction of the 
azimuthal variation, and the remainder term arises from the dominant contribution to the 
quadratic inertial terms in \eqref{vte}, not captured by the Stokes balance and of 
$O(\br^{\si_n+\si_q-3})$. Therefore, $n$ and $q$ stand for the lowest indices $i$ such that 
$f_n$, $f_q$ are both non-trivial and their coupling produces a non-trivial inhomogeneity.

\subsection{Discussion of eigensolutions and their inertially induced response}

The expansion \eqref{fexp} casts the biharmonic problem into the series of eigenvalue problems
\bseqa
 & \mcS_i\{f_i\}=f_i(0)=f_i''(0)=f_i(\upi)=f_i'(\upi)=0, & 
 \slabel{evp}\\[2pt]
 & \mcS_i\{Q\}\deq
 \bigl[(\si_i-2)^2+\ud^2\!/\ud\vth^2\bigr]\bigl(\si_i^2+\ud^2\!/\ud\vth^2\bigr)\{Q\} &
 \slabel{so}
 \label{sevp}
\eseqa
for any function $Q$. The reduced Stokes operator $\mcS_i$ acting on $\vth$ and parametrised 
by the discrete eigenvalues is already seen in \eqref{jh} as this replaces \eqref{fexp} for 
\m{\si_i=0}. One readily finds that \eqref{sevp} has no solution in the degenerate cases 
\m{\si_i=1} and \m{\si_i=2}. In any other case, the first two BCs in \eqref{evp} yield 
\bseq
\beq
 f_i=a_i\sin(\si_i\vth)+b_i\sin[(\si_i-2)\vth],
 \label{fi}
\eeq
where the constants $a_i$ and $b_i$ are functions of $\Lao$, are determined by the global 
solution to \eqref{bgov} and must not all be zero. Notably, the $\sin(\si_i\vth)$-term refers 
to a potential-flow contribution. The eigenvalue relation \m{\sin(2\si_i\upi)=0} equivalent to 
the last two BCs implies
\beq
 \si_i=(1+i)/2.
 \label{si} 
\eeq
In turn, \eqref{fi} holds for some real $a_i$ and $b_i$ satisfying
\beq
 b_i=-a_i\;\;\; (i=0,2,4,\ldots),\q (2-\si_i)b_i=\si_i a_i\;\;\; (i=1,3,5,\ldots),
 \label{abi}
\eeq
 \label{sfabi}%
\eseq
which confirms that \m{f_1\equiv f_3\equiv 0} and \m{a_2 f_0\equiv a_0 f_2} and, a-posteriori, 
the validity of \eqref{fexp}. \hl{Here we refer to the subsequent discussion in 
\S\,\ref{ss:e2}.} As $\si_i$ take on integer values for $i$ being odd, it is readily seen that 
exactly this category refers to regular eigensolutions $\bpsi_i$. Their series, ordered by 
ascending integer powers in $\bx$ and $\by$,
\beq
 \bpsi_5=-4a_5\by^3,\; \bpsi_7=-8a_7\bx\by^3,\; 
 \psi_9=40a_9\by^3(\by^2\!/15-\bx^2\!/3),\;\ldots,
 \label{sxy}
\eeq
ensues systematically from expressing $\bDe^2$ accordingly and using the BCs. We also infer 
from \eqref{bps}, \eqref{sevp} and \eqref{fi} that
\beq
 \bp-\bp_0\sim\sum\nolimits_{i=0,\,i\neq 3}^\infty 
 \br^{\si_i-2}p_i(\vth)+O\bigl(\br^{\si_n+\si_q-2}\bigr),\;\;\;
 p_i=4b_i(\si_i-1)\cos[(\si_i-2)\vth].
 \label{pexp}
\eeq
The constant $\bp_0$ is again to be extracted from the complete NS solution.

The symmetry of $f_i$ in $i$ around \m{i=1} and its azimuthal symmetry/antisymmetry with 
respect to \m{\vth=\upi/2} for odd/even values of $\si_i$ (\ie odd values of $i$) deserve a 
comment. The first eight members of the series $f_i$ with $a_i$ set to unity are plotted in 
figure~\ref{f:fi}. One then finds that \m{f_0\geq 0}, \m{f_5\leq 0}, and $f_i(\vth)$ changes 
its sign $(i-2)/2$ times if \m{i=4,6,\ldots} and \m{(i-5)/2} times if \m{i=7,9,\ldots} over 
the interval $(0,\upi)$. Therefore, exactly three alternatives can accommodate the 
forward-flow condition \eqref{ff}:
\benum
 \item[(A)] \m{a_0>0};
 \item[(B)] \m{a_0=0} and \m{a_2>0};
 \item[(C)] \m{a_0=a_2=a_4=0} and \m{a_5<0}.
\eenum

\bfig
 \centering
 \scalebox{0.7}{\input{fi.pdf_t}}
 \caption{Eigenfunctions $f_i$ of Stokes operator and their symmetry properties, see 
  \eqref{sfabi}: \m{a_i=1}, labels indicate \m{i=0,2,4,6,8} (solid), \m{i=5,7,9} (dashed).}
 \label{f:fi} 
\efig

We \hl{now proceed to demonstrate that case (C) is the appropriate choice. To this end,} we
conveniently restate \eqref{fexp} in general more precisely as
\beq
 \bpsi\sim\sum\nolimits_{i=0}^\infty \br^{\si_i}g_i(\vth),\q g_0\deq f_0,
 \label{gexp}
\eeq
where the functions $g_i$ represent the solutions to the hierarchy of inhomogeneous Stokes 
problems provoked by the inertia terms in \eqref{vte}, which cause the remainder term in 
\eqref{fexp}. As we now demonstrate, the forcing of these eigensolutions of the Stokes 
operator by the higher-order, convective terms controls the selection of the leading non-zero 
coefficient $a_i$ of the homogeneous contribution $f_i$ to $g_i$. Substituting \eqref{gexp} 
into \eqref{vte} and collecting powers of $\br$ results in the inhomogeneous extension of 
\eqref{sevp} for \m{i>0}:
\bseqa
 &\disp \mcS_i\{g_i\}=
 I_i(\vth)\deq\sum\nolimits_{j=0}^{i-1} I_{i,j}(\vth), &
 \slabel{gieq}\\[2pt]
 & I_{i,j}(\vth)\deq\bigl[(\si_j-2)g_k'-\si_k g_k\,\ud/\ud\vth\bigr](\si_j^2 g_j+g_j''),\q
 k\deq i-j-1, &
 \slabel{giin}\\[6pt]
 & g_i(0)=g_i''(0)=g_i(\upi)=g_i'(\upi)=0, &
 \slabel{gibc}
 \label{gipr}
\eseqa
where \eqref{giin} is consistent with the identity \m{\si_j+\si_k\equiv\si_{j+k+1}}, see 
\eqref{si}. The self-adjointness of the homogeneous Stokes operator defined by \eqref{sevp} 
gives
\beq
 0=\int_0^\uppi \mcS_i\{g_i\}f_i(\vth)\,\ud\vth=
 S_i\deq\int_0^\uppi I_i(\vth)f_i(\vth)\,\ud\vth.
 \label{sc}
\eeq
This describes the well-known three alternatives: the solution of \eqref{gipr} is unique if 
\m{f_i\equiv 0}; it is non-unique if $f_i$ is non-trivial and \m{S_i=0}; it does not exist 
otherwise. Thus \eqref{sc} establishes the following analysis of the possible cases concerning 
the solvability of~\eqref{gipr}.
\benum
 \item[\tit{Case:} \m{a_0\neq 0}.] In this least-degenerate scenario including case (A) above, 
\eqref{gexp} is specified as \m{\bpsi\sim\br^{1/2}f_0+\br g_1+\br^{3/2}g_2+O(\br^2)}; 
\eqref{sfabi} \hl{implies} $f_0=a_0[\sin(\vth/2)+\sin(3\vth/2)]$ and \eqref{giin} 
\m{I_1=3 a_0^2[\sin(2\vth)/2+\sin(3\vth)]}. Since \m{f_1\equiv 0}, we construct for \m{i=1} 
the unique solution \m{g_1=a_0^2[37\sin(\vth)+32\sin(2\vth)+9\sin(3\vth)]/192} of 
\eqref{gipr}. In turn, we specify \eqref{gipr} and \eqref{sc} for \m{i=2}. A tedious but 
straightforward calculation involving $g_1$ and \m{f_2\equiv a_2 f_0/a_0} yields 
\m{S_2=-25\upi a_0^2 a_2/128} so that $g_2$ does not exist. We are thus left with the 
following less singular situation.
 \item[\tit{Case:} \m{a_i=0} (\m{0\leq i<n}), \m{a_n\neq 0}.] Here $n$ identifies the index of 
the dominant non-trivial eigensolution so that \m{g_n=f_n}. Accordingly, \eqref{giin} produces 
non-trivial $I_{i,j}$ for \m{j,k\geq n} only. This confirms that \m{I_i\equiv 0} 
(\m{0\leq i\leq 2n}) as the self-coupling of $f_n$ yields the potential lowest-order 
inhomogeneity
\beq
 I_{2n+1}=I_{2n+1,n}=
 -4b_n(\si_n-1)(\si_n-2)\bigl\{a_n\si_n\sin(2\vth)+b_n\sin[(2\si_n-4)\vth]\bigr\},
 \label{i2n1}
\eeq
which corresponds to the case \m{n=q} in \eqref{fexp}. It is emphasised that $I_{2n+1}$ 
vanishes identically only for \m{n=1} (\m{\si_1=1}), \m{n=3} (\m{\si_3=2}) and \m{n=5} 
(\m{\si_5=3}). Furthermore, \eqref{sc} gives after some standard manipulations, involving 
$\si_n$ and and $\si_{2n+1}$ specified by \eqref{si},
\beq
 S_{2n+1}=\lb\{\!\barray{ll} 0 & (n\neq 2), \\[2pt] 3\upi a_2^2 a_5/2 & (n=2). \earray\rb.
 \label{s2n1}
\eeq
The last statement requires \m{a_2=0}. This renders the forward-flow case (B) also not 
possible. Hence, the scenario (C) motivates the following discussion of the special case 
\m{n=5}.
 \item[\tit{Case:} \m{a_i=0} (\m{0\leq i<5}), \m{a_5\neq 0}.] The result \eqref{s2n1} includes 
that the here dominant eigensolution of the Stokes operator $\bpsi_5$ given by \eqref{sxy}, 
describing a non-degenerate flow profile at separation, trivially generates a vanishing 
inhomogeneity $I_{11}$. That said, \eqref{gexp} then degenerates and reads more accurately, 
with the help of \eqref{si},
\beq
 \bpsi\sim\bpsi_5+\sum\nolimits_{i=q}^\infty \br^{(1+i)/2}f_i+\br^{(7+q)/2}g_{6+q}+ 
 o\bigl(\br^{(7+q)/2}\bigr)\q (q>5,\;\; a_q\neq 0).
 \label{gexp5}
\eeq
At first, any index \m{q>5} is conceivable. If \eqref{gexp5} initiates the solution to the 
full NS problem, such an index indicating the non-trivial follow-up term to $\br^3 f_5$ must 
exist. As a central observation, the lowest-order inhomogeneity in \eqref{gieq} specified by 
\eqref{giin} is \m{I_{6+q}=I_{6+q,5}+I_{6+q,q}} and produces $g_{6+q}$, where the 
eigenfunction $f_{6+q}$ corresponds to to the eigenvalue \m{\si_{q+6}=(7+q)/2}. Inserting 
these findings into \eqref{sc} yields indeed \m{S_{6+q}\equiv 0} as for \eqref{s2n1} but for 
any $q$, where we skip the technical details. This guarantees the existences of $g_{q+6}$ and, 
in turn, of~\eqref{gexp5}.
\eenum

As an important step, the above analysis determines the least singular (most-degenerate) local 
representation of $\bpsi$ to be given by case (C) extended by \eqref{gexp5} as the sole 
reliable option satisfying~\eqref{ff}. It should be emphasised that the impact of the 
interactive flow on flow detachment on the NS scales is condensed into the aforementioned 
(pending) dependence of the coefficient $a_5$ on~$\Lao$.

The shear rate at the plate immediately upstream of detachment observed on the global scale 
reads \m{u_y|_{y=0}\sim\bpsi_{\vth\vth}(\br,\upi)/\br^2}, according to \eqref{bexp} and 
\eqref{bbcpi}. Since \eqref{sfabi} entails \m{f_i''(\upi)=0} for odd $i$, \cf \eqref{sxy}, and 
\m{f_i''(\upi)=2a_i(-1)^{i/2}(1-i)} for even $i$, it is dominantly fixed either by the 
eigenfunction $f_j$ of the smallest even index, \m{j\geq 6}, that enters \eqref{gexp5} or 
$g_{6+q}$. With dots abbreviating smaller terms, that shear rate tends to zero in the form
\beq
 u_y|_{y=0}\sim 
 2a_j(-1)^{j/2}(1-j)(-\bx)^{(j-1)/2}+(-\bx)^{(3+q)/2}g''_{6+q}(\upi)+\cdots\;\;\; (\bx\ttz-).
 \label{uy0}
\eeq
Since \eqref{ff} requires \m{u_y|_{y=0}>0} here, we expect either \m{a_j>0} (\m{a_j<0}) for 
\m{j=6,10,14,\ldots} (\m{j=8,12,16,\ldots}) or $f_{6+q}''(\upi)$ behaves such that 
\m{g_{6+q}''(\upi)>0}. Typically, the adverse pressure gradient predicted by \eqref{pexp} in 
the form \m{\bp-\bp_0\sim -24a_5\bx+O\bigl(\br^{(q-3)/2}\bigr)} initiates flow detachment. 
\hl{Finally, this together with \eqref{gexp5} and \eqref{bbc0} results in \eqref{bphexp}.}

\subsection{Eigensolutions of the Stokes operator having weakly non-algebraic radial 
            variation?}\label{ss:e2}

Given the absence of a reference length and velocity of the Stokes limit considered, typical
dimensional reasoning predicts, in general, algebraic--logarithmic variations of the gauge 
functions in \eqref{fexp} with $\br$. Nonetheless, the following analysis confirms that 
factors with sub-algebraic (logarithmic) dependence on $\br$ do indeed not contribute to 
\eqref{fexp}.

Seeking eigensolutions of the biharmonic operator in the limit \m{\br\ttz} first leaves one 
with a generalisation of the expansion \eqref{fexp} into eigenfunctions:
\beq
 \bpsi\sim\sum\nolimits_{i=j=0}^\infty \br^{\si_i}\chi_{i,j}(\br)f_{i,j}(\vth),\q 
 \chi_{i,j+1}=o(\chi_{i,j}).
 \label{fjexp}
\eeq
Herein, $f_{i,j}$ indicates the double series of eigenfunctions, with \m{f_{i,0}=f_i} as we 
found so far, due to the corresponding sought gauge functions $\chi_{i,j}$ exhibiting 
sub-algebraic variation, including the previous situation \m{\chi_{i,0}\equiv 1} and 
\m{f_{i,j}\equiv\chi_{i,j}\equiv 0} for \m{j>0}. Following the analysis of \cite{Sc14} of the 
Laplace operator, for any such function \m{\Xi_{i,j}\deq\br\chi_{i,j}'} is of $o(\chi_{i,j})$ 
and again belongs to this family of functions. With this relation in mind, we obtain after 
some rearrangements
\beq
 \bDe^2(\br^{\si_i}\chi_{i,j}f_{i,j})\sim\Xi_{i,j}J_{i,j}(\vth)+o(\Xi_{i,j}),\;\;\;
 J_{i,j}(\vth)\deq 4(\si_i-1)[f_{i,j}''+\si_i(\si_i-2)f_{i,j}]
 \label{fij}
\eeq 
(agreeing with the symmetry of $f_{i,j}$ in $\si_i$ with respect to \m{\si_i=1}). In turn, we 
specify \m{\chi_{i,1}=-\Xi_{i,0}}. If $\chi_{i,1}$ does not vanish identically, the 
homogeneous problem determining $f_i$ yields the inhomogeneous follow-up problem fixing 
$f_{i,1}$ according to \eqref{sevp}
\beq
 \mcS_i\{f_{i,1}\}=J_{i,0}(\vth),\q f_{i,1}(0)=f_{i,1}''(0)=f_{i,1}(\upi)=f_{i,1}'(\upi)=0.
 \label{f1pr}
\eeq
Since the homogeneous operator in \eqref{f1pr} and defined by \eqref{sevp} is self-adjoint,
\beq
 0=\int_0^\uppi \mcS_i\{f_{i,1}\}f_i(\vth)\,\ud\vth=
 \int_0^\uppi J_{i,0}(\vth)f_i(\vth)\,\ud\vth=8\upi(-1)^i(\si_i-1)a_i b_i
 \label{scf}
\eeq
with the aid of \eqref{sfabi}. This contradiction implies \m{\chi_{i,0}\equiv 1}, 
\m{\chi_{i,1}\equiv 0} and, by iteration, \m{\chi_{i,j}\equiv 0} for all \m{j>0}. 
Consequently, the appearance of sub-algebraic factors in \eqref{fexp} is indeed ruled out.

\end{document}